\documentclass[twocolumn]{aastex61}
\usepackage[utf8]{inputenc}
\usepackage{hyperref}
\usepackage{rotating}
\usepackage{xcolor}

\begin{document}

\title{The Art of Measuring Physical Parameters in Galaxies: A Critical Assessment of Spectral Energy Distribution Fitting Techniques}

\author{Camilla Pacifici}\affil{Space Telescope Science Institute, 3700 San Martin Drive, Baltimore, MD 21218, USA}
\author{Kartheik G. Iyer}\affil{NASA Hubble Fellow}\affil{Dunlap Institute for Astronomy and Astrophysics, University of Toronto, 50 St George St, Toronto, ON M5S 3H4, Canada}\affil{Columbia Astrophysics Laboratory, Columbia University, 550 West 120th Street, New York, NY 10027, USA}
\author{Bahram Mobasher}\affil{Department of Physics and Astronomy, University of California, Riverside, 900 University Avenue, Riverside, CA 92521,
USA}
\author{Elisabete da Cunha}\affil{International Centre for Radio Astronomy Research, University of Western Australia, 35 Stirling Hwy, Crawley, WA 6009, Australia}\affil{Research School of Astronomy and Astrophysics, Australian National University, Canberra, ACT 2611, Australia}\affil{ARC Centre of Excellence for All Sky Astrophysics in 3 Dimensions (ASTRO 3D)}

\author{Viviana Acquaviva}\affil{Physics Department, CUNY NYC College of Technology, Brooklyn, NY 11201, USA}\affil{Center for Computational Astrophysics, Flatiron Institute, New York, NY 10010, USA}
\author{Denis Burgarella}\affil{Aix Marseille Univ, CNRS, CNES, LAM, Marseille, France}
\author{Gabriela Calistro Rivera}\affil{European Southern Observatory, Karl-Schwarzschild-Str. 2, D-85748, Garching, Germany}
\author{Adam C. Carnall}\affil{SUPA, Institute for Astronomy, University of Edinburgh, Royal Observatory, Edinburgh EH9 3HJ, UK}
\author{Yu-Yen Chang}\affil{Department of Physics, National Chung Hsing University, 40227, Taichung, Taiwan}\affil{Institute of Astronomy \& Astrophysics, Academia Sinica, Taipei, 10617, Taiwan}
\author{Nima Chartab}\affil{Department of Physics and Astronomy, University of California, Riverside, 900 University Avenue, Riverside, CA 92521,
USA}
\author{Kevin C. Cooke}\affil{AAAS S\&T Policy Fellow hosted at the National Science Foundation, 1200 New York Ave, NW, Washington, DC, US 20005}
\author{Ciaran Fairhurst}\affil{Astronomy Centre, Department of Physics and Astronomy, University of Sussex, Brighton, BN1 9QH, UK}
\author{Jeyhan Kartaltepe}\affil{School of Physics and Astronomy, Rochester Institute of Technology, Rochester, NY 14623, USA}
\author{Joel Leja}\affil{Department of Astronomy \& Astrophysics, The Pennsylvania State University, University Park, PA 16802, USA}\affil{Institute for Computational \& Data Sciences, The Pennsylvania State University, University Park, PA, USA}\affil{Institute for Gravitation and the Cosmos, The Pennsylvania State University, University Park, PA 16802, USA}
\author{Katarzyna Ma\l{}ek}\affil{National Centre for Nuclear Research, ul. Pasteura 7, 02-093 Warszawa, Poland}\affil{Aix Marseille Univ, CNRS, CNES, LAM, Marseille, France}
\author{Brett Salmon}\affil{Space Telescope Science Institute, 3700 San Martin Drive, Baltimore, MD 21218, USA}
\author{Marianna Torelli}\affil{INAF - Osservatorio Astronomico di Roma, via Frascati 33, I-00078 Monteporzio Catone, Italy}
\author{Alba Vidal-Garc\'ia}\affil{Observatorio Astr\'onmico Nacional (IGN), C. Alfonso XII, 3, 28014 Madrid, Spain}\affil{Laboratoire de Physique de l'\'Ecole Normale Sup\'erieure, ENS, Universit\'e PSL, CNRS, Sorbonne Universit\'e, Universit\'e de Paris, 75005, Paris, France}
\author{M\'ed\'eric Boquien}\affil{Centro de Astronom\'ia (CITEVA), Universidad de Antofagasta, Avenida Angamos 601, Antofagasta, Chile}
\author{Gabriel G. Brammer}\affil{Niels Bohr Institute, University of Copenhagen, Jagtvej 128, K\o{o}benhavn N, DK-2200, Denmark}\affil{Cosmic Dawn Center (DAWN)}
\author{Michael J. I. Brown}\affil{School of Physics and Astronomy, Monash University, Clayton, VIC 3800, Australia}
\author{Peter L. Capak}\affil{Cosmic Dawn Center (DAWN), Denmark}
\author{Jacopo Chevallard}\affil{Sorbonne Universit\'e, CNRS, UMR7095, Institut d'Astrophysique de Paris, F-75014, Paris, France}
\author{Chiara Circosta}\affil{Department of Physics \& Astronomy, University College London, Gower Street, London, WC1E 6BT, UK}
\author{Darren Croton}\affil{Centre for Astrophysics \& Supercomputing, Swinburne University of Technology, Hawthorn, VIC 3122, Australia}\affil{ARC Centre of Excellence for All Sky Astrophysics in 3 Dimensions (ASTRO 3D)}
\author{Iary Davidzon}\affil{Cosmic Dawn Center (DAWN), Denmark}\affil{Niels Bohr Institute, University of Copenhagen, Jagtvej 128, DK-2200 Copenhagen, Denmark}
\author{Mark Dickinson}\affil{Community Science and Data Center/NSF's NOIRLab, 950 N. Cherry Ave., Tucson, AZ 85719, USA}
\author{Kenneth J. Duncan}\affil{SUPA, Institute for Astronomy, University of Edinburgh, Royal Observatory, Edinburgh EH9 3HJ, UK}
\author{Sandra M. Faber}\affil{UCO/Lick Observatory, Department of Astronomy and Astrophysics, University of California, Santa Cruz, CA, USA}
\author{Harry C. Ferguson}\affil{Space Telescope Science Institute, 3700 San Martin Drive, Baltimore, MD 21218, USA}
\author{Adriano Fontana}\affil{INAF - Osservatorio Astronomico di Roma, via Frascati 33, I-00078 Monteporzio Catone, Italy}
\author{Yicheng Guo}\affil{Department of Physics and Astronomy, University of Missouri, Columbia, MO 65211, USA}
\author{Boris Haeussler}\affil{European Southern Observatory, Alonso de Cordova 3107, Vitacura, Santiago, Chile}
\author{Shoubaneh Hemmati}\affil{IPAC, California Institute of Technology, 1200 E California Blvd, Pasadena, CA 91125, USA}
\author{Marziye Jafariyazani}\affil{IPAC/Caltech, 1200 E. California Blvd. Pasadena, CA 91125, USA}\affil{Department of Physics and Astronomy, University of California, Riverside, 900 University Avenue, Riverside, CA 92521, USA}
\author{Susan A. Kassin}\affil{Space Telescope Science Institute, 3700 San Martin Drive, Baltimore, MD 21218, USA}
\author{Rebecca L. Larson}\affil{The University of Texas at Austin, Department of Astronomy, Austin, TX, United States}
\author{Bomee Lee}\affil{Korea Astronomy and Space Science Institute, Daedeokdae-ro 776, Yuseong-gu, Daejeon 34055, Republic of Korea}
\author{Kameswara Bharadwaj Mantha}\affil{Minnesota Institute for Astrophysics, University of Minnesota, 116 Church St SE, Minneapolis, MN 55455, USA}\affil{School of Physics and Astronomy, University of Minnesota, 116 Church St SE, Minneapolis, MN 55455, USA}
\author{Francesca Marchi}\affil{INAF - Osservatorio Astronomico di Roma, Via di Frascati 33, 00078, Monte Porzio Catone, Italy}
\author{Hooshang Nayyeri}\affil{Department of Physics and Astronomy, University of California, Irvine, CA92697, USA}
\author{Jeffrey A. Newman}\affil{Department of Physics and Astronomy, University of Pittsburgh, Pittsburgh, PA 15260, USA}\affil{Pittsburgh Particle Physics, Astrophysics, and Cosmology Center (PITT PACC), University of Pittsburgh, Pittsburgh, PA 15260, USA}
\author{Viraj Pandya}\affil{UCO/Lick Observatory, Department of Astronomy and Astrophysics, University of California, Santa Cruz, CA, USA}\affil{Department of Astronomy, Columbia University, 550 West 120th Street, New York, NY 10027, USA}\affil{NASA Hubble Fellow}
\author{Janine Pforr}\affil{Scientific Support Office, Directorate of Science and Robotic Exploration, European Space Research and Technology Centre (ESA/ESTEC), Keplerlaan 1, 2201 AZ Noordwijk, The Netherlands}
\author{Naveen Reddy}\affil{Department of Physics and Astronomy, University of California, Riverside, 900 University Avenue, Riverside, CA 92521, USA}
\author{Ryan Sanders}\affil{Department of Physics, University of California, Davis, One Shields Ave, Davis, CA 95616, USA}\affil{NASA Hubble fellow}
\author{Ekta Shah}\affil{Department of Physics, University of California, Davis, One Shields Ave, Davis, CA 95616, USA}\affil{School of Physics and Astronomy, Rochester Institute of Technology, 84 Lomb Memorial Drive, Rochester NY 14623, USA}\affil{LSSTC DSFP Fellow}
\author{Abtin Shahidi}\affil{Department of Physics and Astronomy, University of California, Riverside, 900 University Avenue, Riverside, CA 92521, USA}
\author{Matthew L. Stevans}\affil{Department of Astronomy, University of Texas at Austin, Austin, TX 78712, USA}
\author{Dian Puspita Triani}\affil{Centre for Astrophysics \& Supercomputing, Swinburne University of Technology, Hawthorn, VIC 3122, Australia}\affil{ARC Centre of Excellence for All Sky Astrophysics in 3 Dimensions (ASTRO 3D)}\affil{Research School of Astronomy and Astrophysics, Australian National University, Weston Creek, ACT 2611, Australia}
\author{Krystal D. Tyler}\affil{School of Physics and Astronomy, Rochester Institute of Technology, Rochester, NY 14623, USA}
\author{Brittany N. Vanderhoof}\affil{School of Physics and Astronomy, Rochester Institute of Technology, Rochester, NY 14623, USA}
\author{Alexander de la Vega}\affil{Department of Physics and Astronomy, Johns Hopkins University, Bloomberg Center, 3400 N. Charles St., Baltimore, MD 21218, USA}
\author{Weichen Wang}\affil{Department of Physics and Astronomy, Johns Hopkins University, Bloomberg Center, 3400 N. Charles St., Baltimore, MD 21218, USA}
\author{Madalyn E. Weston}\affil{Department of Physics and Astronomy, University of Missouri-Kansas City, Kansas City, MO 64110, USA}

\begin{abstract}
The study of galaxy evolution hinges on our ability to interpret multi-wavelength galaxy observations in terms of their physical properties. To do this, we rely on spectral energy distribution (SED) models which allow us to infer physical parameters from spectrophotometric data. In recent years, thanks to the wide and deep multi-waveband galaxy surveys, the volume of high quality data have significantly increased. Alongside the increased data, algorithms performing SED fitting have improved, including better modeling prescriptions, newer templates, and more extensive sampling in wavelength space. We present a comprehensive analysis of different SED fitting codes including their methods and output with the aim of measuring the uncertainties caused by the modeling assumptions. We apply fourteen of the most commonly used SED fitting codes on samples from the CANDELS photometric catalogs at $z\sim1$ and $z\sim3$. We find agreement on the stellar mass, while we observe some discrepancies in the star formation rate (SFR) and dust attenuation results. To explore the differences and biases among the codes, we explore the impact of the various modeling assumptions as they are set in the codes (e.g., star formation histories, nebular, dust, and AGN models) on the derived stellar masses, SFRs, and $A_V$ values. We then assess the difference among the codes on the SFR-stellar mass relation and we measure the contribution to the uncertainties by the modeling choices (i.e., the modeling uncertainties) in stellar mass ($\sim0.1$~dex), SFR ($\sim0.3$~dex), and dust attenuation ($\sim0.3$~mag). Finally, we present some resources summarizing best practices in SED fitting. 
\end{abstract}

\section{Introduction}
Recent ground and space based galaxy surveys have generated rich data products  that shed light on the physics of nearby and distant galaxies. Photometric surveys generally provide a large wavelength baseline at coarse resolution, e.g., COSMOS \citep{scoville2007,weaver2022}, CANDELS \citep{grogin2011,koekemoer2011}, KINGFISH \citep{kennicutt2011}, UltraVISTA \citep{mccracken2012,muzzin2013}, and ZFOURGE \citep{straatman2016}. Spectroscopic surveys on the contrary provide generally a smaller wavelength range at higher resolution, e.g., SDSS \citep{york2000}, GAMA \citep{driver2011}, AGES \citep{kochanek2012}, 3D-HST \citep{brammer2012,momcheva2016}, DEEP2 \citep{newman2013}, MOSDEF \citep{kriek2015}, LEGA-C \citep{vanderwel2016}, VANDELS \citep{mclure2018,pentericci2018}, and VIPERS \citep{scodeggio2018}. Some spectroscopic surveys are also accompanied by photometric catalogs using ancillary data and dedicated campaigns, e.g. SDSS, GAMA, and 3D-HST. While each of these surveys has been designed with specific science goals in mind, they share the same general set of tools necessary to translate observational measurements into physical parameters. Historically, single colors or luminosities were directly translated to physical parameters (e.g.. \citealt{kennicutt1989}, \citealt{kennicutt1998}, \cite{bell2001}, \citealt{rujopakarn2013}, \cite{du2020}). When more data are available, the mapping from multiple colors to physical parameters is not as obvious. Full spectral energy distribution  (SED) modelling, being physically motivated, is more widely applicable and does not rely on empirical calibrations that may only be valid for specific samples of galaxies.

SED fitting (see review by \citealt{conroy2013}) consists of comparing observed SEDs of individual galaxies to existing templates based on various models to ultimately estimate the physical properties of such galaxies. This method has many advantages: it combines all information in a consistent way; it gives the flexibility to adapt the modeling to the data type; and it allows for customization of the ingredients that go into the creation of the model templates. SED fitting has also some limitations: it relies on assumptions that can be well established or still relatively unknown; and thus requires careful assessment of such assumptions (see e.g., the introduction by \citealt{chevallard2016} on dust attenuation). To mitigate these limitations, we need to measure and account for the modeling uncertainties when deriving galaxy physical properties. Existing work in this area consists in varying modeling assumptions using individual codes and fitting simulated data. The conclusions of such studies are however hard to generalize because the results are normalized to a specific set of assumptions (i.e., the parameters assumed in the simulations). A systematic code comparison based on a real dataset is instead not dependent on a specific set of assumptions and thus its results can quantify the modeling uncertainties without any specific normalization. This is the aim of this work.

SED fitting is used widely in the literature. In its early days, the targeted galaxies were at fairly low redshift and generally very luminous. Simple assumptions were enough to model the few datapoints acquired (e.g., models based on the evolutionary population synthesis technique: \citealt{spinrad1969}, \citealt{bruzual1983},  \citealt{guiderdoni1987}, \citealt{worthey1994}, \citealt{bruzual1993}, \citealt{leitherer1995}, \citealt{maraston1998}, \citealt{vazdekis1999}). Nowadays, data have become much more complex, sampling across redshift at high signal-to-noise ratios, for both massive and dwarf galaxies, at both integrated and spatially resolved scales. The spanned wavelength ranges have also increased with datasets that can cover the whole space from the far ultraviolet (FUV) to the fir infrared (FIR) and beyond. These expansions in the datasets require improvements in the tools used to translate data points to physical parameters. 

At the time of writing, there are many state-of-the-art SED fitting tools. To give some examples among the most recently developed, we cite AGNfitter \citep{calistrorivera2016}, BEAGLE \citep{chevallard2016}, Pipe3D \citep{sanchez2016}, Prospector \citep{leja2017,johnson2021}, Dense Basis \citep{iyer2017,iyer2019}, FIREFLY \citep{wilkinson2017}, Lightning \citep{eufrasio2017}, Mr-Moose \citep{drouart2018}, BAGPIPES \citep{carnall2018}, FortesFit \citep{rosario2019}, PEGASE.3 \citep{fioc2019}, X-CIGALE \citep{yang2020}, MCSED \citep{bowman2020}, MIRKWOOD \citep{gilda2021}, piXedfit \citep{abdurrouf2021}, ProSpect \citep{thorne2021}, and Starduster \citep{qui2021}. Machine learning techniques are also advancing rapidly to provide redshift and physical parameter estimates for galaxies across a large redshift range, e.g. \cite{davidzon2019,simet2021}. All the above packages aim at covering the large parameter space spanned by multiwavelength observations of galaxies.

In order to cover a large parameter space, the SED fitting tools need to include a certain number of ingredients with enough freedom to modify the modeling assumptions (see Figure 1 and Table 2 in \citealt{thorne2021}). A first basic ingredient is the assumption of star formation and eventual metal enrichment histories (SFH and MEH). The most common ways to model the histories of galaxy growth are analytic and non-parametric functions (e.g. \citealt{carnall2019a}; \citealt{leja2019}). \cite{lower2020} provide a comprehensive introduction on the various SFH models used in SED fitting (see also \citealt{lee2018}). Along with the SFHs, accounting for metal formation histories is important, especially when interpreting photometric datasets spanning a large wavelength range or high resolution spectroscopic datasets. For example, \cite{bellstedt2020,bellstedt2021} show that SFHs derived from SED fits that account for close-box metal enrichment can better match the cosmic star formation history \citep{madau2014} compared to SFHs derived from SED fits with constant metallicity. 

In addition to SFH and MEH components, simple stellar population (SSP) and nebular emission models are required to build the final model spectra. The papers by \cite{baldwin2018} and \cite{han2019} are examples of comparisons of different SSP models when fitting spectra of nearby galaxies. Additionally, the importance of including nebular emission in the spectra is highlighted by \cite{pacifici2012, pacifici2015, smit2014, salmon2015} among others. Dust attenuation and emission are ingredients necessary to interpret the SEDs of galaxies from UV to NIR (near IR) and to FIR, respectively (see review by \citealt{salim2020}). Multiple studies point to the need for flexible attenuation curves with varying parameters (e.g., \citealt{kriek2013,wang2018,buat2018,salim2018,barisic2020}), while differences among IR models are reported by e.g., \cite{hunt2019}. 

Finally, some SED fitting tools also account for the contribution of a potential Active Galactic Nucleus (AGN), to model the UV-to-IR emission and especially the mid-IR SED (e.g., \citealt{calistrorivera2016} among others). The Inter Galactic Medium (IGM) absorption on the bluest side of the SED is also very important especially for high redshift sources (e.g., the Lyman-break or \textit{dropout} technique to detect galaxies; \citealt{steidel1996}).

With all the SED fitting codes available, each with differences, it becomes difficult to compare results that use different codes, and also to quantify how much spread in physical parameters is introduced by the use of different SED fitting approaches. All the studies mentioned above dive into the differences between specific models, generally focusing on a single ingredient. 

Here, we provide a comprehensive comparative analysis of the output of fourteen SED-fitting codes applied to the same observational photometric dataset. We do not focus on any specific modeling ingredient and we assume that all modeling choices are reasonable and appropriate for the dataset in question.\footnote{We note that some codes make similar assumptions and thus they can be biased in similar ways. In the same way, not all possible choices are explored in this work.} We thus can measure the contribution of the modeling assumptions to the uncertainty in the physical parameters. Knowing this contribution to the uncertainty is necessary when comparing results derived using different modeling approaches (see \citealt{goddard2017}) and when comparing results to predictions from cosmological simulations. We stress that, in this work, we do not plan to identify any best code or best set of assumptions. The analysis presented here can be applied to any specific relation (e.g., the stellar mass function, the age-mass relation, the SFR density vs time). Here, we focus on the SFR-stellar mass relation (for star-forming galaxies, this is often referred to as the ``star-formation main sequence", e.g., \citealt{brinchmann2004,noeske2007,whitaker2012}), assess the different outputs by the different codes, and compute a ``median'' SFR-stellar mass relation, which accounts for the uncertainties in the modeling.

The Paper is organized as follows: in Section~\ref{sec:definitions} we provide some important working definitions in SED fitting; in Section~\ref{sec:data} we present the three photometric datasets we use; in Section~\ref{sec:list} we introduce the fourteen SED fitting codes used in this study; in Section~\ref{sec:compare} we compare the results of the codes and assess the effects of including/excluding IR measurements and AGN models; in Section~\ref{sec:impact} we assess the impact of the different modeling choices on the derived SFR-stellar mass relation and estimate modeling uncertainties for stellar mass, SFR, and dust attenuation; in Section~\ref{sec:lessons} we list our lessons learned on SED fitting in general; finally in Section~\ref{sec:conclusion} we provide our summary and conclusions.

Throughout this paper we use a cosmology with $\Omega_M=0.3$,  $\Lambda=0.7$, and $h=H_0/100$~km~s$^{-1}$~Mpc$^{-1}=0.7$. We present all magnitudes in the AB system.

\section{The SED Fitting Approach}
\label{sec:definitions}

The overall aim of the SED fitting process is to constrain physical parameters, and the way this is achieved is by comparing the observed SEDs with a matrix of the template model SEDs spanning a wide range of physical parameters. How the best matching models are selected varies from code to code. The code can return a single best-fitting model following a Frequentist approach, or it can return a probability distrubution following a Bayesian approach. Many codes nowadays adopt the latter. In Bayes' theorem, we must not only assess how well the model matches the data but also think critically about the reliability of the models themselves given all our prior knowledge \citep{kass1995}. For example from SED fitting, we could find that a stellar population model of age 10~Gyr with a certain combination of dust attenuation and metallicity matches the observed data of a $z\sim2$ galaxy with a high likelihood. However, we know that stars cannot be older than the age of the Universe, which at $z\sim2$ is $\sim3.2$~Gyr. Given that knowledge, we ought to down-weight that model as a solution for these data. Therefore, \emph{priors} on the models themselves are introduced to weight the likelihood. This is the basis of Bayesian statistics.

In technical terms, Bayes theorem  states the probability of an event (a model being a good representation of the data) given the prior knowledge of the conditions that might be related to that event (model). This is mathematically expressed as
\begin{equation}
\mathcal{L}(\Theta|D) = \frac{\mathcal{L}(D|\Theta)\times \mathcal{L}(\Theta)}{\mathcal{L}(D)}
\end{equation}
where $\mathcal{L}(\Theta|D)$ is the likelihood that the model $\Theta$ is true given data $D$. $\mathcal{L}(\Theta)$ contains the \textit{prior information} on the model, the prior likelihood distribution. $\mathcal{L}(D|\Theta)$ is the likelihood that data $D$ is expressed by model $\Theta$, so-called the likelihood function, and  $\mathcal{L}(D)$ is the unconditional marginal likelihood of the data.\footnote{$\mathcal{L}(D)$ is a constant, derived by integrating over all parameters $\Theta$, and tells us the likelihood of all the models given the data. It is commonly called the ``Bayesian evidence" and has many uses in Bayes factors and Bayesian inference criteria \citep{salmon2016}.}. The terms on the right are collectively referred to as the \textit{posterior distribution}, and can be thought of as a likelihood function that has accounted for prior knowledge. 

The likelihood function is defined as the confidence for a given model $\Theta$ expressed in terms of some parameters, representing real data $D$. The likelihood for independent Gaussian errors is commonly expressed by the \textit{chi-squared} metric and for SED fitting is defined as
\begin{equation}
\chi^2 = \sum_{i=1}^{N} \frac{\big(f_{\Theta}(\lambda_i) - f(\lambda_i)\big)^2}{\sigma_{i}^2}
\end{equation}
where each observed flux $f$ is measured at some wavelength $\lambda_i$ with an uncertainty $\sigma_i$, compared to the flux from the model $f_\Theta$. The chi-squared is generally calculated using fluxes instead of magnitudes because the photometric errors are calculated on the flux and thus are symmetric in linear space and not in logarithmic space. For Gaussian error distributions, this is redefined as a likelihood via $\mathcal{L}=\exp({-\chi^2/2})$. If normalized such that it integrates to unity, this is called the probability density function (PDF). By this definition, at a low $\chi^2$ there is a high probability that the model is a good representation of the data, given the optimized model parameters.

A ``Bayesian approach" or ``Bayesian framework" means to specifically design the model space with priors in mind, using prior knowledge to define which models to use, or to weight them directly. The prior is generally defined as the distribution of values of the parameters that describe a specific physical model. For example, a flat distribution  of $0<\tau_V<4$ is the prior for the parameter $\tau_V$ in a specific dust attenuation physical model. However, we argue here that the choice of the specific physical model (or lack of such model) should also be considered a prior, even when the relative model parameters are fixed to a single value (e.g., the lack of emission line modeling or a fixed initial mass function; see the various choices of physical models in Table~\ref{tab:codes_all} and the work by \citealt{curtislake2021}).

Within a Bayesian approach, the scientific use-case determines whether a particular sampling technique is preferred over others while performing the SED fit. Sampling in this context refers to the way the SED fitting code traverses the likelihood space $\mathcal{L}(D|\Theta)$, especially when the number of dimensions in $\Theta$ is large and cannot be covered easily with simple grid-search methods. Brute-force Bayesian methods (e.g. \citealt{dacunha2008,dacunha2015,pacifici2012,iyer2017,abdurrouf2021}) that use a pre-computed atlas of galaxies drawn from pre-defined prior distributions are preferred when scaling to large datasets from upcoming surveys using observatories such as the Nancy Grace Roman Space Telescope that will contain about $10^8$ galaxies. These are particularly suited to large datasets since they trade space- for time-complexity, and are easily parallelizable. On the other hand, Markov Chain Monte Carlo (MCMC) and related techniques \citep{foreman2013} are well suited for exploring the likelihood space of high S/N objects, and accurately characterizing their uncertainties. Finally, nested sampling based techniques \citep{skilling2004,skilling2006,buchner2016} are suited to problems where the likelihood space has pathological or highly multi-modal features, or for model comparison where the Bayesian evidence needs to be estimated.

\section{Datasets}
\label{sec:data}

\begin{figure}
\begin{center}
\includegraphics[width=0.45\textwidth]{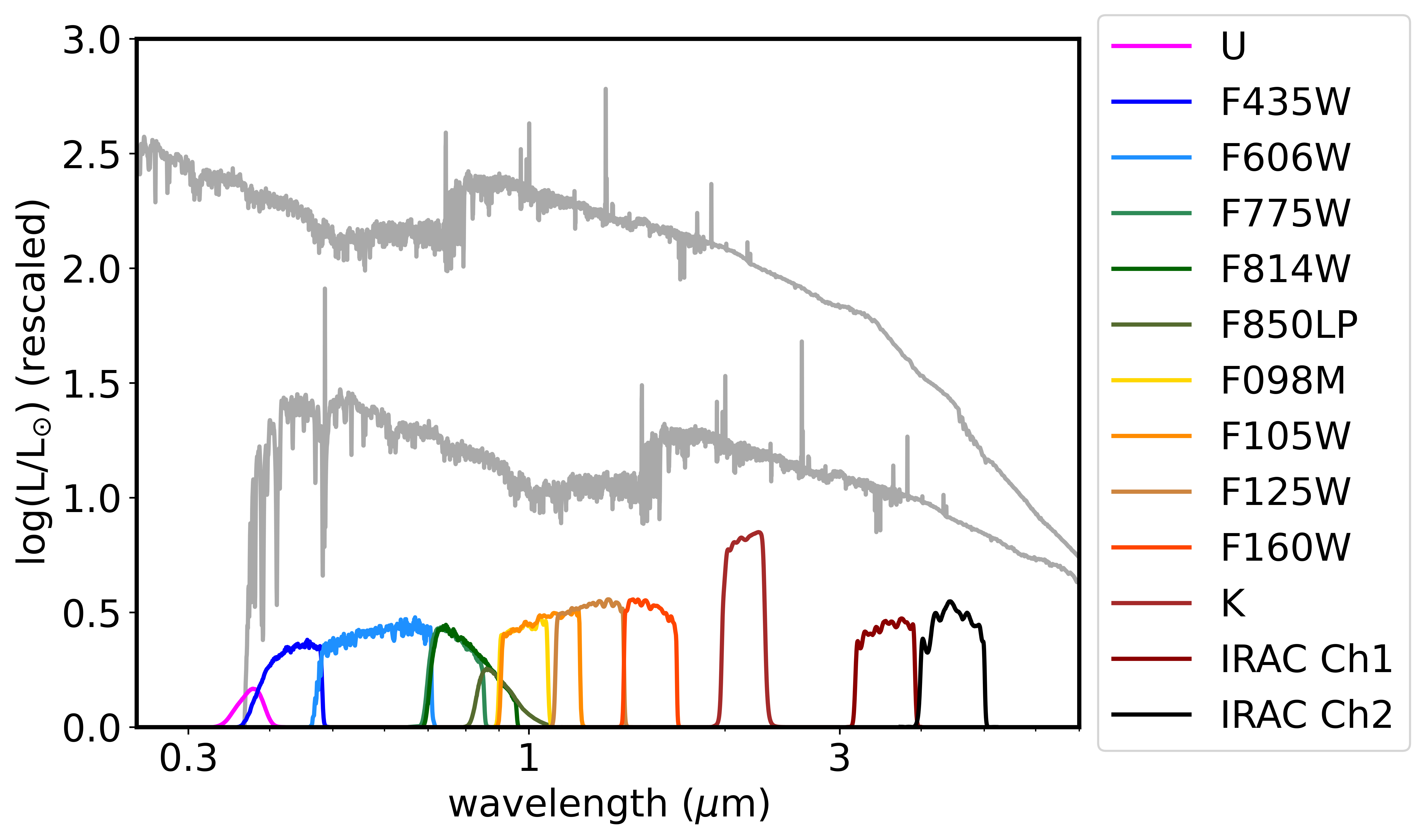}
\caption{A model SED plotted at $z=1$ (upper) and $z=3$ (lower), along with the throughput curves of the filters in the photometric catalog (color code in the legend).}
\label{fig:sedfilters}
\end{center}
\end{figure}

For our comparative analyses, we use three photometric datasets. These are chosen for two reasons: 1) the photometric bands cover a wavelength range from the UV to the NIR to sample the light from both old and young stars (the normalization of the NIR is necessary to infer the stellar mass, while the light in the UV is necessary to measure the current SFR of the galaxies) and also to have a large enough wavelength baseline to assess the effects of dust attenuation; 2) the additional IR photometry is measured at good-enough S/N ratios, for accurate measurements of the SFR. The three datasets are:

\begin{enumerate}
    \item a $z\sim1$ sample with measured photometry spanning the rest-frame NUV to the NIR (observed wavelength range between $\sim0.35$ and $4.5\mu$m);
    \item a subset of sample 1 that includes targets observed with IR photometry (24 to 250$\mu$m) with S/N in MIPS/24$\mu$m larger than 3 and IR flags to avoid heavily blended targets;
    \item a $z\sim3$ sample with measured photometry spanning the rest-frame FUV to the NIR.
\end{enumerate}

All samples are extracted from CANDELS (\citealt{grogin2011,koekemoer2011}) in the GOODS-South field \citep{guo2013}. Photometric redshifts are provided by the CANDELS collaboration \citep{kodra2022} by combining photometric redshift estimates obtained using different codes. The full catalog of redshifts includes also publicly available spectroscopic redshifts and grism redshifts from the 3D-HST survey \citep{momcheva2016}. A \textit{best redshift} ($z_{best}$) is selected for every galaxy using, in the following order of priority, an available spectroscopic redshift, or an available grism redshift, or the photometric redshift. In order to make the comparison of the derived absolute quantities (stellar mass and SFR) more meaningful, the tools assume $z_{best}$ as the true redshift when fitting. We do not look for potential Active Galactic Nuclei (AGN) in any way, thus the sample may or may not include AGN.

The $z\sim1$ sample includes 339 galaxies selected at $1.0<z<1.1$, with good photometric flags (no contamination by bright stars and targets not on the edge of the detector), and $H<24$ mag. We matched this catalog with IR measurements from the catalog by Inami et al. (in prep), following the methodology by \citet{magnelli2013} (see their Section 4) and flags by \citet{barro2019}. The fiducial IR sample includes 107 galaxies with S/N larger than 3 in MIPS 24$\mu$m accounting for flags to avoid heavily blended targets. Out of these 107 galaxies, 30 have photometry at 70$\mu$m, 66 have photometry at 100$\mu$m, 55 have photometry at 160$\mu$m, and 28 have photometry at 250$\mu$m.

The $z\sim3$ sample includes 127 galaxies selected at $3.0<z<3.1$, with good photometric flags (the same as for the $z\sim1$ sample), and $H<26$. In all samples, objects with stellarity larger than 0.8 have been excluded.

Figure~\ref{fig:sedfilters} shows a model SED plotted at $z=1$ and $z=3$ along with the response curves of the filters available in the catalogs. This shows the rest-frame wavelength range covered by the filters (excluding the IR) at the two redshifts. In both cases, UV, optical, and NIR are anchored, offering a good baseline to constrain both the old stellar populations (NIR) and the young and unobscured stellar populations (UV).

The catalogs can be found on GitHub. By visual inspection of the photometry, we flagged and excluded 31 galaxies in the $z\sim1$ sample and 34 galaxies in the $z\sim3$ sample because of bad photometric measurements (e.g., photometry in a particular band is inconsistent with the photometry in adjacent bands). Flags files are published at the same URL.

\section{The SED Fitting Tools}
\label{sec:list}
We present here, in alphabetical order, the codes that take part in this study. Table~\ref{tab:codes_all} summarizes their main characteristics. All codes except 1 (AGNFitter is optimized to work only with IR data points, hence it is run only on dataset 2) are run on datasets 1 and 3. The codes that can process IR data points (BAGPIPES, CIGALE, MAGPHYS, Prospector, and SED3FIT) are also run on dataset 2. All fits were performed with the latest versions of each code available between 2018 and 2019.

\begin{table*}
\begin{tabular}{l | l l l l l l l l}
\textbf{Code} & \textbf{Sampler} & \textbf{B/F} & \textbf{SFH} & \textbf{SSP} & \textbf{Neb. em.} & \textbf{Dust att.} & \textbf{Dust em.} & \textbf{AGN} \\
\hline
\hline
AGNFitter$^{2}$     & MCMC          & B & Exp. decl.       & BC03      & No    & Single    & Multiple  & Yes \\
BAGPIPES$^{1,2,3}$  & Nested sam.   & B & Flex.            & Multiple  & C17   & Multiple  & Single    & No  \\
BEAGLE$^{1,3}$      & Nested sam.   & B & Flex. param.     & BC03(16)  & C13   & 2 comp.   & No        & No \\
CIGALE$^{1,2,3}$    & Grid          & B & Flex.            & Multiple  & C13   & Multiple  & Multiple  & Yes \\
Dense Basis$^{1,3}$ & Atlas         & B & Non-param.       & FSPS      & C     & Multiple  & Single    & No  \\
FITSED$^{1,3}$      & Grid          & B & Exp. inc./decl.  & BC03      & C     & Single    & No        & No \\
Interrogator$^{1,3}$& MCMC          & B & Flex. param.     & Multiple  & C17   & 2 comp.   & No        & No  \\
LePhare$^{1,3}$     & Grid          & F & Exp. decl.       & BC03      & No    & Single    & No        & No \\
MAGPHYS$^{1,2,3}$   & Atlas         & B & Flex. param.     & BC03      & No    & 2 comp.   & Multiple  & Yes  \\
P12$^{1,3}$         & Atlas         & B & Non-param.       & BC03(11)  & C     & 2 comp.   & No        & No   \\
Prospector$^{1,2,3}$& Nested sam.   & B & Non-param.       & FSPS      & C13   & 2 comp.   & Single    & Yes \\
SED3FIT$^{1,2,3}$   & Grid          & B & Exp. decl.       & BC03      & No    & 2 comp.   & Single    & Yes \\
SpeedyMC$^{1,3}$    & MCMC          & B & Exp. inc./decl.  & BC03      & Empir.& Multiple  & No        & No  \\
zPhot$^{1,3}$       & Grid          & F & Exp. inc./decl.  & BC03      & Empir.& Multiple  & No        & No \\
\hline
\end{tabular}
\caption{This table summarizes the possible choices in the SED fitting tools. The superscript in column [Code] identifies the sample from Sec.~\ref{sec:data} to which the code has been applied. The [Sampler] identifies the algorithm used to explore the parameter space of the priors: by Markov Chain Monte Carlo ``MCMC'', by nested sampling ``Nested sam.'', on a ``Grid'', or on a pre-computed atlas ``Atlas''. ``B'' signifies Bayesian and ``F'' signifies Frequentist (i.e., a single best fitting result). The [SFH] prior can include exponentially declining, increasing, or both functions (``Exp. decl.'', ``Exp. inc./decl.''), multiple parametric functions (``Flex. param.''), non parametric functions only (``Non-param''), or multiple parametric and non parametric functions (``Flex.''). The codes can select among different Simple Stellar Population [SSP] models (``Multiple'') or use a single SSP model (e.g., ``BC03'' or ``FSPS''). The number in parenthesis for BC03 indicates the year the model was updated. The column [Neb. em.] stands for nebular emission and reports whether a code includes a nebular component calculated with the software Cloudy (``C'' followed by the year of update), using empirical relations (``Empir.'') or does not include a nebular component (``No''). The dust attenuation model [Dust att.] can be in the form of a ``Single'' attenuation law, ``Multiple'' attenuation laws, or a two-component attenuation law (``2 comp.''). The dust emission [Dust em.] modeling can be done using a ``Single'' or ``Multiple'' models, or can be skipped (``No''). An [AGN] component can be included or not. More information about the various models can be found in the text.}
\label{tab:codes_all}
\end{table*}

\subsection{\bf AGNFitter}
\label{sec:AGNFitter}
AGNfitter \citet{calistrorivera2016} is a Python package to fit the SEDs of active galactic nuclei (AGN) and galaxies from the UV to the sub-mm with a fully Bayesian MCMC method. AGNFitter uses exponentially declining SFHs at constant metallicity and combines them with \citet{BC03} (BC03) stellar population syntesis models. It does not include nebular emission. Dust attenuation is calculated with the \cite{calzetti2001} attenuation curve. The AGNfitter dust and AGN models consist of three physical emission components: an accretion disk, a torus of AGN heated dust, and cold dust in star forming regions.

\subsection{\bf BAGPIPES}
\label{sec:BAGPIPES}

Bayesian Analysis of Galaxies for Physical Inference and Parameter EStimation, or BAGPIPES \citep{carnall2018}, is a Python package for modeling galaxy spectra and fitting spectroscopic and photometric observations in the FUV to FIR wavelength range. The code incorporates highly customisable models for emission from the stellar, nebular and dust components of galaxies, and an efficient Bayesian fitting approach using nested sampling. BAGPIPES includes the option to fit flexible empirical models for systematic uncertainties in spectroscopic data, as described in \cite{carnall2019b}. The documentation can be found at \href{https://bagpipes.readthedocs.io}{https://bagpipes.readthedocs.io}.

\subsection{\bf BEAGLE}
\label{sec:BEAGLE}

BayEsian Analysis of GaLaxy sEds, or \textsc{beagle} \citep{chevallard2016}, enables the modeling of any combination of photometric and spectroscopic galaxy observable (e.g. full spectra, spectral indices, emission line fluxes) with a flexible and fully self-consistent physical model from the UV to the NIR. It is used in combination with the \textsc{multinest} algorithm to derive probabilistic (Bayesian) constraints on several galaxy physical parameters, and to create mock galaxy catalogs \citep[e.g.][]{chevallard2017, williams2018, curtislake2021}. \textsc{beagle} incorporates population synthesis models and models describing the emission of gas photoionized by young stars (lines and continuum). Several prescriptions to account for dust attenuation and IGM absorption are available. Galaxy star formation and chemical enrichment history are included in a flexible parametric way.  Documentation and instructions can be found at \url{https://www.iap.fr/beagle}.

\subsection{\bf CIGALE}
\label{sec:CIGALE}
Code Investigating GALaxy Emission, or CIGALE \citep{burgarella2005,boquien2019} is a Python-based code developed to interpret galaxy SEDs from the X rays to the radio. SFHs can be parametric or ingested from a file. The code can use different stellar population models including ionised gas (continuum and emission lines) and a variety of dust models. An AGN component can be included and the newest version of the code can also accept X-ray data \citep{yang2020,yang2022}. The results are derived using a Bayesian approach on a grid of model SEDs and can take upper limits into account. In addition to fluxes, it can also fit intensive and extensive physical properties (e.g., a line equivalent width or the dust luminosity). CIGALE can be found at \url{https://cigale.lam.fr}.

\subsection{\bf Dense Basis SED fitting}
The Dense Basis SED fitting code \citep{iyer2017,iyer2019} is a Python-based code that creates and uses an atlas of galaxy SEDs from a physically motivated basis of flexible, non-parametric SFHs. The code uses Flexible Stellar Population Synthesis (FSPS, \citet{conroy2009}) to generate spectra adopting a variety of metallicities and dust attenuation values. Gaussian process regression is used to compute the SFHs and the lookback times at which a galaxy formed different fractions of its observed mass. These SFHs allow us to estimate quantities previously inaccessible through SED fitting, such as the number and duration of star formation episodes in a galaxy's past. The code can be found at \url{https://github.com/kartheikiyer/dense_basis}.

\subsection{\bf FITSED}
\textsc{FITSED} \citep{papovich2001,salmon2015} is an IDL-based code to infer physical properties of galaxies from broadband photometry using a Bayesian framework.  The code is set up to accept parametric SFHs, \citet{BC03} stellar population synthesis models, emission line models, and variable dust attenuation laws \citep{salmon2016}. \textsc{FITSED} operates in two steps: it first creates a look-up-table grid of photometry from SED models and then calculates the likelihood between the data and every model in the grid.

\subsection{\bf Interrogator}
\label{sec:Interrogator}

\textsc{Interrogator} (Fairhurst private communication, \url{http://users.sussex.ac.uk/~sw376/Interrogator/}) is a Python based SED fitting code exploiting the power of MCMC and Bayesian statistics to fit galaxy photometry and spectroscopy. The code takes a modular approach to modelling the stars (including \citealt{BC03} and \citealt{eldridge2016}), gas, and dust of a galaxy. In addition to data, users provide a choice of model for these various constituents, and can overwrite any of the default uninformative priors with arbitrary priors of their choice. The current version does not include models in the IR and so it interprets SEDs from the UV to the NIR.

\subsection{\bf LePhare}
\label{sec:LePhare}
LePhare \citep{arnouts1999, ilbert2006} consists of a set of Fortran commands to compute photometric redshifts and to perform SED fitting. It works with a grid of model templates created with PEGASE \citep{fioc2019} and \citet{BC03} population synthesis models, exponentially declining SFHs, and the Calzetti attenuation curve \citep{calzetti2001}. Since the physical parameters are estimates on a grid, uncertainties are derived by bootstrapping the data.

\subsection{\bf MAGPHYS}
\label{sec:MAGPHYS}
Multi-wavelength Analysis of Galaxy Physical Properties, or MAGPHYS \citep{dacunha2008,dacunha2015} is a Fortran-based code  to model the emission by stellar populations and dust in galaxies and infer galaxy properties using a Bayesian approach. It uses a variety of exponentially-declining and delayed exponentials SFHs along with superimposed random bursts, to account for stochasticity in star formation. The dust attenuation is modeled using the two-component model of \cite{charlot2000}, and dust emission is self-consistently modeled using an energy balance technique. Recent updates include a radio component, an AGN component (see \citealt{chang2017a,chang2017b}), the inclusion of a UV bump in the dust attenuation curve \citep{battisti2020}, and a photometric redshift extension \citep{battisti2019}. The code simultaneously fits the UV to radio emission by comparing each observed SED with all the models. MAGPHYS can be found at \url{www.iap.fr/magphys}.

\subsection{\bf P12}
\label{sec:pacifici}
P12 \citep{pacifici2012} is a Fortran-based code to model and interpret any set of photometric and spectroscopic observations from the FUV to the NIR with a Bayesian algorithm. Model galaxies are generated assuming star formation and metal enrichment histories from a semi-analytical model of galaxy formation. The model SEDs are created combining consistently the emission by the stars from simple stellar population models (the 2011 version of \citealt{BC03}) and the emission by the gas using a photo-ionization code \citep{charlot2000}. Dust attenuation is included with a two-component model where multiple parameters are allowed to vary (similar to \citealt{charlot2001}.

\begin{figure*}
\begin{center}
\includegraphics[width=0.9\textwidth]{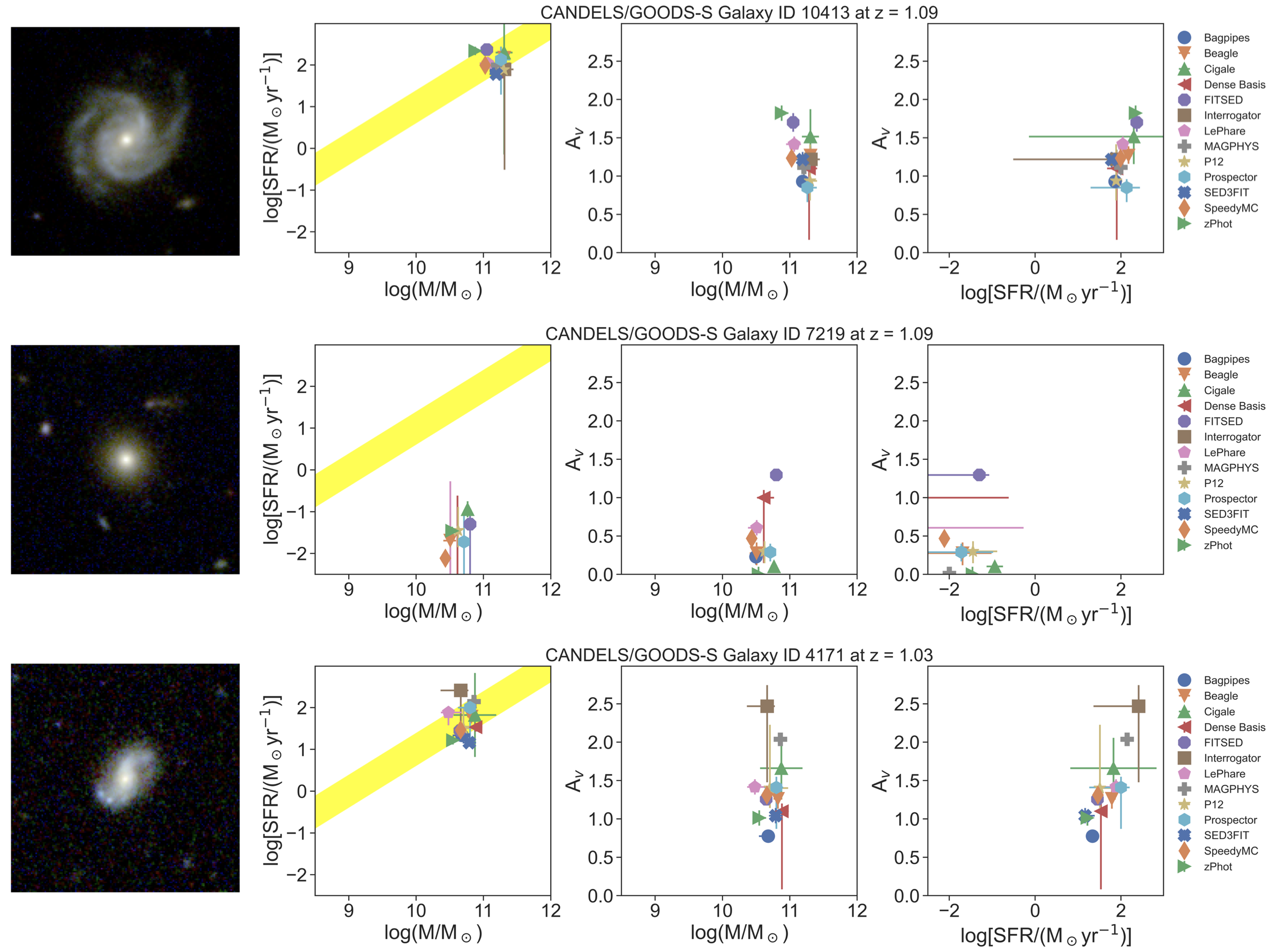}
\caption{Individual measurements from the thirteen codes for three galaxies in the $z\sim1$ sample. We select a typical star forming galaxy (top row), a typical quiescent galaxy (middle row), and a galaxy where we find significant disagreement between the different codes, especially in their estimates of SFR and $A_V$ (bottom row). On the left, RGB images of the three galaxies at the same spacial scale. The three relations are: SFR vs stellar mass (the yellow shade marks an illustrative linear relation 0.8~dex wide, see text for details), $A_V$ vs stellar mass, and $A_V$ vs SFR.}
\label{fig:singlegal}
\end{center}
\end{figure*}

\subsection{\bf Prospector}
\label{sec:Prospector}
Prospector \citep{leja2017,johnson2021} is a Python-based SED fitting code to interpret photometric and/or spectroscopic data across the full FUV-FIR wavelength range. Sampling can be performed with simple minimization, MCMC, or nested sampling techniques. The SED models are built using either parametric or nonparametric SFHs, the FSPS code \citep{conroy2009}, and various dust attenuation models. Prospector can also forward-model data calibration parameters such as spectrophotometric calibrations and wavelength solutions and incorporates uncertainties in these components in the final parameter uncertainties. Installation instructions and Jupyter notebooks are available at \url{https://github.com/bd-j/prospector}.

\subsection{\bf SED3FIT}
\label{sec:SED3FIT}
SED3FIT \citep{berta2013} is a publicly available SED fitting package which is a modification to the original MAGPHYS fitting routine.  SED3FIT adds an AGN component to both the optical and FIR fitting steps.  The AGN library in the original release of SED3FIT includes 10 AGN SED models which span viewing angles 0$^{\circ}$ to 90$^{\circ}$.  The stellar, dust, and AGN components are moderated by energy balance at all times during the fitting process.

\subsection{\bf SpeedyMC}
\label{sec:SpeedyMC}
SpeedyMC \citep{acquaviva2012} is a MCMC algorithm for SED fitting, which builds upon the GalMC \citep{acquaviva2011} Bayesian framework for SED fitting, but is optimized for fitting large galaxy catalogs. The computational speed-up in SpeedyMC comes by pre-computing galaxy spectra on a grid of locations exploring the entire parameter space, and then running the MCMC using multi-linear interpolation between the pre-computed spectra. It assumes exponentially increasing and declining SFHs, stellar population models \citep{BC03}, and various dust attenuation curves.

\subsection{\bf zPhot}
The zPhot code was born as a tool to measure galaxy photometric redshifts \citep{fontana2000} and has later evolved to also measure galaxy physical properties (see e.g., \citealt{merlin2019}). It uses a grid of template SEDs generated with exponentially declining, exponentially rising or constant SFH, coupled with \cite{BC03} stellar models, nebular emission models, and a selection of dust attenuation laws. The fitting range covers the UV to the NIR. Since the physical parameters are estimates on a grid, uncertainties are derived by bootstrapping the data.

\section{Performance of the codes on the different galaxy samples}
\label{sec:compare}
In this Section, we present the physical parameters obtained from the various codes when run on the datasets described in Section~\ref{sec:data}. Each code has been run by their users with sufficient tuning in order to produce robust results for the datasets in question. Leaving the users free to choose their code's setup is representative of what normally happens in the scientific community and thus allows us to address that component of the uncertainty that is due to the modeling choices (i.e. the modeling uncertainty). In this exercise, we do not aim at identifying the ``best'' SED fitting code, but we aim at assessing the biases and uncertainties caused by the modeling assumptions. The modeling uncertainties need to be taken into account when comparing results to other studies or to theoretical simulations. The outputs of the codes have been checked (e.g., evaluating the distribution of the $chi^2$ or the convergence of the fits) and considered acceptable.

\subsection{Code comparison on individual galaxies}
\label{sec:degen}

We present here a comparison of the code outputs on three example individual galaxies in the parameter space of stellar mass, SFR (averaged over 100~Myr), and $A_V$ (Figure~\ref{fig:singlegal}). The best-estimate values are conventionally defined as the medians of the PDFs (see Section~\ref{sec:def_results} for a discussion on other ways in which the best-estimate values can be defined). We select a star forming galaxy (top row), a quiescent galaxy (middle row), and a galaxy where we find significant disagreement between the different codes, especially in their estimates of SFR and $A_V$ (bottom row). All galaxies are taken from the $z\sim1$ sample and the results are from fits to the UV-to-NIR photometry. All measurements are returned with uncertainties. For the Bayesian codes, this is done by calculating the full PDF and extracting either the half 16th-to-84th percentile interval, or the asymmetric 16th-to-50th and 50th-to-84th percentile intervals, depending on the code. For the codes that return only a single best-fitting model, uncertainties are calculated by fitting the dataset multiple times after bootstrapping it about the photometric uncertainties and then measuring the dispersion in the derived parameters. The yellow shade marks an illustrative linear relation at constant specific SFR (SSFR; SFR divided by the stellar mass) of 1 Gyr$^{-1}$ and 0.8~dex width similar to the observed SFR-stellar mass relation (e.g.,  \citealt{brinchmann2004,whitaker2012,speagle2014}).

For the star-forming and quiescent galaxies, the results from the individual codes agree within their uncertainties (i.e., the observational uncertainties propagated to the physical parameters). Although the derived dust attenuation values have small uncertainties in the majority of the codes (suggesting that the parameter is well constrained), they show the largest difference in the estimates because of the different assumed attenuation laws. The third galaxy shows some discrepancies and the differences between the estimates can be as high as one order of magnitude, while the uncertainties are generally smaller. The most evident correlation is between SFR and $A_V$ (Figure~\ref{fig:singlegal}, right-hand side panel), where the higher the inferred dust attenuation, the higher the SFR. This happens when there is a ``degeneracy'', i.e., when more than one parameter has the same effect on the observables and thus the SED is matched to different sets of parameters with the same likelihood. One way to break these degeneracies is to include new observables that depend on physical parameters in different ways. For example, adding IR observations can help break the degeneracy between SFR and $A_V$ (see e.g., \citealt{dacunha2008,dacunha2015}).

\subsection{Distribution of results for the $z\sim1$ sample}

\begin{figure*}
\begin{center}
\includegraphics[width=0.30\textwidth]{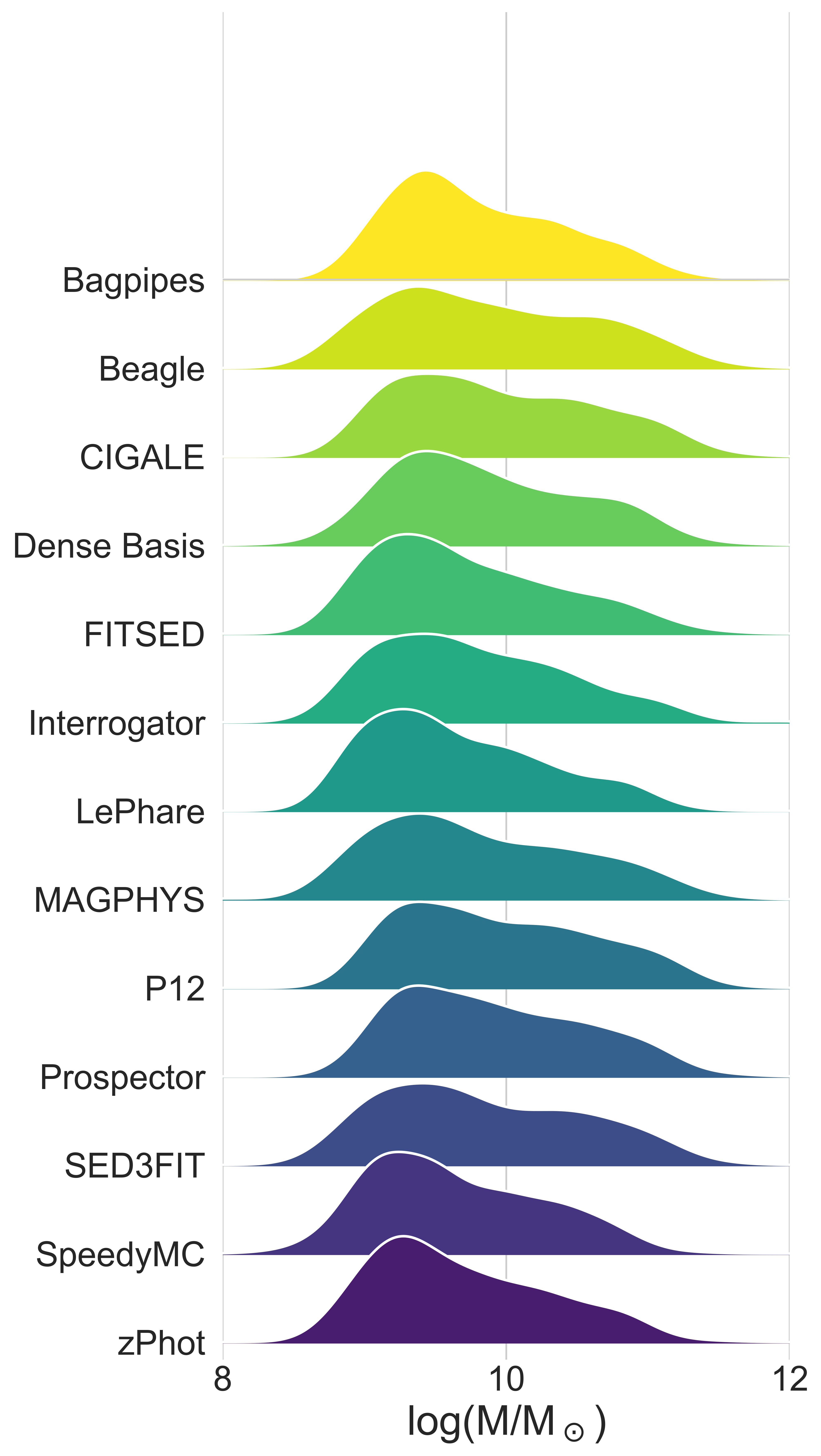}
\includegraphics[width=0.30\textwidth]{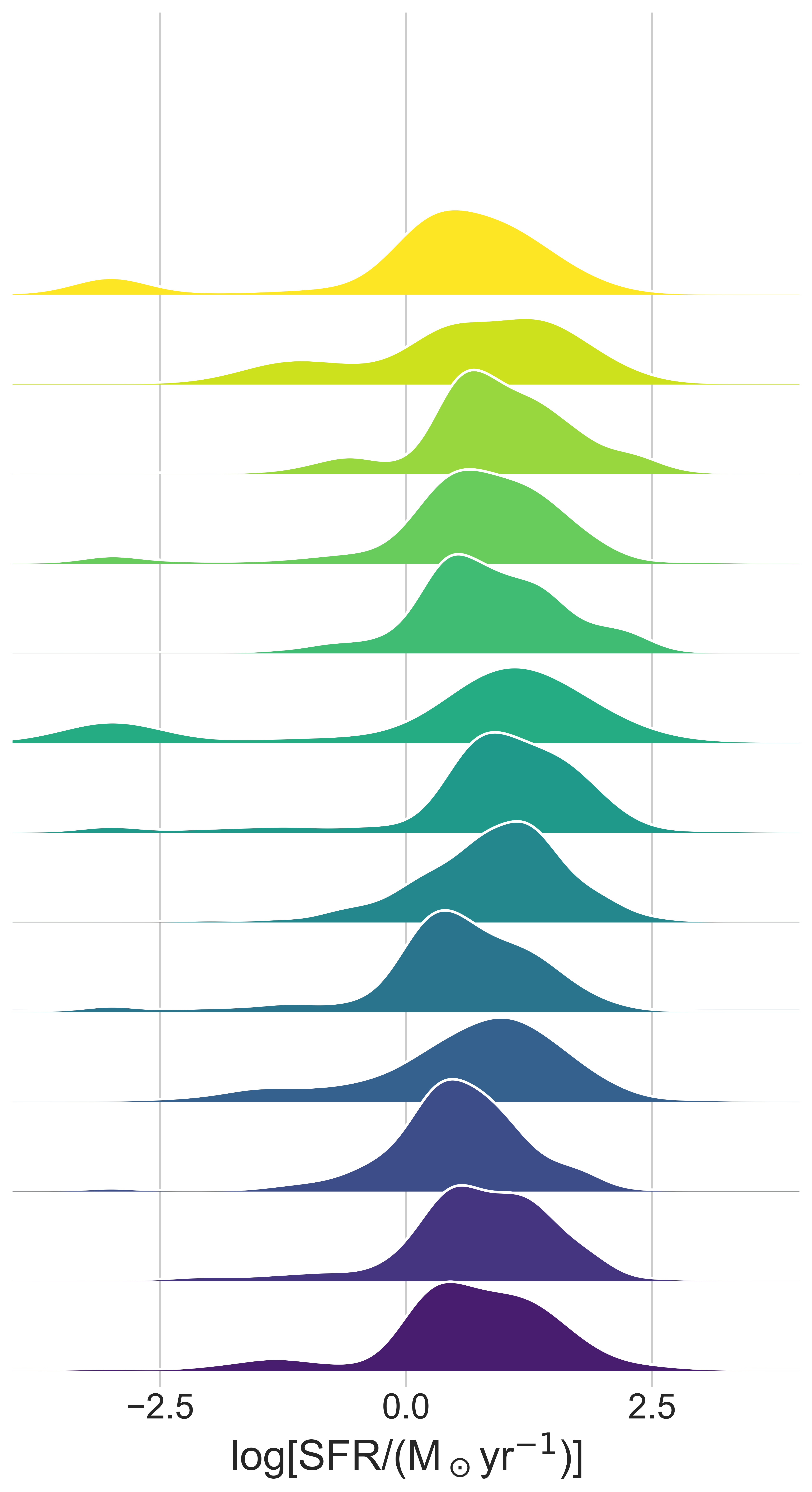}
\includegraphics[width=0.30\textwidth]{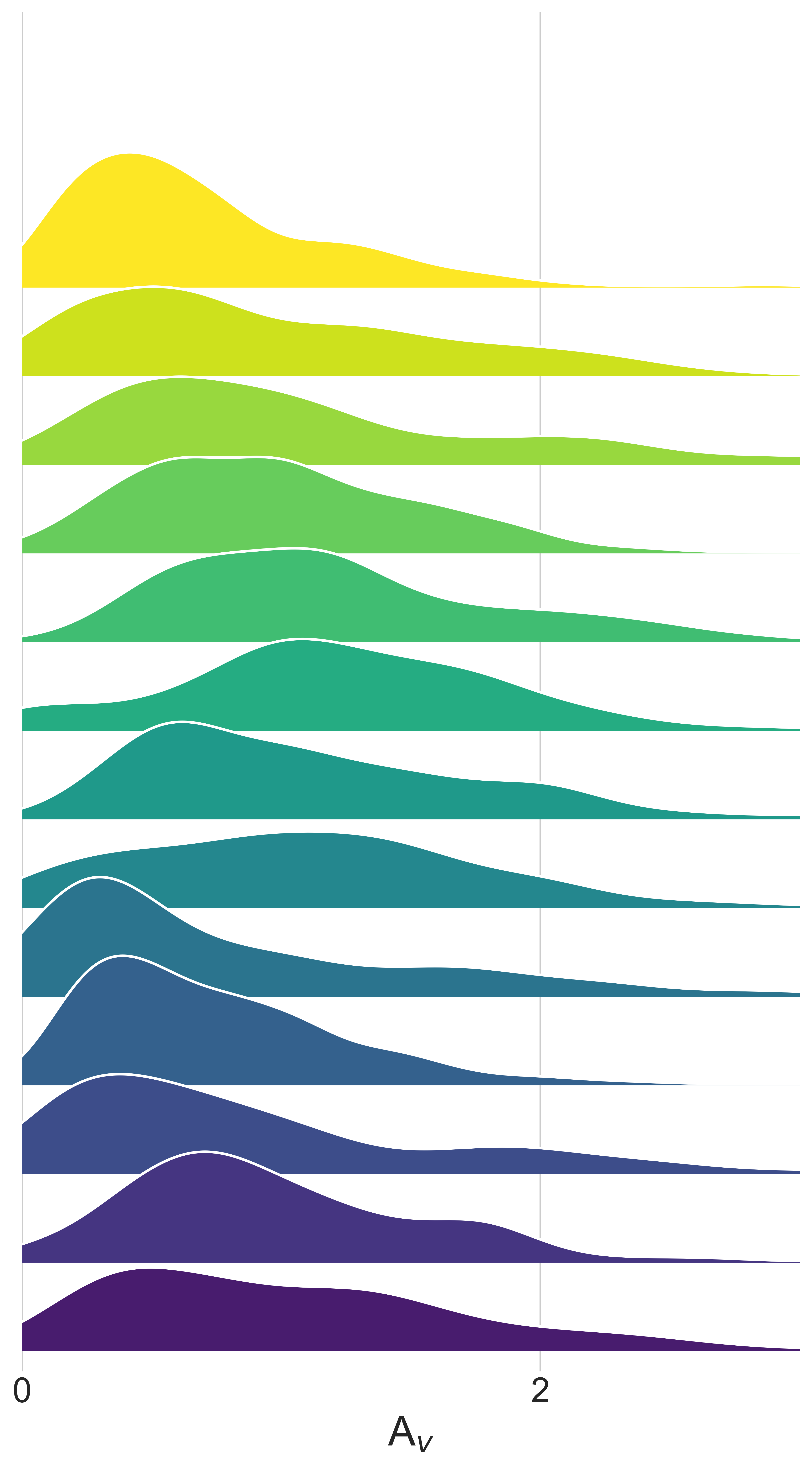}
\caption{Each shade represents the distribution of the best-estimate results (defined as the median of the probability density function) of a specific parameter for all the galaxies in the $z\sim1$ sample. Columns show stellar mass, SFR, and $A_V$. Each row represents the results from an individual code. The color code is purely graphical to ease the distinction from one row to the next. The differences in the distributions depend entirely on the the assumptions in the SED fitting tools and sampling of their parameter space.}
\label{fig:z1mass}
\end{center}
\end{figure*}

We now explore the distributions of the best-estimate results (defined as the median of the PDF) for the three physical parameters of interest (stellar mass, SFR, and dust attenuation) from the various codes. The way the distributions of the physical parameters vary code by code given the same set of observations depends entirely on the assumptions in the SED fitting tools and sampling of their parameter space. The stellar mass distributions, as shown on the left-hand side in Figure~\ref{fig:z1mass}, are very similar among codes. This suggests that the different modeling assumptions do not have a large impact on this parameter (see also \citealt{santini2015}) and thus it could be considered the most robust result that can be obtained with SED fitting. The differences are mainly in the normalizations of the distributions. Generally, systematic offsets in stellar mass (i.e., offsets that affect the sample on the entire range) can be caused by different IMF assumptions, differences in the stellar evolution models, or differences in the dust attenuation models. For this exercise, the IMF was fixed to \cite{chabrier2003}, thus the systematic differences are imputable to the different stellar evolution models (see e.g., \citealt{baldwin2018}) and to assumptions for the dust attenuation law, (see e.g. \citealt{malek2018}). We will not explore the details of the difference stellar population models in this work. The SFH choice can also affect the stellar mass estimates, and generally affects low-mass and high-mass galaxies in different ways. For example, low-mass or high-redshift galaxies are better represented by rising SFHs, while high-mass or low-redshift galaxies can be approximated by declining SFH functions (see e.g., \citealt{pacifici2013}). Thus, it is important to adopt appropriate SFHs for the sample under analysis.

The SFR distributions show more variety than the stellar mass ones and thus the constraint of this parameter is generally less robust than stellar mass. This is because the SFR estimates are affected by more modeling assumptions. For example, along with the assumed SFHs (and especially the SSFR prior) and stellar evolution models, the SFR estimates can be largely affected by the dust attenuation models. Usually, large dust attenuation measurements are coupled with large SFRs. Figure~\ref{fig:z1mass} shows that when the distribution of $A_V$ is broad or peaks around $A_V=1$ (e.g., for CIGALE, FITSED, Interrogator, LePhare, and MAGPHYS) the distribution of the SFR is also broad or peaks at large values. By contrast, the $A_V$ distributions that peak at $A_V<1$ are coupled with SFR distributions that peak at values close to $\log(SFR/(M_{\odot}~yr^{-1})=0$. The location of the quiescent galaxies ($\log[SFR/(M_{\odot}/yr)]<-1$) varies from code to code with BAGPIPES, Dense Basis, and Interrogator showing a clear peak at $\log[SFR/(M_{\odot}/yr)]\sim-3$, while the others show a tale instead.

\subsection{Distribution of results for the $z\sim1$ sample with IR}

\begin{figure*}
\begin{center}
\includegraphics[width=0.3\textwidth]{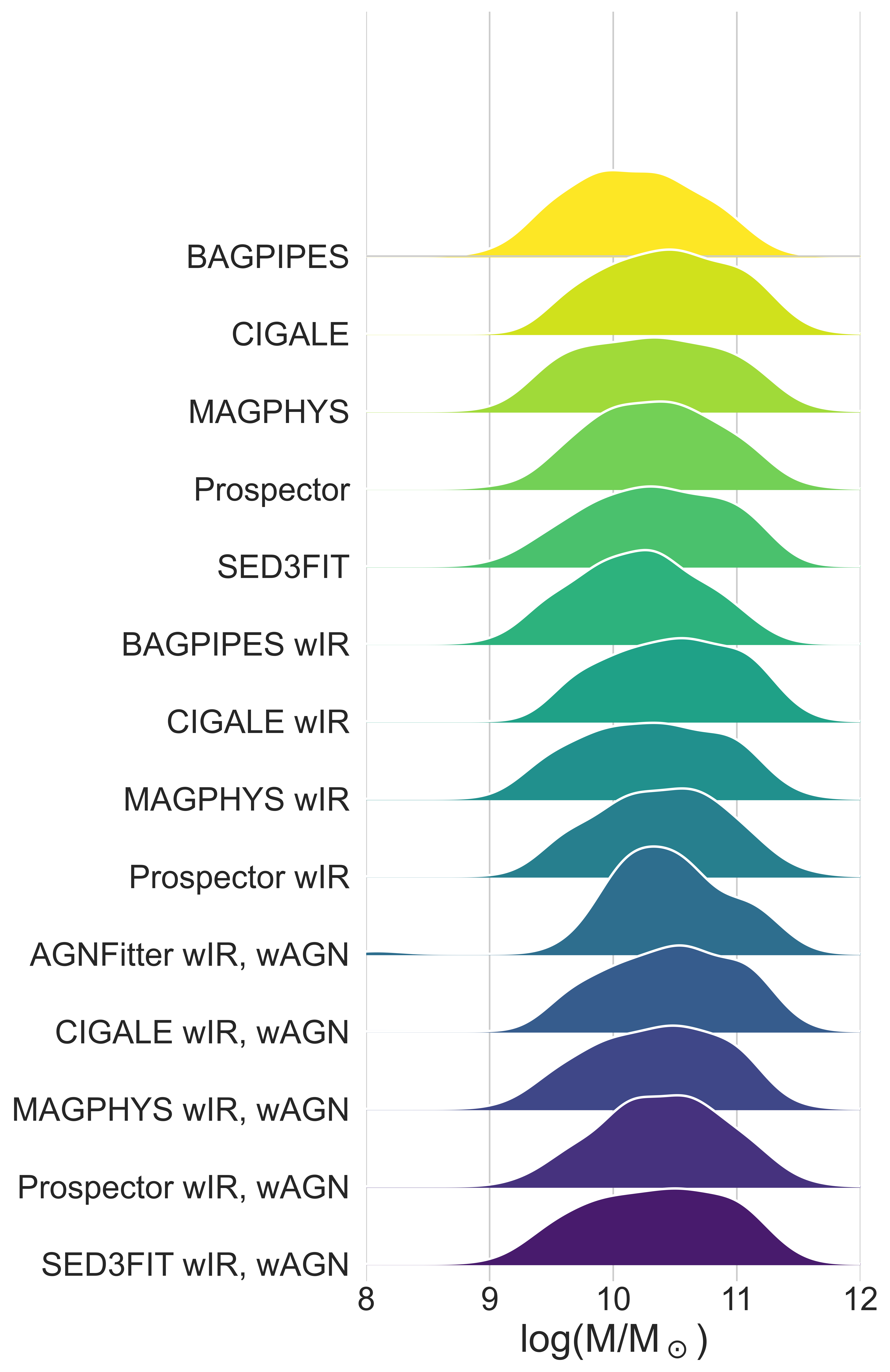}
\includegraphics[width=0.3\textwidth]{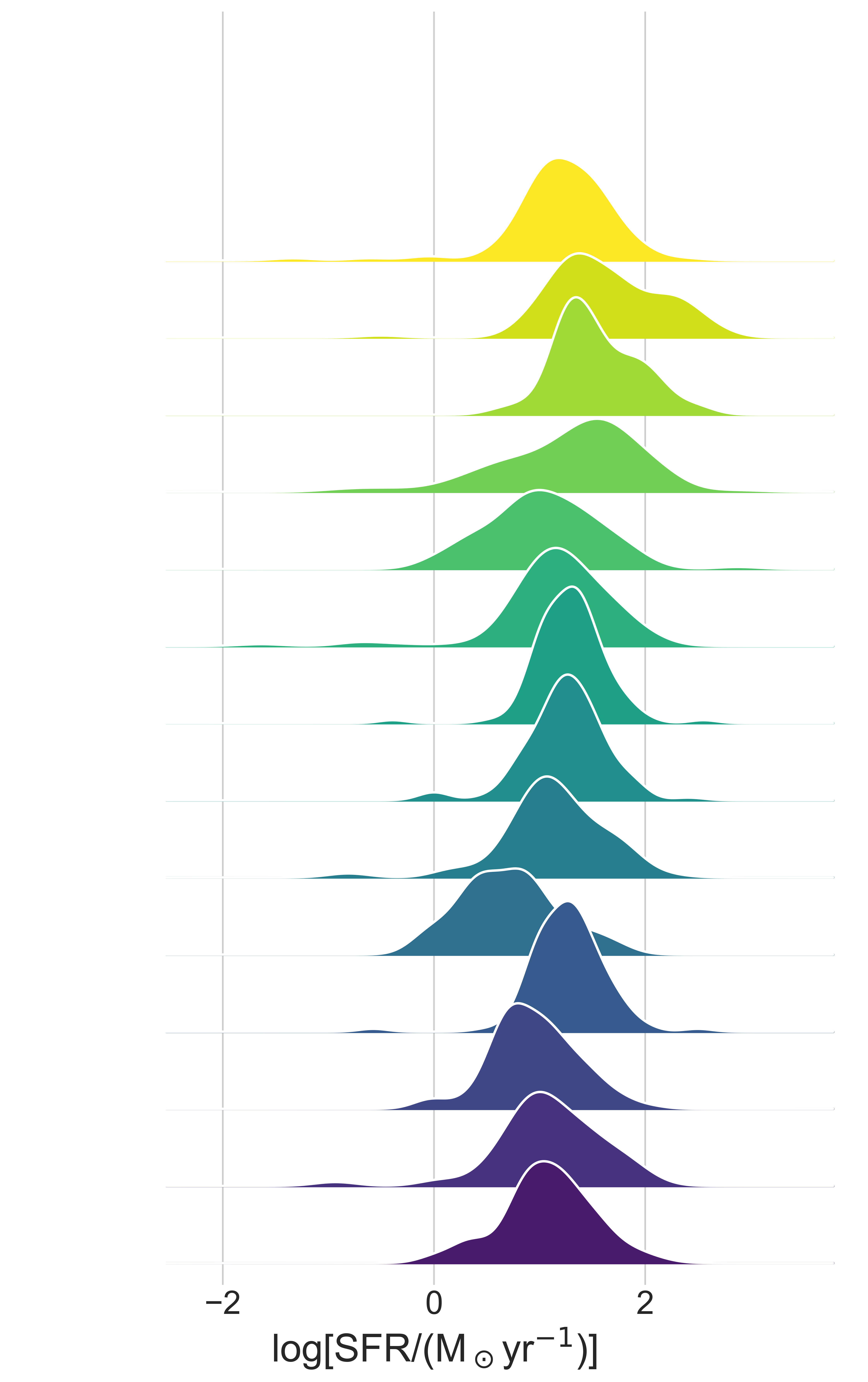}
\includegraphics[width=0.3\textwidth]{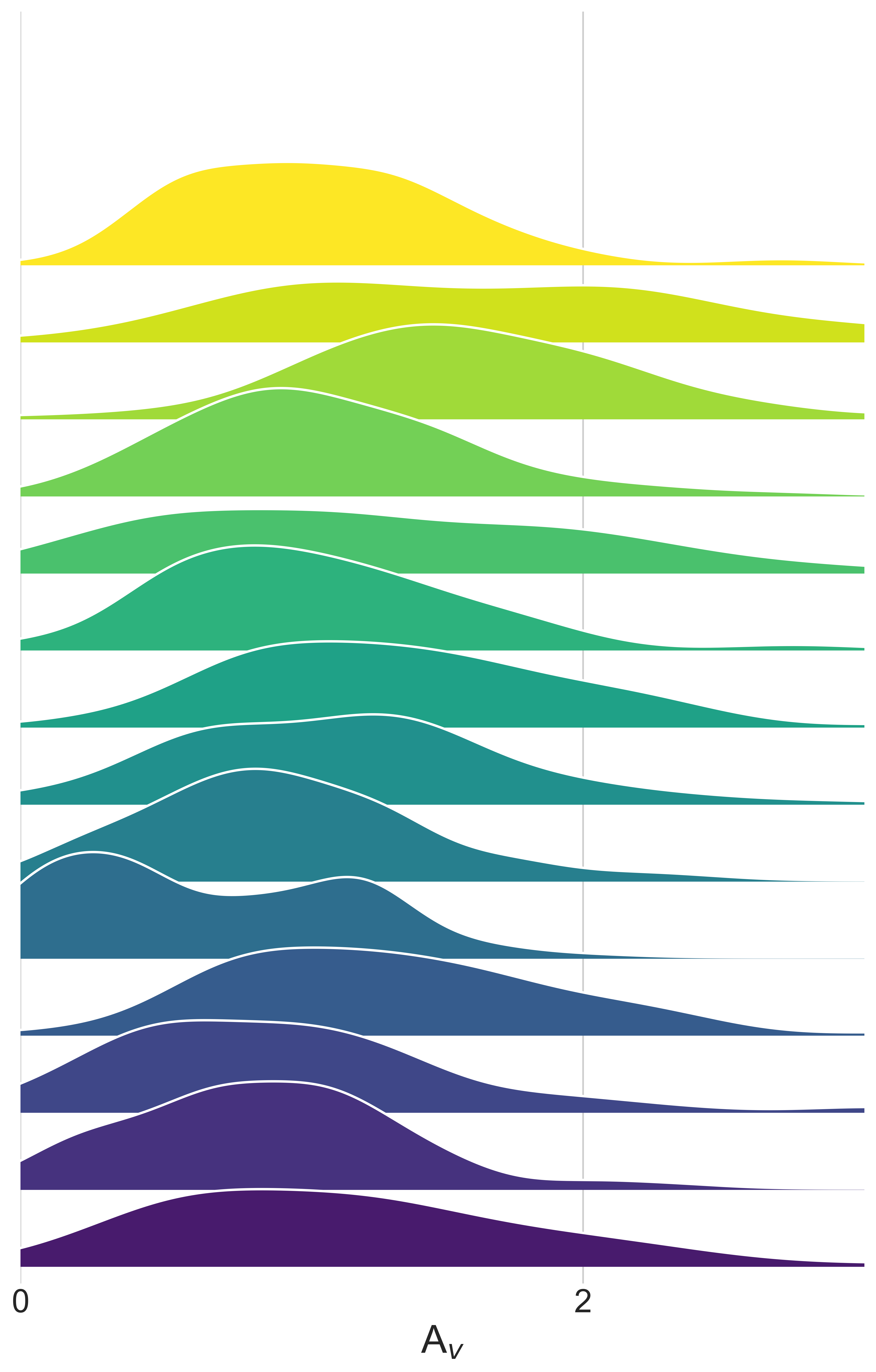}
\caption{Distribution of the results obtained by all the codes (one code per row with different configurations) when fitting the $z\sim 1$ with IR measurements. The name of the code alone means the fits do not include IR measurements. "wIR" means the fits include IR measurements. "wAGN" means the code allows for an AGN component in the fit. Columns show stellar mass, SFR, and $A_V$. The color code is purely graphical to ease the distinction from one row to the next.}
\label{fig:z1IRmass}
\end{center}
\end{figure*}

Here, we discuss how crucial is the inclusion of IR data when measuring the physical parameters of galaxies, and outline some important caveats. The IR is especially important to constrain dust emission, and thus dust attenuation and SFRs (see e.g. \citealt{buat2005,dacunha2008,davies2017}, and the detailed introduction by \citealt{pappalardo2021}). However, the large point spread functions of some IR facilities (e.g., \textit{Spitzer} and \textit{Herschel}) can hamper our ability to properly separate individual sources and can easily cause biases in the photometric measurements at $z>1$. Although precious work has been done to deblend IR data in order to produce reliable galaxy catalogues (e.g., \citealt{liu2018}) and similar problems are getting resolved with the JWST, it is important to be aware of possible issues and make decisions to identify the best dataset for the specific science questions.

Six codes among the fourteen we have presented can process IR data along with the UV to NIR wavelength range. Using these codes, we can compare the outputs obtained including or not including the IR portion of the SED in the fit.

We present in Figure~\ref{fig:z1IRmass} the distribution of the results obtained running the different SED fitting codes that model the IR emission on our sample number 2 (see Section~\ref{sec:data}). The sample includes 107 galaxies with S/N in MIPS/24$\mu$m larger than 3 and covers the mid-IR and IR with 7 bands (5.8, 8.0, 24, 70, 100, 160, and 250~$\mu$m). We show the results when including and not including the IR emission in the fit. The distributions for the fits without IR are different than those presented in Figure~\ref{fig:z1mass} because sample 2 is a subset of sample 1. AGNFitter, CIGALE, MAGPHYS, and Prospector can include the contribution by AGN light. We note that AGNFitter is specifically designed to fit the AGN contribution in the IR and is not optimal to fit the UV-to-NIR alone, thus its default output is with IR and with AGN. 

We observe a trend in SFR and $A_V$: fits without the IR photometry seem to return larger values of these parameters (by about a factor of 2) compared to fits that include the IR photometry, suggesting that the correction for dust attenuation was overestimated when fitting the UV-to-NIR alone. This is the case for galaxies with moderate SFRs and moderate dust attenuation (i.e., not luminous or ultra luminous infrared galaxies - LIRGs or ULIRGs) where the dust attenuation prior can easily bias the results towards large dust attenuation values rather than towards small dust attenuation values. Such effect can be mitigated by choosing an exponential prior for the dust attenuation (large weight at low $A_V$ values and small weight at high $A_V$ values) instead of a flat prior (same weight for all $A_V$ values). The codes that return more similar SFR values with and without fitting the IR bands are those with more flexible SFHs (BAGPIPES and Prospector). In these two cases, the flexible SFHs allow the fit to find solutions with old stellar populations rather than adding more dust to match a \textit{red} SED. Thus, more flexible SFHs also help mitigate the effects caused by a lack of IR coverage. The inclusion of AGN models further lowers the estimates, which is expected since part of the light is now interpreted as dust emission heated by AGN rather than young stars.

\begin{figure*}
\begin{center}
\includegraphics[width=\textwidth]{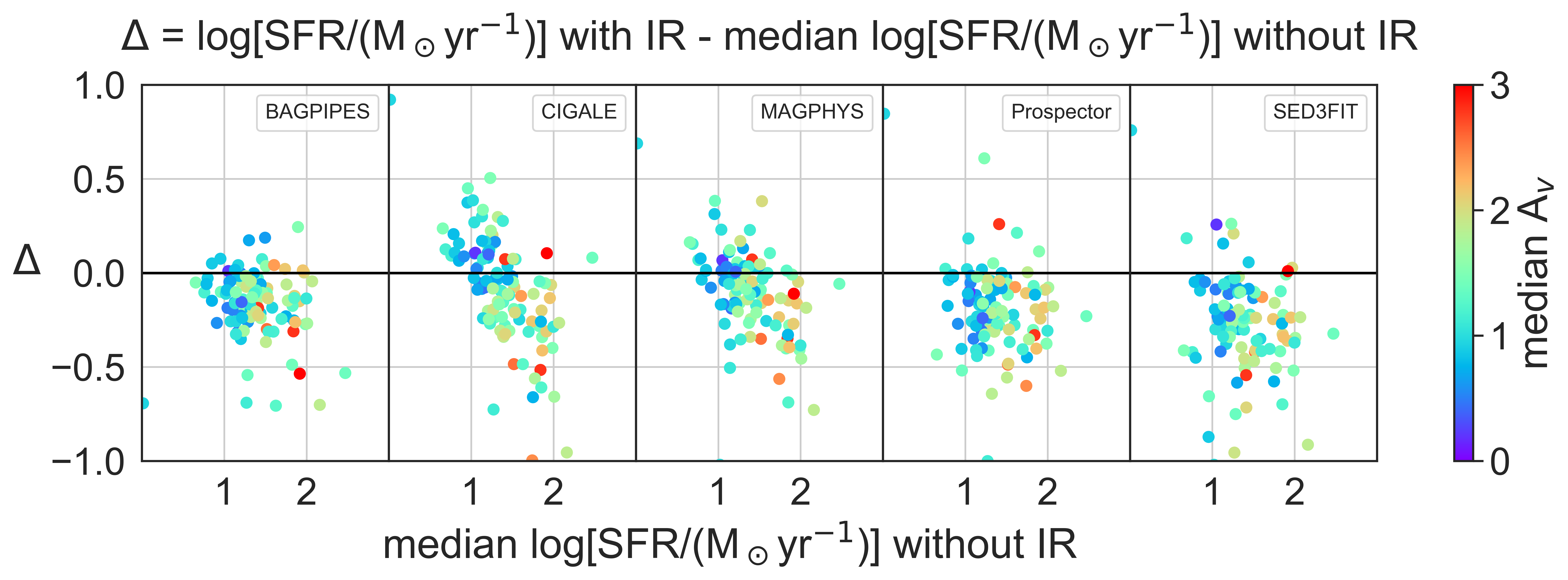}
\caption{For the five codes that can fit IR measurements, we plot the difference between the SFR estimate including IR measurements and the median of the distribution of the SFR best-estimates obtained with all codes without including IR measurements vs the median of the distribution of the SFR best-estimates obtained with all codes without including IR measurements. Points show individual galaxies in the $z\sim1$ sample with IR measurements and are color coded by the median of the distribution of the $A_V$ best-estimates obtained with all codes without including IR measurements. The black line marks zero, i.e. no difference between the two SFRs.}
\label{fig:noIRvsIR}
\end{center}
\end{figure*}

Figure~\ref{fig:noIRvsIR} shows, for each galaxy, the measurement of SFR obtained by each individual code including the IR photometry versus the median of the distribution of the SFR best-estimates (i.e., medians of the PDFs) obtained with all codes fitting only the UV to NIR wavelength range. We find a slight bias towards higher SFRs when the IR data are not included in the fit. This is due to an overestimate of the dust attenuation which is otherwise better constrained when IR measurements are included. This bias is of the order of 0.2 dex and in the case of CIGALE it is a strong function of the SFR and $A_V$. This bias can be reduced by allowing for more freedom in the dust attenuation curve (e.g. allowing for variations in the slope of the attenuation law) and by including enough freedom in the SFHs to make the spectrum ``red'' by adding old stars instead of by increasing the dust attenuation. In this context, SFHs with a rising-and-falling shape and freedom in the time when the peak of star formation happens are preferred. The codes that allow these kind of variations (BAGPIPES, Dense Basis, P12, and Prospector) return small values of $A_V$ without fitting the IR wavelength range (see Figure~\ref{fig:z1mass}). 

\begin{figure*}
\begin{center}
\includegraphics[width=\textwidth]{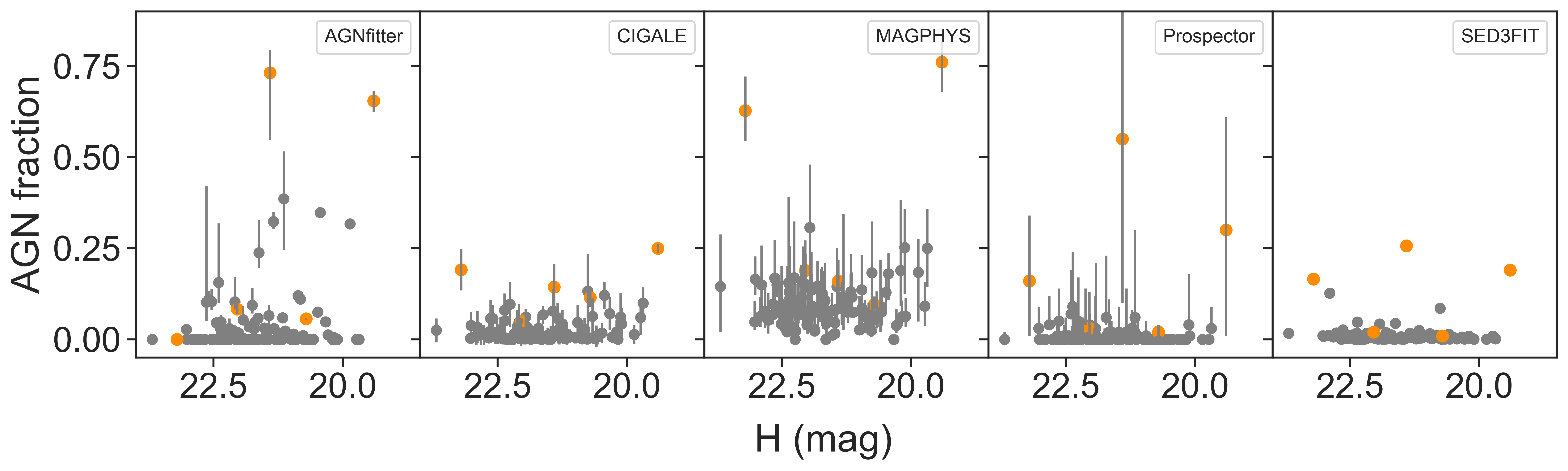}
\caption{For the five codes that can include an AGN component in the fit, we plot the AGN fraction vs the H band magnitude. Points are individual galaxies in the $z\sim1$ sample with IR measurements. Orange circles mark galaxies identified as AGN according to their IRAC colors \citep{lacy2004,lacy2007}. The AGN fraction is defined as the luminosity due to the AGN over the total luminosity in the wavelength range between 8~$\mu$m and 1000~$\mu$m.}
\label{fig:agnfrac}
\end{center}
\end{figure*}

Figure~\ref{fig:agnfrac} shows the AGN fraction as a function of H band magnitude for the five codes that can include an AGN component. The AGN fraction is defined as the luminosity due to the AGN over the total luminosity in the wavelength range between 8~$\mu$m and 1000~$\mu$m. Five galaxies are identified as AGN according to their IRAC colors \citep{lacy2004,lacy2007} (orange marks). The AGN fractions derived by the five different codes are fairly comparable and generally smaller than 25\%. The few galaxies for which each code infers AGN fractions larger than the bulk of the sample also show AGN signatures in their IRAC colors. The codes that predict large AGN fractions measure slightly lower SFRs compared to the other codes. This demonstrates the degeneracy between AGN and star formation contribution in the mid-IR. The number of IR bands fitted and their S/N also affect the measured AGN fractions and their accuracy. More in depth discussions about the effects of choosing specific AGN templates or of the quality of the IR photometry are outside the scope of this work, but we can conclude that the inclusion of AGN components in SED fits is generally beneficial and enables more robust studies of galaxies with strong AGN features, as opposed to including them solely for the purpose of constraining AGN contamination fractions.

\subsection{Distribution of results for the $z\sim3$ sample}

\begin{figure*}
\begin{center}
\includegraphics[width=0.3\textwidth]{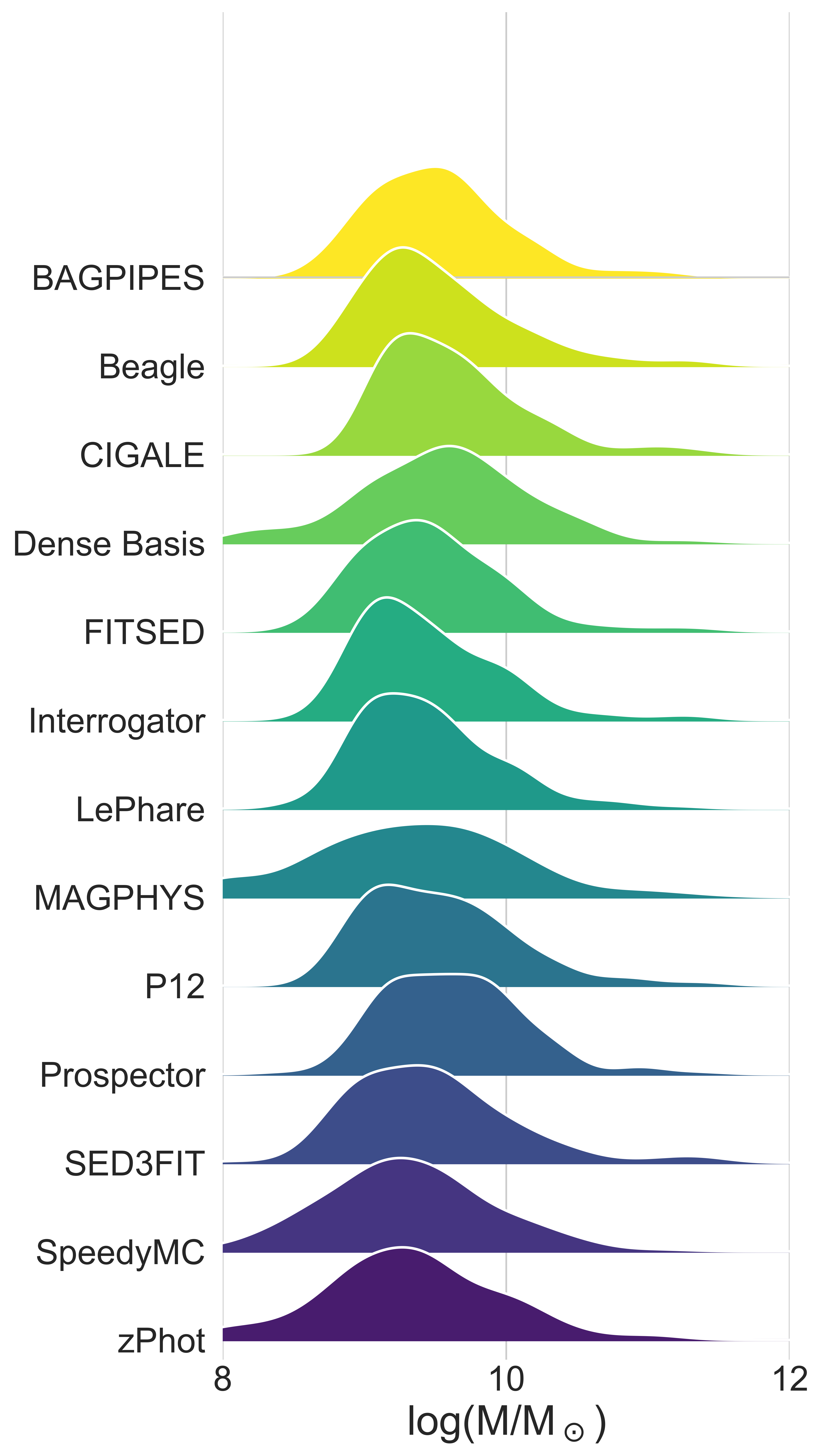}
\includegraphics[width=0.3\textwidth]{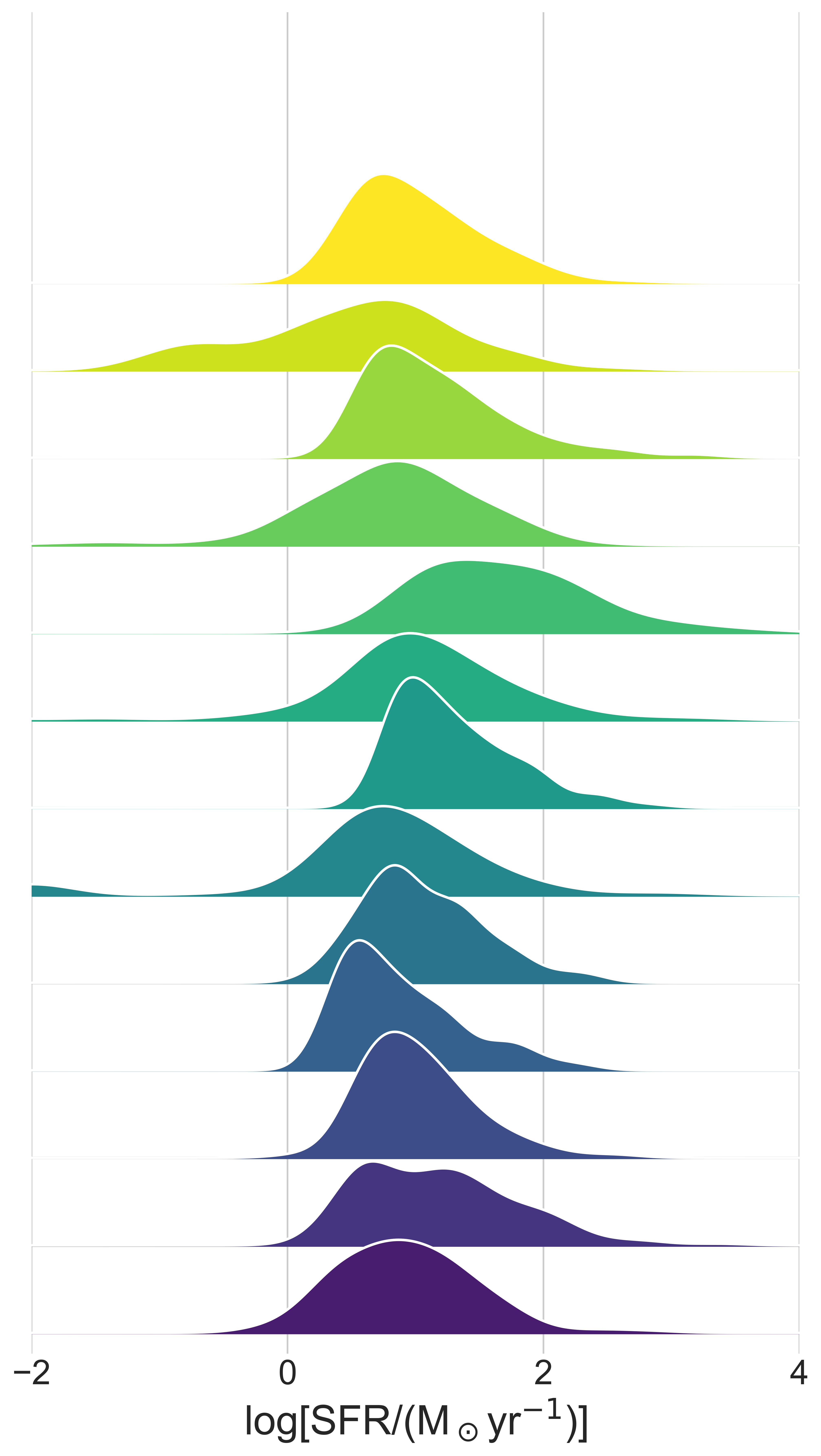}
\includegraphics[width=0.3\textwidth]{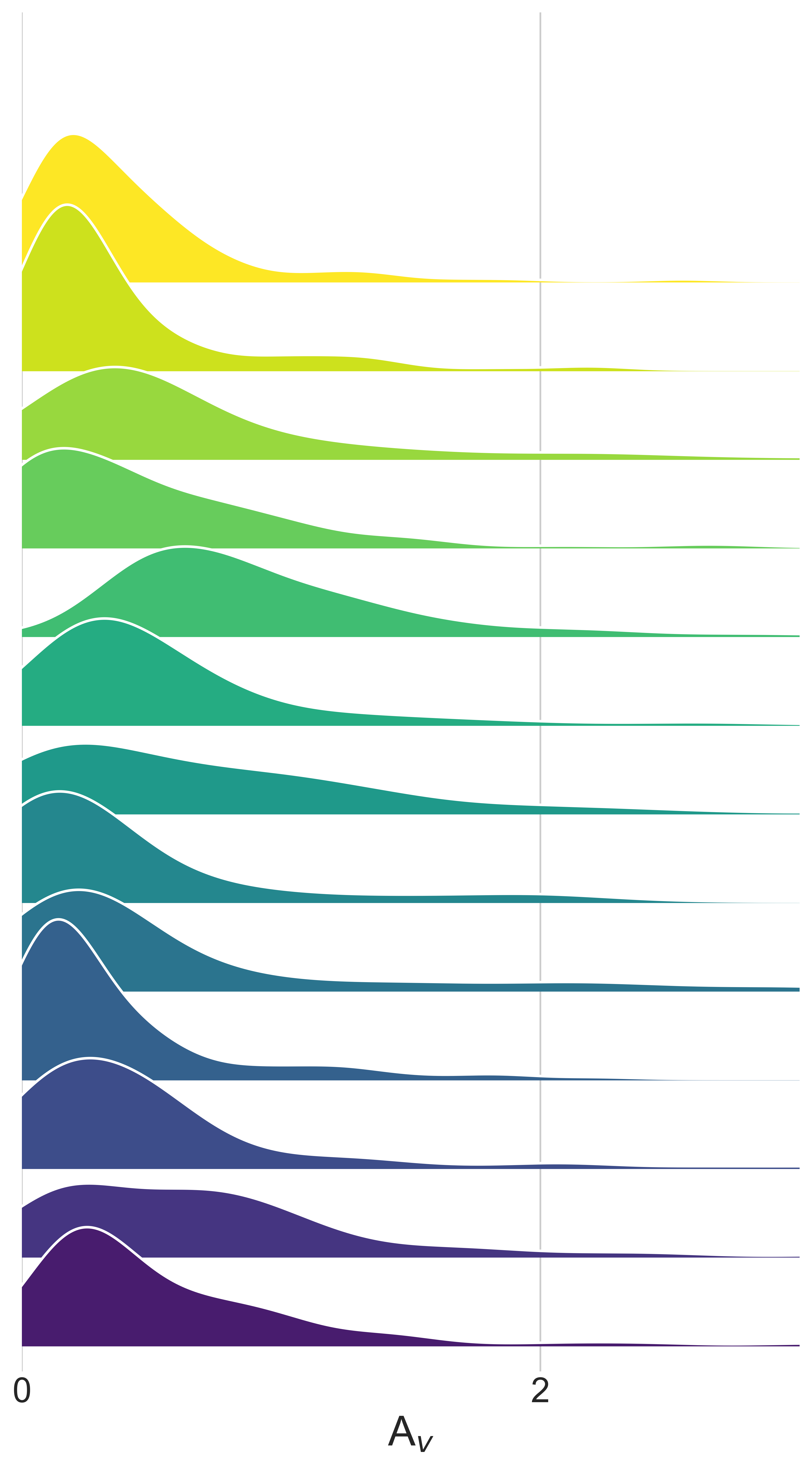}
\caption{Distribution of the results obtained by all the codes (one code per row) when fitting the $z\sim3$ sample. Columns show stellar mass, SFR, and $A_V$. The color code is purely graphical to ease the distinction from one row to the next.}
\label{fig:z3mass}
\end{center}
\end{figure*}

In Figure~\ref{fig:z3mass}, we present the comparison of the different codes run on the $z\sim 3$ sample. The trends are similar to those we pointed out for the $z\sim 1$ sample: the mass distributions are all very similar; and the codes that return larger SFRs also return larger $A_V$ values. Although at $z\sim 3$ the rest-frame wavelength range covered by the photometry only extends to $\sim$1.2$\mu$m (while at $z\sim1$ it extends to $2.2\mu$m), the Universe is much younger. The emission by young stellar populations is less degenerate in age than the emission by old stellar populations, thus it is easier to disentangle the contributions by different stellar populations (i.e., the SFH) to the SED. This reduces the degeneracies among the parameters and with that the possible discrepancies among the different codes. We note that all codes point to a low level of dust attenuation which is expected for H-band detected galaxies at $z\sim3$, because the H band traces the blue side of the optical which becomes too faint for very dusty galaxies.

\section{Impact on science results}
\label{sec:impact}

\subsection{SFR - stellar mass relation}
\label{sec:sfrmass}

\begin{figure*}
\begin{center}
\includegraphics[width=1\textwidth]{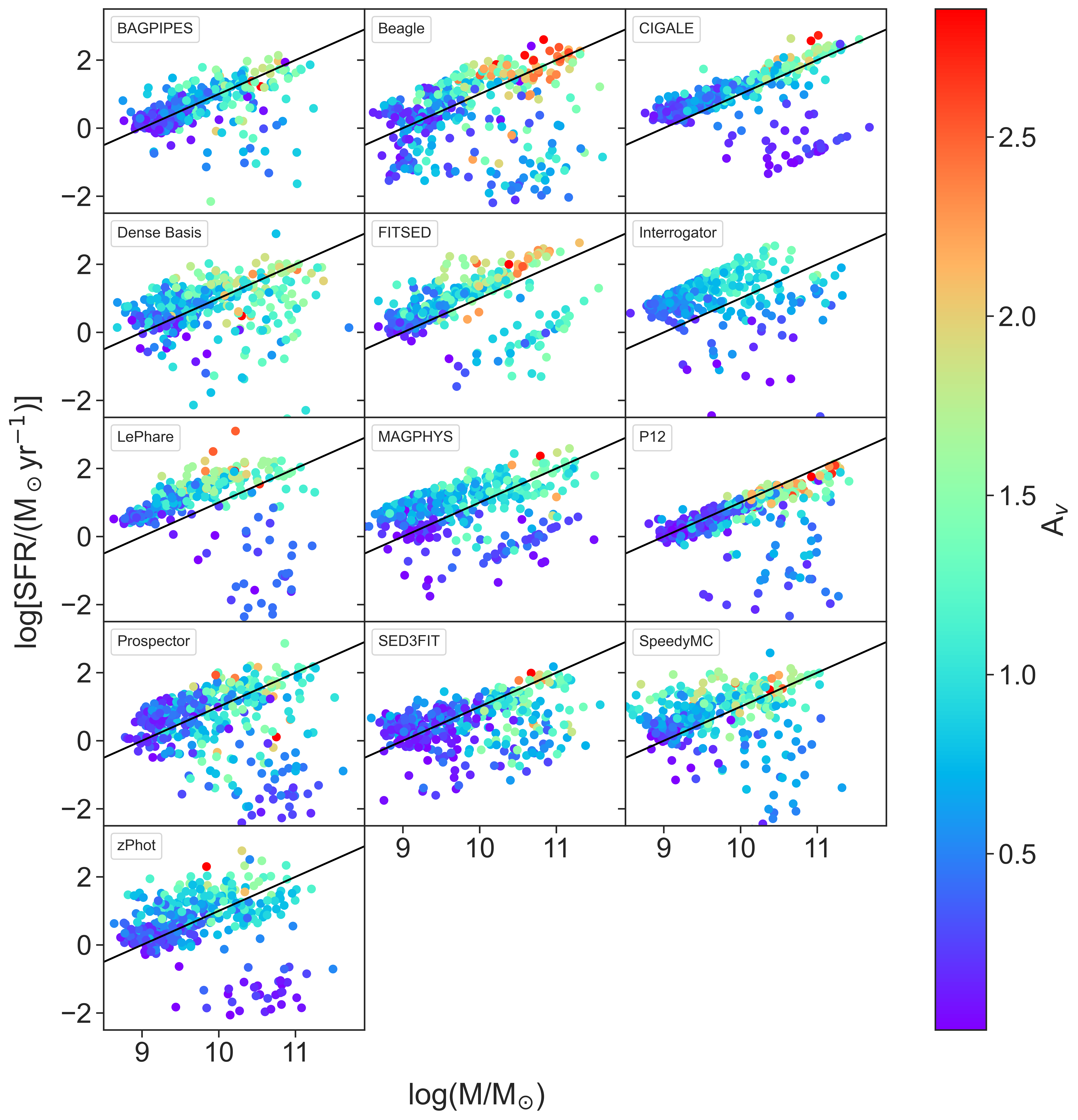}
\caption{SFR-stellar mass relation for all the codes when fitting the $z\sim1$ sample without IR measurements. Points are individual galaxies and the color code is $A_V$. The black line marks constant specific SFR of 1 Gyr$^-1$ in every plot and it is not a fit to the points.}
\label{fig:msall}
\end{center}
\end{figure*}

The SFR - stellar mass (SFR-$M_{*}$) correlation shows the current growth rate of galaxies \citep{brinchmann2004,noeske2007,whitaker2014,speagle2014,leja2021}. It is generally described by its intercept, slope, and scatter (intrinsic or observed) assuming a linear correlation. The intercept marks the level of star formation activity at a given mass and epoch and is observed to increase with redshift \citep{whitaker2012,schreiber2015}. The slope is used to measure the differential rates at which higher mass galaxies form stars compared to lower mass galaxies. It is generally linked to the different feedback mechanisms that regulate star formation across different populations of galaxies (see e.g. \citealt{schreiber2015,popesso2019a,popesso2019b,davies2021}). The scatter is associated with the burstiness of star formation on short timescales, because galaxies can scatter above and below the relation on timescales of 5-100 Myr. A large scatter can also be caused by a spread of galaxy ages on long timescales in the sense that galaxies can keep high or low SFR levels for their entire life and thus assemble their stellar mass faster or slower compared to the average population \citep{tacchella2016,ciesla2017,pandya2017,boogaard2018,matthee2018,katsianis2019,berti2021,curtislake2021}. A small scatter would instead be consistent with all galaxies having very similar SFHs. These three measurements are often used to constrain growth and feedback mechanisms in simulations of galaxy evolution. It is thus important to understand the potential effects of different modeling and prior assumptions on the slope, intercept, and scatter of measurements of the SFR-$M_{*}$ relation. 

Figure~\ref{fig:msall} shows the relations at $z\sim1$ derived by the individual codes. The color code is set by the dust attenuation derived by each code. The black lines mark a SSFR of 1 Gyr$^{-1}$ for comparison purposes. Globally, all codes measure a correlation between SFR and stellar mass. All codes measure more dust attenuation at higher masses and little to no dust attenuation for the quiescent galaxies that fall well below the relation.

Conversely, the normalization of the relation as well as the measured scatter can differ by up to 0.5dex. Also the locus of the galaxies that fall below the relation can vary considerably. We note that intermediate and low SFR galaxies generally show the highest uncertainties in this measurement (see Section~\ref{sec:uncert}) because the signal in the SED by SFRs lower than $10^{-1} M_{\odot}/yr$ is not strong enough to be detected with high confidence.

The similarities and differences we observe can be attributed to the specific choices of modeling assumptions and priors in the individual codes. The choice of stellar population model, the prescription for the amount of mass returned to the interstellar medium in aging stellar populations, and the assumptions for the dust attenuation law can change the stellar mass shifting galaxies horizontally in Figure~\ref{fig:msall}. The choice of SFH can affect the derived ages and thus the derived SFRs, causing, to first order, shifts up and down in the plots. Massive galaxies are generally older than low-mass galaxies, thus the shift can affect differently low- and high-mass galaxies. In addition, SFH priors can also change the extent to which the oldest stellar populations contribute to the stellar mass, by up to a factor of ~0.2 dex \citep{leja2019}. 

The prior on the SFHs affects the prior on the SSFRs which can cause artificial boundaries and affect the measurement of the scatter in the relation. We note here that the definition of SFR (essentially the timescale on which it is calculated) and the data used to measure it can affect systematically the derived SFR values and thus, to make meaningful comparisons, it is imperative to keep the definition and the datasets consistent. For example, \cite{caplar2019} shows there can be a nearly 0.3 dex difference between the H$\alpha$ and UV measured scatter for different types of SFHs. In this paper, the dataset is the same for all codes and the SFR is defined as the average SFR in the last 100 Myr which is appropriate for photometric measurements. This is because the light in the rest-frame UV is produced by O and B stars whose lifetimes range between $\sim$10 and $\sim$300 Myr.

\begin{figure}
\begin{center}
\includegraphics[width=0.45\textwidth]{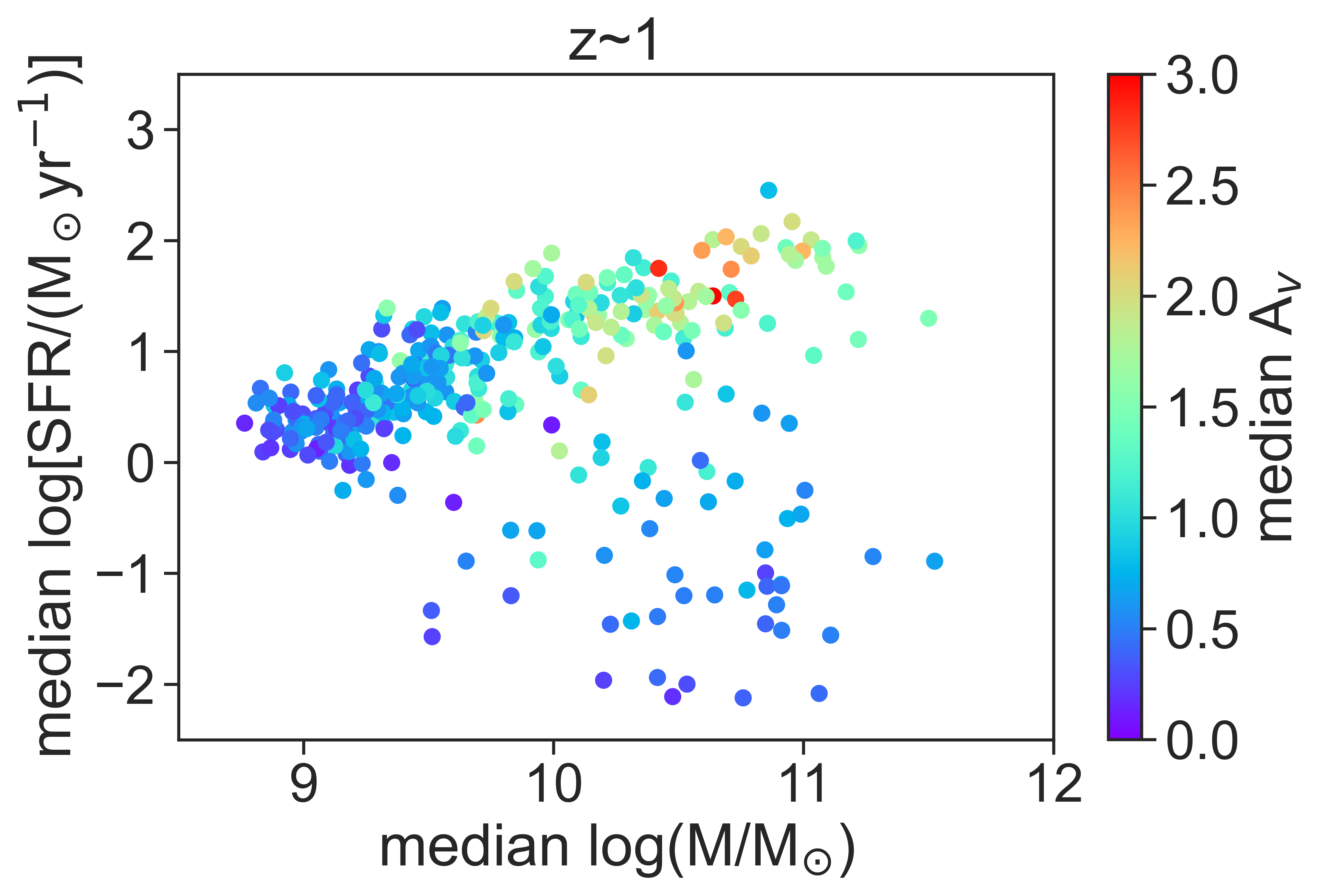}
\includegraphics[width=0.45\textwidth]{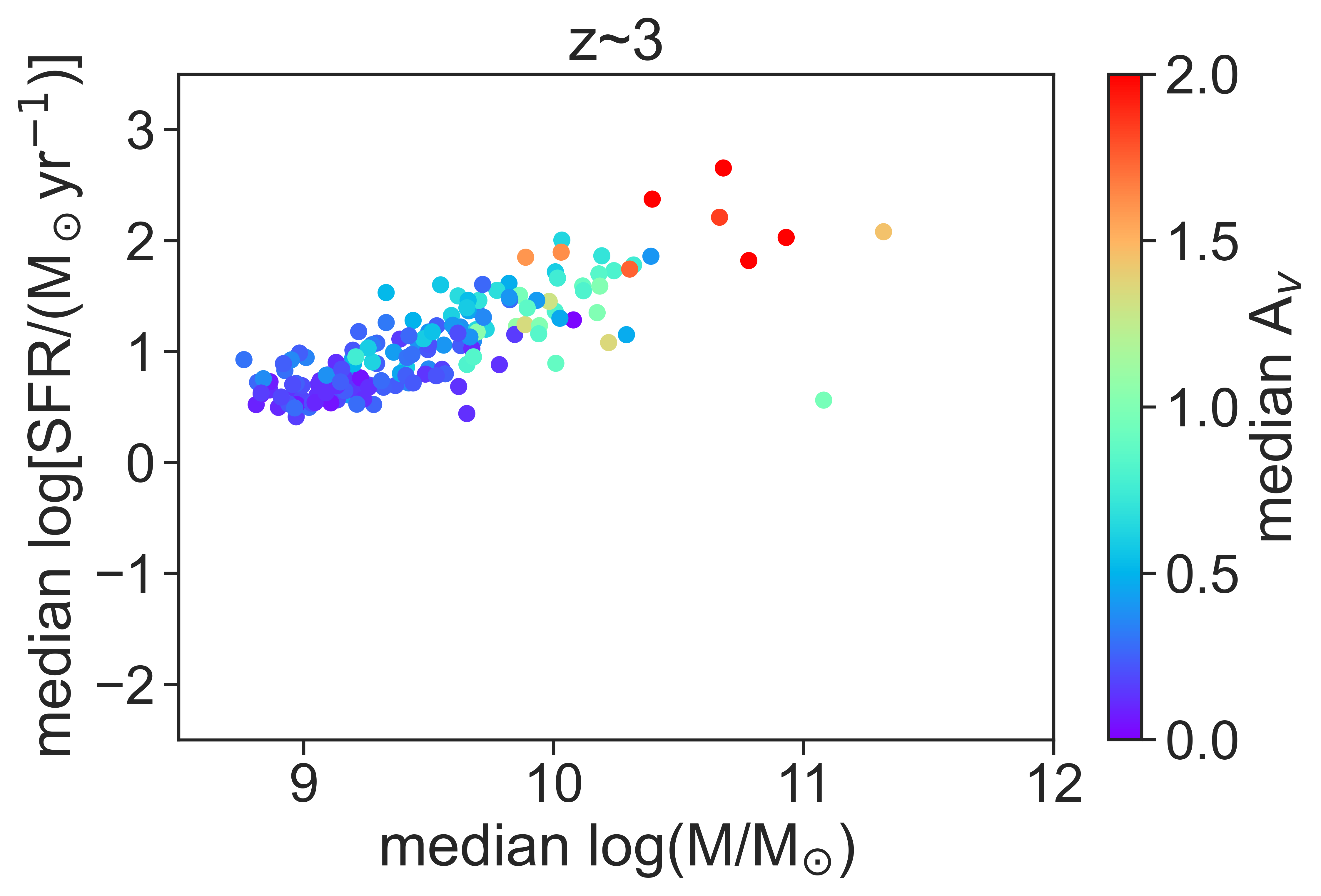}
\caption{All plots show the SFR-$M_{*}$ relation using the median stellar mass and median SFR from all codes for every galaxy in the $z\sim1$ sample (top) and $z\sim3$ sample (bottom). Points are color coded by the median $A_V$ from all the codes.
}
\label{fig:zzms}
\end{center}
\end{figure}

In this exercise, all codes are programmed with reasonable assumptions for the majority of the priors given this specific dataset and we encouraged the participants to keep track of the ``goodness'' of the fits and deliver results only if ``good''. The ``goodness'' of a fit can be assessed in many ways (i.e., by checking the $\chi^2$ values, by comparing the posteriors to the priors, etc.) and this task was left in the hands of the participants. Comparing ``goddness'' of fits among codes is beyond the scope of this exercise. With these assumptions, it is not possible for us to decide which SFR-$M_{*}$ relation is the \textit{true} one, but it is fair to say that different scientific conclusions could be derived on the modes of galaxy formation from any individual SFR-$M_{*}$ relation. This is a first hint of how important it is to account for modeling uncertainties.

For every galaxy in the sample, there is a distribution of estimates from all the codes and therefore a median and a dispersion of such distribution can be calculated. Figure~\ref{fig:zzms} shows the SFR-$M_{*}$ relation at $z\sim1$ and $z\sim3$ using the median values of stellar mass, SFR, and A$_V$ (for the color code) for each galaxy (we will address the dispersion on these measurements in the next section). The median estimates of stellar mass and SFR take into account the difference in priors and modeling assumptions in deriving these quantities. As such, these estimates (to some extent) marginalize over these different assumptions and therefore provide more robust estimates of the mass and SFR.

\subsection{Observational and modeling uncertainties of the physical parameters}
\label{sec:uncert}

\begin{table*}
\begin{center}
\begin{tabular}{l | c c c | c c c}
    & \multicolumn{3}{c}{$z\sim1$} & \multicolumn{3}{c}{$z\sim3$} \\
    & $\log(M_{*}/M_{\odot})$ & $\log(SFR/(M_{\odot} yr^{-1})$ & $A_V$ & $\log(M_{*}/M_{\odot})$ & $\log(SFR/(M_{\odot} yr^{-1})$ & $A_V$  \\
    & (dex) & (dex) & (mag) & (dex) & (dex) & (mag) \\
\hline
\hline
BAGPIPES    & 0.09 & 0.12 & 0.08 & 0.14 & 0.11 & 0.08\\
BEAGLE      & 0.07 & 0.15 & 0.14 & 0.08 & 0.33 & 0.09\\
Cigale      & 0.17 & 1.07 & 0.27 & 0.14 & 0.36 & 0.12\\
Dense Basis & 0.10 & 0.11 & 0.15 & 0.10 & 0.18 & 0.15\\
FITSED      & 0.07 & 0.13 & 0.14 & 0.22 & 0.22 & 0.16\\
Interrogator& 0.04 & 0.24 & 0.25 & 0.11 & 0.25 & 0.27\\
LePhare     & 0.21 & 0.42 & NA   & 0.39 & 0.43 & NA  \\
MAGPHYS     & 0.02 & 0.05 & 0.08 & 0.00 & 0.00 & 0.00\\
P12         & 0.06 & 0.14 & 0.16 & 0.08 & 0.16 & 0.14\\
Prospector  & 0.09 & 0.38 & 0.13 & 0.17 & 0.18 & 0.15\\
SED3FIT     & 0.10 & 0.10 & 0.01 & 0.06 & 0.09 & 0.10\\
SpeedyMC    & 0.04 & 0.10 & 0.09 & 0.00 & 0.14 & 0.11\\
zPhot       & 0.10 & 0.10 & NA   & 0.10 & 0.10 & NA  \\
\hline
\end{tabular}
\caption{Median observational uncertainties from all codes for the galaxies in the $z\sim1$ and $z\sim3$ samples, fitting the UV-to-NIR SEDs.}
\end{center}
\label{tab:unc}
\end{table*}

\begin{figure*}
\begin{center}
\includegraphics[width=0.47\textwidth]{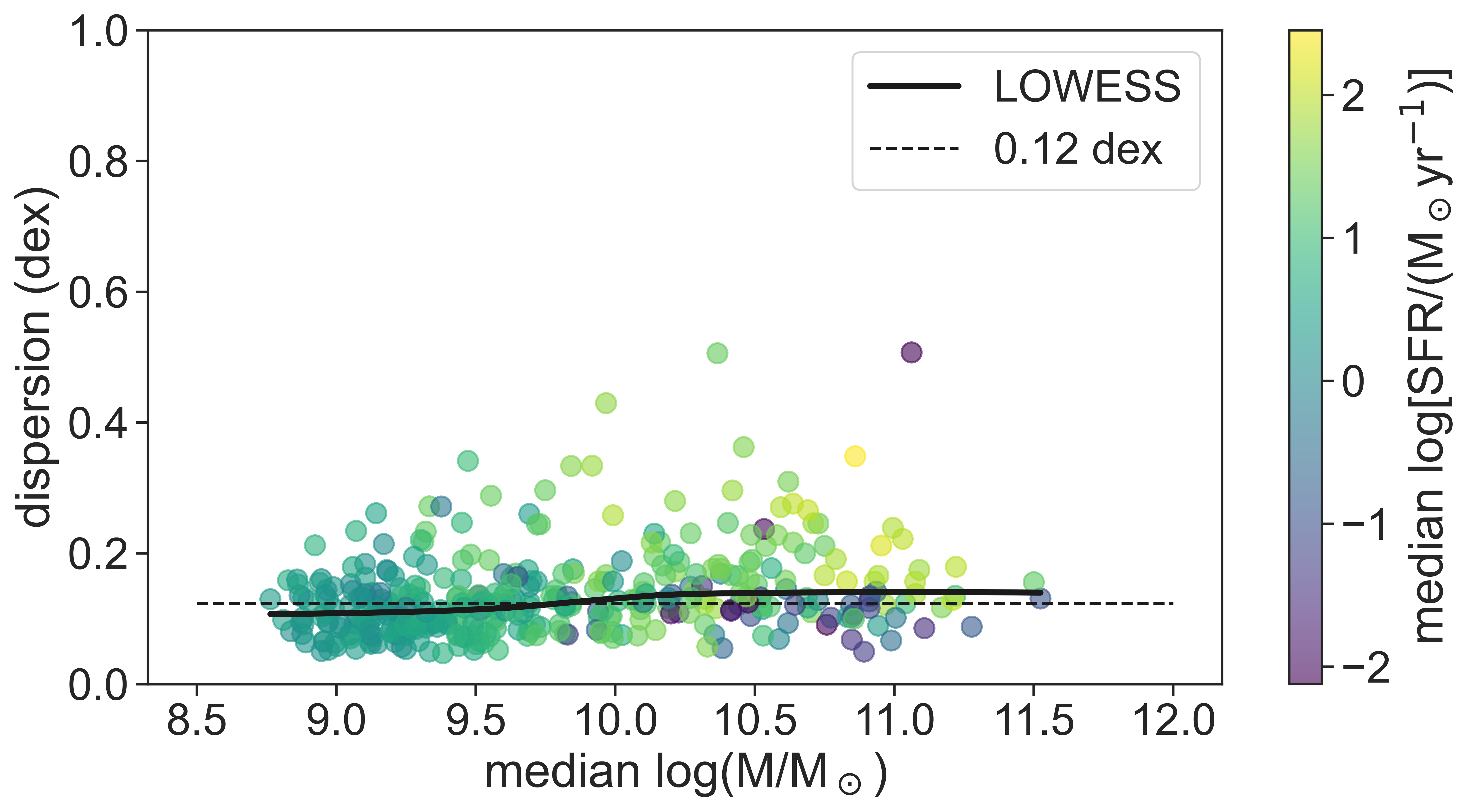}
\includegraphics[width=0.47\textwidth]{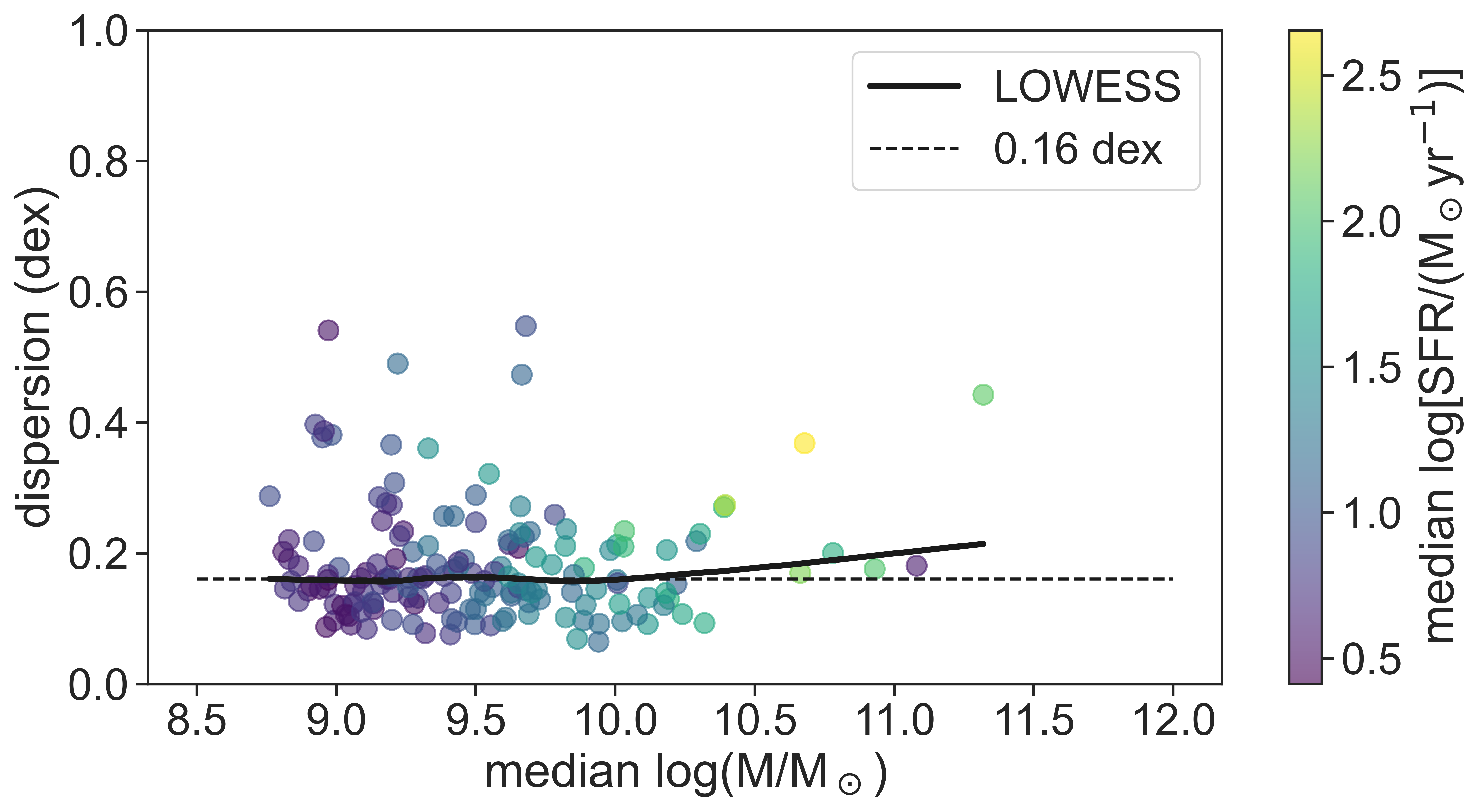}
\includegraphics[width=0.47\textwidth]{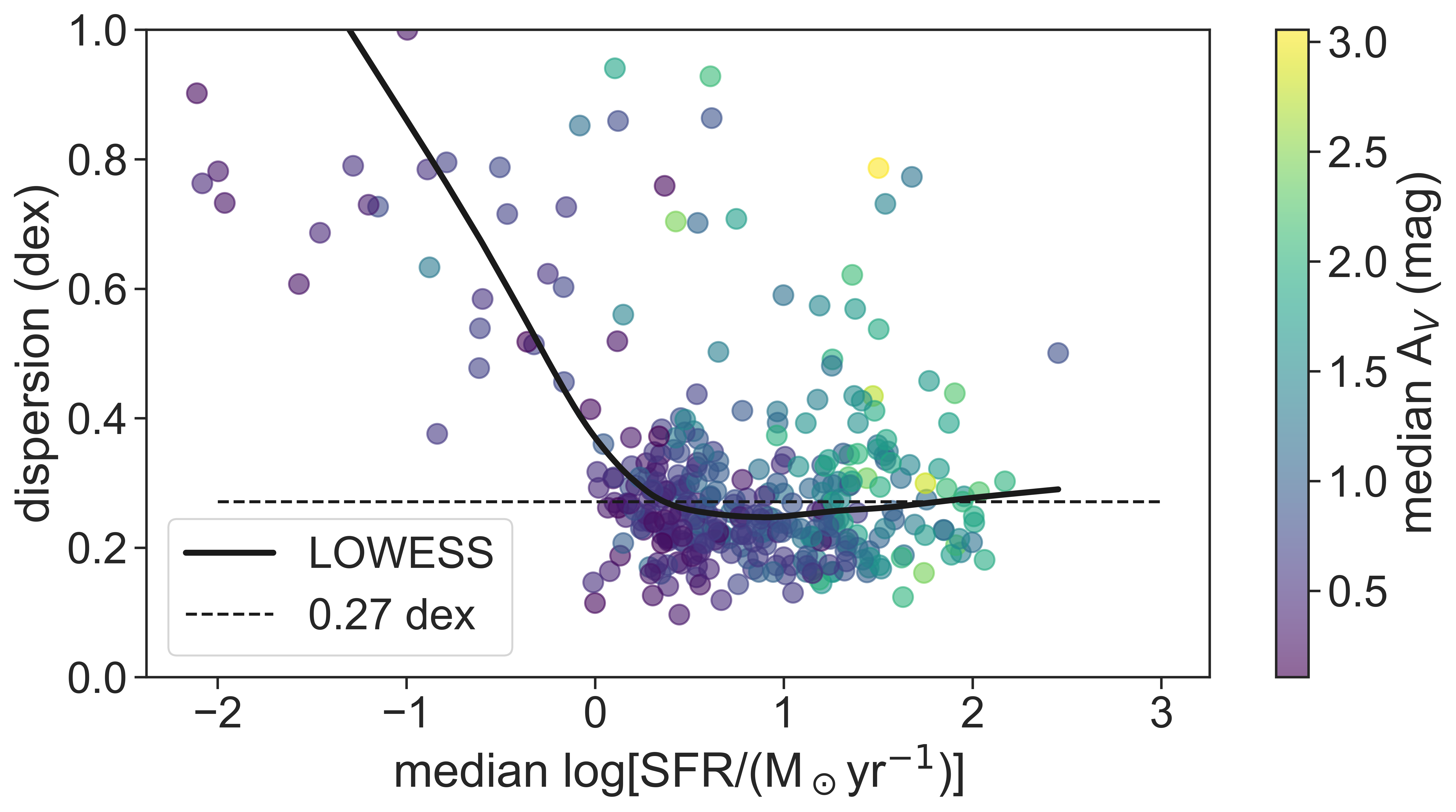}
\includegraphics[width=0.47\textwidth]{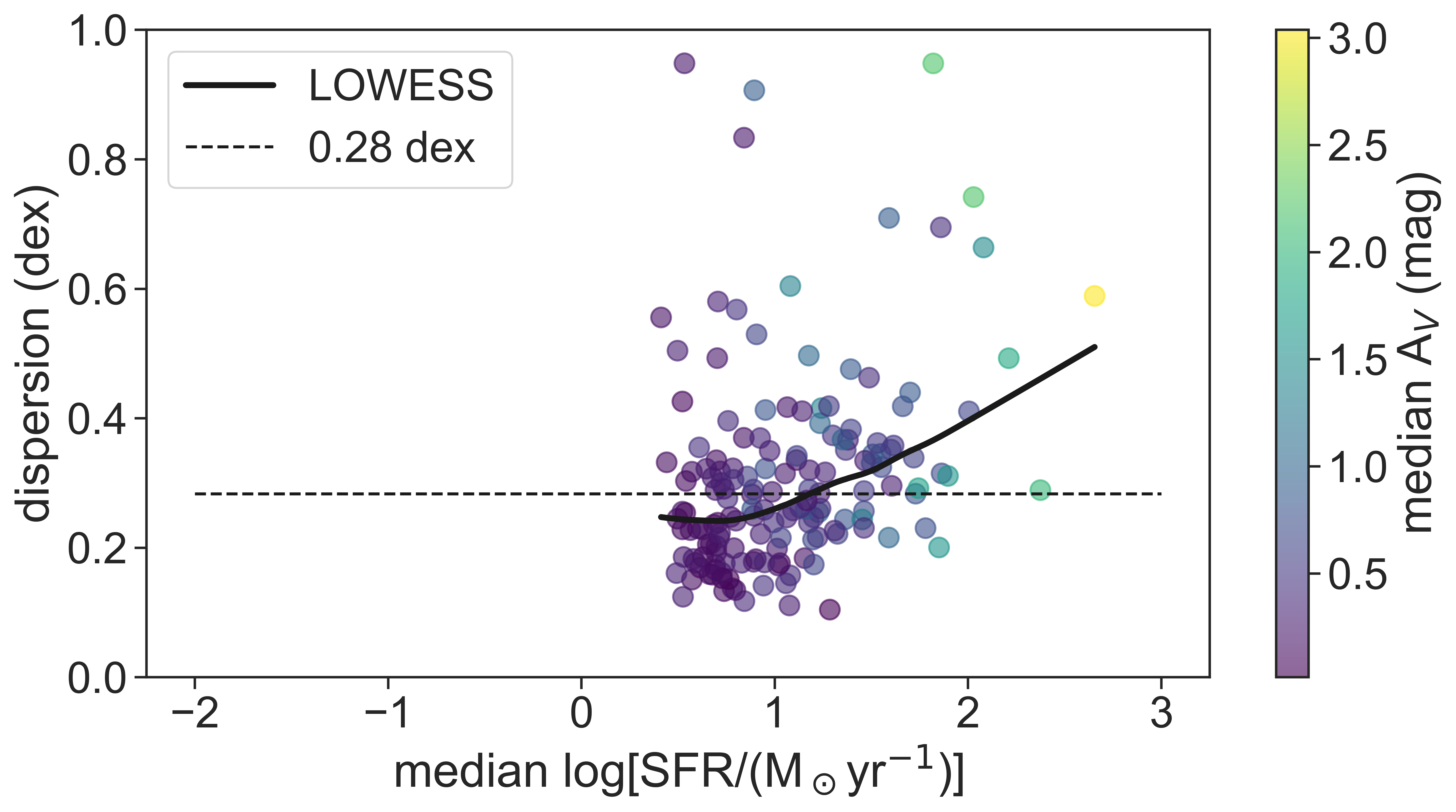}
\includegraphics[width=0.47\textwidth]{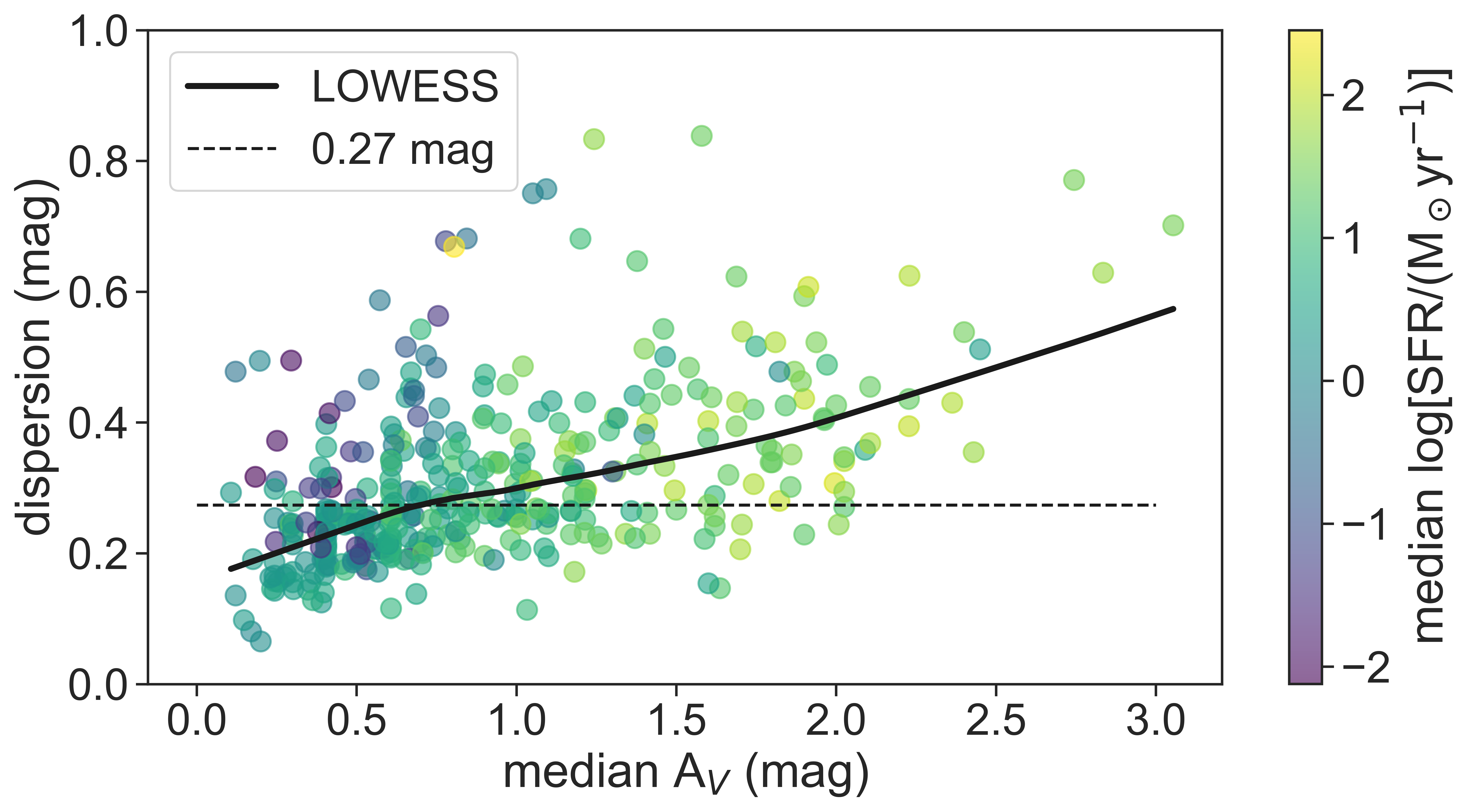}
\includegraphics[width=0.47\textwidth]{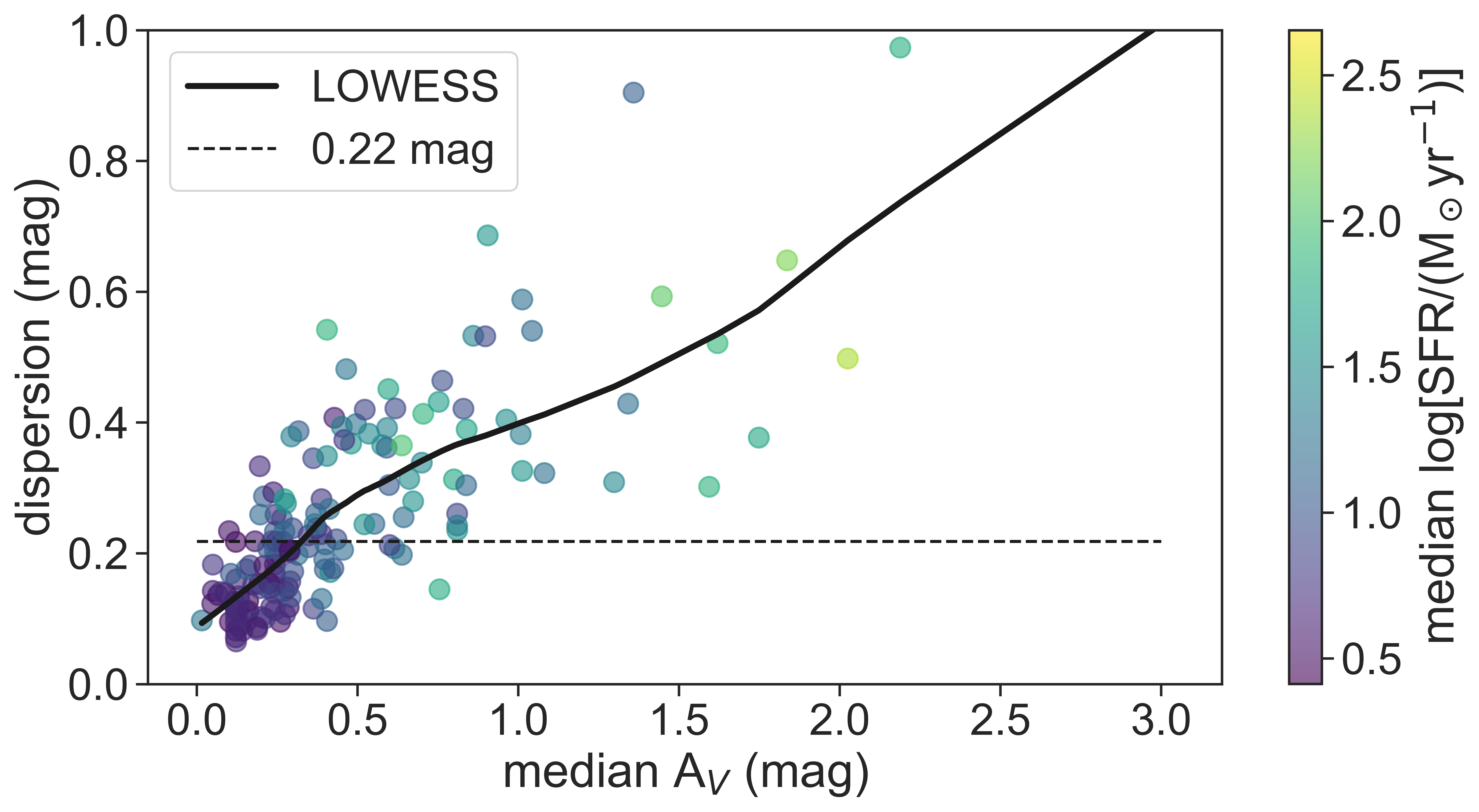}
\caption{In all plots, each point represents the dispersion in the estimate of the physical parameter from all the codes vs the median parameter from all the codes. The dispersion is defined as the half difference between the 16th and 84th percentile of the distribution of the parameter derived from all the codes. Stellar mass is plotted on top, SFR in the middle, and $A_V$ at the bottom. The left side uses measurements from the $z\sim1$ sample, while the right side uses measurements from the $z\sim3$ sample. In each plot, the dotted line marks median value for all the points, while the solid line marks the Locally Weighted Scatterplot Smoothing (LOWESS) to the data.}
\label{fig:z1variability}
\end{center}
\end{figure*}

With the available measurements, we can quantify more realistic uncertainties that includes both \textit{observational} (i.e. propagated from the uncertainties in the photometric measurements) and \textit{modeling} (i.e., caused by the specific choices in the models) contributions. To calculate the total uncertainty for each galaxy, the observational uncertainty needs to be added in quadrature to the modeling uncertainty. Table~\ref{tab:unc} summarized the median observational uncertainties returned by the codes for the three parameters of interest, for the sample at $z\sim1$ and $z\sim3$, fitting the UV-to-NIR SEDs.

For the $z\sim 1$ sample, Figure~\ref{fig:z1variability} shows the variability of stellar masses, star formation rates, and $A_V$, along with a median value for all points and the Locally Weighted Scatterplot Smoothing (LOWESS, a non-parametric smoothing algorithm) to the data. To compute this we use the estimates of the given parameter for individual galaxies from each code and estimate the dispersion, here quantified as $(X_{84} - X_{16})/2$ where $X$ is the distribution of either $\log(M_{*}/M_{\odot})$ or $\log(SFR/(M_{\odot} yr^{-1})$ or $A_V$ and the subscripts denote the 16th and 84th percentiles respectively. The total uncertainty (including the observational and modeling components) should thus be the observational uncertainty each individual code returns added in quadrature to the modeling uncertainty (i.e., dispersion) measured here.

For the stellar mass, we find that the median of the dispersion is $0.12$~dex at $z\sim1$ and is largely independent of stellar mass itself, but can be mildly correlated with the SFR, such that the modeling uncertainty is larger for high SFRs. At $z\sim3$, the median of the dispersion is $0.16$~dex and is similarly not a function of stellar mass itself.

For the SFR, we find that the amount of variability depends on the inferred SFR value itself, and thus it depends on the flux in the rest-frame UV (these fits do not include IR measurements). Galaxies with low SFRs show faint UV luminosity with low S/N. This hamper our ability to measure accurate SFRs, in the sense that it is generally possible to infer that the SFR is low (or effectively consistent with no star formation), but it is hard to constrain the exact value. We measure the median of the dispersion to be on average 0.27~dex at $z\sim1$ and 0.28~dex at $z\sim3$, with low-SFR galaxies with dispersions as high as 1~dex.

The modeling uncertainty in $A_V$ is a strong function of the value itself. Highly attenuated galaxies are, not surprisingly, the harder to interpret. At $z\sim1$, the median of the dispersion is 0.27~mag and it is larger than 0.4~mag for $A_V>2$. At $z\sim3$, we find a similar correlation between the dispersion and the $A_V$ value itself. The median of the dispersion is 0.22~mag with values as high as 1~mag for large $A_V$ values.

\section{Lessons learned}
\label{sec:lessons}

In this Section, we will provide some suggestions on how to approach SED fitting and understanding the power and limitations of this technique. We will list a series of good practices that should be obvious to the experienced users, but might be of help for those who are just starting working with SED fitting. As we have seen, it is important to be aware of the tool characteristics and make a few appropriate decisions (e.g., type of tool, availbale ingredients, flexibility of the priors, etc.). We will start this Section with a flowchart to check whether the tool of choice is appropriate for the selected sample. We will then parse a list of ``knobs'' that can potentially be adjusted in a SED fitting tool. Finally, we will talk about the outputs of the tools and the definition of ``results''.

\subsection{SED Fitting Flowchart}

\begin{figure*}
\begin{center}
\includegraphics[width=0.9\textwidth]{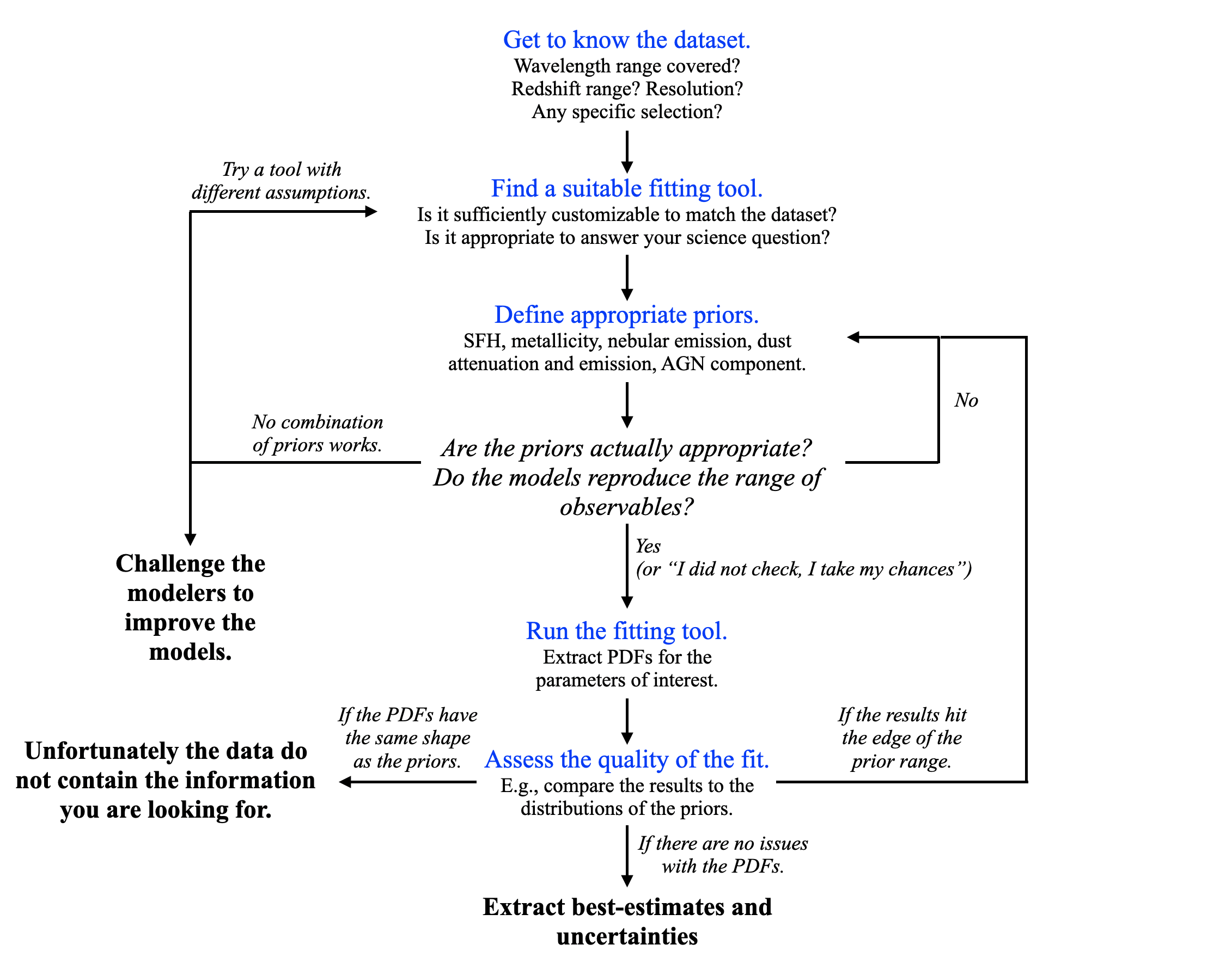}
\caption{This chart shows the basic flow of the SED fitting process, that requires understanding the dataset, setting the priors, testing the assumptions, running the selected tool, and evaluating the results.}
\label{fig:chart}
\end{center}
\end{figure*}

Running a SED fitting tool requires a few steps and necessary customization as seen in Figure~\ref{fig:chart}. First of all, it is critical to choose the appropriate SED fitting tool for the selected dataset. There needs to be a match in wavelength coverage and supported spectral resolution. The second critical step is the customization of the priors. This is rarely a one-time activity, and is often an iterative step. We note that over-customizing the tool can make the comparison to other results more challenging, thus it should be done only when necessary. A common technique to check whether the selected priors are appropriate for the sample is the comparison between observed and model-predicted colors (see e.g., \citealt{pacifici2015}). It is always more appropriate to bring the models to the observed redshift than to modify the observations converting them to rest frame. We will go into more details in Section~\ref{sec:setpriors}. The last step is the analysis of the results and this too can be used to assess the appropriateness of the priors. We will talk more about results in Section~\ref{sec:def_results}.

\subsection{Setting up the appropriate models}
\label{sec:setpriors}

A set of ingredients is needed to generate SED templates that are used to extract physical parameters (photometric redshifts, stellar mass, star formation rates, age, extinction, etc) from the multi-waveband data. The priors on these ingredients are of crucial importance.

SED fitting tools can be generally divided into two groups: those where a library of models is created beforehand spanning a certain parameter space and all models are compared to each observation (e.g., Dense Basis, CIGALE, FITSED, LePhare, MAGPHYS, P12, SED3FIT, zPhot), and those where the library of models is built on the fly exploring only the range of the parameter space that matches the data more closely (e.g., AGNFitter, BAGPIPES, BEAGLE, Interrogator, Prospector, SpeedyMC). In both cases, it is the user who chooses that parameter space, deciding the range over which the individual ingredients can vary (or not) (e.g., IMF, type of star formation and metal enrichment history, stellar population models, nebular emission models, dust attenuation and emission models, AGN models).

The user has to specify a physically reliable range for these parameters and test that the produced templates match the selected observed sample. Such set of priors allows the user to rule out possible solutions that are nonphysical or simply very unlikely. For example:
\begin{itemize}
\item Low-redshift, massive galaxies are likely in a declining SFR phase, while high-redshift galaxies are likely in a rising SFR phase (see e.g. \citealt{madau2014}).
\item High-redshift galaxies tend to be more metal poor than low-redshift ones (see e.g. \citealt{maiolino2008}).
\item The SFR seems to be correlated with the stellar mass at all redshifts (see e.g. \citealt{whitaker2012,schreiber2015}).
\end{itemize}
If fitting a large sample of galaxies, it is important to account for all these possibilities because we do not necessarily know what types of galaxies we are looking at and restricting the priors might lead to missing a very interesting population of outliers. If instead the aim is to refine the fits of a specific population of galaxies, it is appropriate to restrict the priors to account for existing knowledge. Any model can return a $\chi^2$ and a likelihood, but it is in the hands of the user to set up the appropriate models for the specific observations.

There are at least two ways to validate the choice of the priors and they should both be adopted. The first is to compare observable quantities (e.g., emission line fluxes and ratios, spectral indices, photometric colors) between the model and the observations. An appropriate library of models will span a similar (if not larger) space in those observable quantities compared to the galaxies in the observed sample. The second way requires running the fitting tool first and analyze the distribution of the results. We will talk about this in the next Section (\ref{sec:def_results}).

This brings us to the concept that SED fitting can often be an iterative process where the user refines the priors to find the appropriate set for the specific set of observations. When the sample to be analyzed includes all sorts of galaxies, it is usually better to start with priors that allow for all possible combination of parameters and thus a very large parameter space. Alternatively, if the sample is selected in a specific way (e.g., spectroscopy data tell us that all the galaxies have low Balmer breaks), the parameter space can be tailored to match the observations (e.g., the dust attenuation parameter space can be reasonably small, i.e., there is no need to include model galaxies with large $A_V$ in the prior).

\subsection{Outputs and results}
\label{sec:def_results}
Many commonly used codes return as outputs the physical parameters associated with the best-fitting model. For physical parameters that are sensitive mainly to the normalization of the SED, this could be a sensible approximation. However, small variations in the fluxes generating the SEDs could result in very different colors and hence, very different physical parameters \citep{ocvirk2006}. In this case, models with different physical parameters can match the data to the same likelihood (same $\chi^2$) resulting in degeneracies. This needs to be taken into account in the SED fitting process.

To move away from the single best-fitting model, we can generate PDFs for each of the parameters of interest using the likelihood of each model to match the data. In case of PDF being a Gaussian function, we can define a peak (most likely value for the parameter) and a width (the uncertainty in the parameter). In cases of skewed or non-symmetric distributions, we define the 50th percentile as the ``best estimate'' and the distance between the 84th and 16th percentiles as the confidence interval.

The case of bi-modal PDFs is tricky and thus it is in the hands of the user to decide what to report as result because neither a median nor a single percentile can be acceptable values. This means there is a strong degeneracy that needs to be understood (see for example the redshift - dust attenuation degeneracy in the work by \citealt{dacunha2015}). In such a case, we can suggest carrying the full PDF. The attention of the user is also required when the prior distribution and the PDF of a certain physical parameter have the same shape and extension. This should tell the user that the specific parameter is not constrained. This does not mean that the fit is useless. It simply means that the information relative to that specific parameter is not in the data. Alternatively, when the PDFs hit the edges of the allowed range, one needs to extend  the parameter space sampled in the model library.

In summary, using a single best-fitting model can easily introduce unnecessary scatter in the correlations we wish to study. Exploring the full PDFs is the safest choice, and it can tell us whether the parameter we are trying to measure can actually be constrained by the data we have and whether the model library we are using is appropriate for our project.

\section{Summary and Conclusion}
\label{sec:conclusion}

In recent years, data have improved dramatically in quality and quantity, and SED fitting techniques and codes have started to upgrade accordingly. There is a large number of publicly available SED fitting codes. They are generally customizable and adaptable to different datasets. Yet, when used to interpret the same dataset, they provide different results. Here, we explore the differences among the results (stellar mass, SFR, and dust attenuation) of fourteen different SED fitting codes, run on three datasets: galaxies at $z\sim1$ with photometry spanning rest-frame UV to NIR wavelengths; a subset of the first dataset including IR measurements out to 200$\mu$m; and galaxies at $z\sim3$ with photometry spanning rest-frame UV to NIR wavelengths.

The aim of this exercise is twofold: assessing the differences and eventual biases between codes to demonstrate the impact of different modeling choices on the results; and measuring the modeling uncertainties that need to be taken into account when comparing studies that use different assumptions and when comparing observational results with theoretical predictions. We stress that we do not attempt to identify any ``best'' code or set of assumptions. We use the SFR-$M_{*}$ relation as a case example to determine the similarities and differences among the codes.

When assessing the differences between the codes, we find that:
\begin{itemize}
    \item Qualitatively, all codes return similar distributions in terms of stellar mass. The distributions of the results for the SFR and dust attenuation parameter ($A_V$) show differences in the location of the peaks among the codes.
    \item The codes that tend to return larger SFR values also return larger $A_V$ values indicating a strong age-dust degeneracy for the used datasets and models.
    \item The codes that can process IR measurements return smaller SFR values when the IR measurements are included in the fit compared to when the IR measurements are not included in the fit. This difference is smaller for the codes that assume flexible SFHs, suggesting that the fit is finding \textit{older} solutions instead of \textit{dustier} solutions to match the \textit{red} SED.
    \item All the codes that model also an AGN component return AGN fractions smaller than 20\% for dataset number 2.
    \item All codes return identifiable SFR-$M_{*}$ relations. The slope is very similar in all codes. The normalization can vary by about 0.1-0.2 dex between codes. The scatter varies the most along with the distribution of the galaxies that fall below the relation.
\end{itemize}

When measuring the modeling uncertainties, we find that:
\begin{itemize}
    \item The median modelling uncertainty is around 0.12~dex for stellar mass, 0.27~dex for SFR (dominated by galaxies on the SFR-$M_{*}$ relation), and 0.27~mag for $A_V$ at $z\sim1$. It is around 0.16~dex for stellar mass, 0.28~dex for SFR, and 0.22~mag for $A_V$ at $z\sim3$.
    \item The modeling uncertainty in SFR can increase to 1~dex or more for galaxies with low SFR ($\log(SFR/(M_{\odot} yr^{-1})<0$).
    \item The modeling uncertainty in $A_V$ is a strong function of $A_V$ itself and can be as larger than 0.4~mag for $A_V>2$.
\end{itemize}

Towards the end of the paper, we also provide some examples and best practices to use any SED fitting tool not as a black box, but being cognizant about the modeling assumptions and how those can shape the results. We advise the user to always test different modeling assumptions and measure the appropriate modeling uncertainties to be able to put results in context with other studies or theoretical models.

The datasets, the outputs from the different SED fitting tools, and the Jupyter Notebooks used to produce the figures in this paper are published on GitHub at \url{https://github.com/camipacifici/art_sedfitting} (\url{https://zenodo.org/badge/latestdoi/440978868}).\\

\software{NumPy \citep{harris2020}, Matplotlib \citep{hunter2007}, pandas \citep{pandas2020}, Astropy \citep{astropy2013,astropy2018}, seaborn \citep{waskom2021}, joypy (\url{https://github.com/leotac/joypy}), and statsmodels \citep{seabold2010}.}\\

We thank the anonymous referee for their very constructive report. We thank the University of California Riverside for hosting the workshop where this work started. The workshop was supported by National Science Foundation funding. This paper does not reflect the views or opinions of the National Science Foundation or the American Association for the Advancement of Science (AAAS). We thank Audrey Galametz, Joel Primack, and Meaghann Stoelting for insightful conversations. CP was supported by the Canadian Space Agency under a contract with NRC Herzberg Astronomy and Astrophysics. Support for KI was provided by NASA through the NASA Hubble Fellowship grant HST-HF2-51508 awarded by the Space Telescope Science Institute, which is operated by the Association of Universities for Research in Astronomy, Inc., for NASA, under contract NAS5-26555. Support for VP was provided by NASA through the NASA Hubble Fellowship grant HST-HF2-51489 awarded by the Space Telescope Science Institute, which is operated by the Association of Universities for Research in Astronomy, Inc., for NASA, under contract NAS5-26555. MB acknowledges support from FONDECYT regular grant 1211000 and by the ANID BASAL project FB210003. KM is grateful for support from the Polish National Science Centre via grant UMO-2018/30/E/ST9/00082. For the purpose of open access, the author has applied a Creative Commons Attribution (CC BY) licence to any Author Accepted Manuscript version arising from this submission.

\bibliography{references}

\begin{thebibliography}{}
\expandafter\ifx\csname natexlab\endcsname\relax\def\natexlab#1{#1}\fi
\providecommand{\url}[1]{\href{#1}{#1}}
\providecommand{\dodoi}[1]{doi:~\href{http://doi.org/#1}{\nolinkurl{#1}}}
\providecommand{\doeprint}[1]{\href{http://ascl.net/#1}{\nolinkurl{http://ascl.net/#1}}}
\providecommand{\doarXiv}[1]{\href{https://arxiv.org/abs/#1}{\nolinkurl{https://arxiv.org/abs/#1}}}

\bibitem[{{Abdurro'uf} {et~al.}(2021){Abdurro'uf}, {Lin}, {Hirashita},
  {Morishita}, {Tacchella}, {Akiyama}, {Takeuchi}, \& {Wu}}]{abdurrouf2021}
{Abdurro'uf}, {Lin}, Y.-T., {Hirashita}, H., {et~al.} 2021, arXiv e-prints,
  arXiv:2110.03158.
\newblock \doarXiv{2110.03158}

\bibitem[{{Acquaviva} {et~al.}(2011){Acquaviva}, {Gawiser}, \&
  {Guaita}}]{acquaviva2011}
{Acquaviva}, V., {Gawiser}, E., \& {Guaita}, L. 2011, \apj, 737,
  \dodoi{10.1088/0004-637X/737/2/47}

\bibitem[{{Acquaviva} {et~al.}(2012){Acquaviva}, {Gawiser}, \&
  {Guaita}}]{acquaviva2012}
{Acquaviva}, V., {Gawiser}, E., \& {Guaita}, L. 2012, in IAU Symposium, Vol.
  284, The Spectral Energy Distribution of Galaxies - SED 2011, ed. R.~J.
  {Tuffs} \& C.~C. {Popescu}, 42--45, \dodoi{10.1017/S1743921312008691}

\bibitem[{{Arnouts} {et~al.}(1999){Arnouts}, {Cristiani}, {Moscardini},
  {Matarrese}, {Lucchin}, {Fontana}, \& {Giallongo}}]{arnouts1999}
{Arnouts}, S., {Cristiani}, S., {Moscardini}, L., {et~al.} 1999, \mnras, 310,
  540, \dodoi{10.1046/j.1365-8711.1999.02978.x}

\bibitem[{{Astropy Collaboration} {et~al.}(2013){Astropy Collaboration},
  {Robitaille}, {Tollerud}, {Greenfield}, {Droettboom}, {Bray}, {Aldcroft},
  {Davis}, {Ginsburg}, {Price-Whelan}, {Kerzendorf}, {Conley}, {Crighton},
  {Barbary}, {Muna}, {Ferguson}, {Grollier}, {Parikh}, {Nair}, {Unther},
  {Deil}, {Woillez}, {Conseil}, {Kramer}, {Turner}, {Singer}, {Fox}, {Weaver},
  {Zabalza}, {Edwards}, {Azalee Bostroem}, {Burke}, {Casey}, {Crawford},
  {Dencheva}, {Ely}, {Jenness}, {Labrie}, {Lim}, {Pierfederici}, {Pontzen},
  {Ptak}, {Refsdal}, {Servillat}, \& {Streicher}}]{astropy2013}
{Astropy Collaboration}, {Robitaille}, T.~P., {Tollerud}, E.~J., {et~al.} 2013,
  \aap, 558, A33, \dodoi{10.1051/0004-6361/201322068}

\bibitem[{{Astropy Collaboration} {et~al.}(2018){Astropy Collaboration},
  {Price-Whelan}, {Sip{\H{o}}cz}, {G{\"u}nther}, {Lim}, {Crawford}, {Conseil},
  {Shupe}, {Craig}, {Dencheva}, {Ginsburg}, {Vand erPlas}, {Bradley},
  {P{\'e}rez-Su{\'a}rez}, {de Val-Borro}, {Aldcroft}, {Cruz}, {Robitaille},
  {Tollerud}, {Ardelean}, {Babej}, {Bach}, {Bachetti}, {Bakanov}, {Bamford},
  {Barentsen}, {Barmby}, {Baumbach}, {Berry}, {Biscani}, {Boquien}, {Bostroem},
  {Bouma}, {Brammer}, {Bray}, {Breytenbach}, {Buddelmeijer}, {Burke},
  {Calderone}, {Cano Rodr{\'\i}guez}, {Cara}, {Cardoso}, {Cheedella}, {Copin},
  {Corrales}, {Crichton}, {D'Avella}, {Deil}, {Depagne}, {Dietrich}, {Donath},
  {Droettboom}, {Earl}, {Erben}, {Fabbro}, {Ferreira}, {Finethy}, {Fox},
  {Garrison}, {Gibbons}, {Goldstein}, {Gommers}, {Greco}, {Greenfield},
  {Groener}, {Grollier}, {Hagen}, {Hirst}, {Homeier}, {Horton}, {Hosseinzadeh},
  {Hu}, {Hunkeler}, {Ivezi{\'c}}, {Jain}, {Jenness}, {Kanarek}, {Kendrew},
  {Kern}, {Kerzendorf}, {Khvalko}, {King}, {Kirkby}, {Kulkarni}, {Kumar},
  {Lee}, {Lenz}, {Littlefair}, {Ma}, {Macleod}, {Mastropietro}, {McCully},
  {Montagnac}, {Morris}, {Mueller}, {Mumford}, {Muna}, {Murphy}, {Nelson},
  {Nguyen}, {Ninan}, {N{\"o}the}, {Ogaz}, {Oh}, {Parejko}, {Parley}, {Pascual},
  {Patil}, {Patil}, {Plunkett}, {Prochaska}, {Rastogi}, {Reddy Janga},
  {Sabater}, {Sakurikar}, {Seifert}, {Sherbert}, {Sherwood-Taylor}, {Shih},
  {Sick}, {Silbiger}, {Singanamalla}, {Singer}, {Sladen}, {Sooley},
  {Sornarajah}, {Streicher}, {Teuben}, {Thomas}, {Tremblay}, {Turner},
  {Terr{\'o}n}, {van Kerkwijk}, {de la Vega}, {Watkins}, {Weaver}, {Whitmore},
  {Woillez}, {Zabalza}, \& {Astropy Contributors}}]{astropy2018}
{Astropy Collaboration}, {Price-Whelan}, A.~M., {Sip{\H{o}}cz}, B.~M., {et~al.}
  2018, \aj, 156, 123, \dodoi{10.3847/1538-3881/aabc4f}

\bibitem[{{Baldwin} {et~al.}(2018){Baldwin}, {McDermid}, {Kuntschner},
  {Maraston}, \& {Conroy}}]{baldwin2018}
{Baldwin}, C., {McDermid}, R.~M., {Kuntschner}, H., {Maraston}, C., \&
  {Conroy}, C. 2018, \mnras, 473, 4698, \dodoi{10.1093/mnras/stx2502}

\bibitem[{{Bari{\v{s}}i{\'c}} {et~al.}(2020){Bari{\v{s}}i{\'c}}, {Pacifici},
  {van der Wel}, {Straatman}, {Bell}, {Bezanson}, {Brammer}, {D'Eugenio},
  {Franx}, {van Houdt}, {Maseda}, {Muzzin}, {Sobral}, \& {Wu}}]{barisic2020}
{Bari{\v{s}}i{\'c}}, I., {Pacifici}, C., {van der Wel}, A., {et~al.} 2020,
  \apj, 903, 146, \dodoi{10.3847/1538-4357/abba37}

\bibitem[{{Barro} {et~al.}(2019){Barro}, {P{\'e}rez-Gonz{\'a}lez}, {Cava},
  {Brammer}, {Pandya}, {Eliche Moral}, {Esquej}, {Dom{\'\i}nguez-S{\'a}nchez},
  {Alcalde Pampliega}, {Guo}, {Koekemoer}, {Trump}, {Ashby}, {Cardiel},
  {Castellano}, {Conselice}, {Dickinson}, {Dolch}, {Donley}, {Espino Briones},
  {Faber}, {Fazio}, {Ferguson}, {Finkelstein}, {Fontana}, {Galametz},
  {Gardner}, {Gawiser}, {Giavalisco}, {Grazian}, {Grogin}, {Hathi}, {Hemmati},
  {Hern{\'a}n-Caballero}, {Kocevski}, {Koo}, {Kodra}, {Lee}, {Lin}, {Lucas},
  {Mobasher}, {McGrath}, {Nandra}, {Nayyeri}, {Newman}, {Pforr}, {Peth},
  {Rafelski}, {Rodr{\'\i}guez-Munoz}, {Salvato}, {Stefanon}, {van der Wel},
  {Willner}, {Wiklind}, \& {Wuyts}}]{barro2019}
{Barro}, G., {P{\'e}rez-Gonz{\'a}lez}, P.~G., {Cava}, A., {et~al.} 2019, \apjs,
  243, 22, \dodoi{10.3847/1538-4365/ab23f2}

\bibitem[{{Battisti} {et~al.}(2020){Battisti}, {Cunha}, {Shivaei}, \&
  {Calzetti}}]{battisti2020}
{Battisti}, A.~J., {Cunha}, E.~d., {Shivaei}, I., \& {Calzetti}, D. 2020, \apj,
  888, 108, \dodoi{10.3847/1538-4357/ab5fdd}

\bibitem[{{Battisti} {et~al.}(2019){Battisti}, {da Cunha}, {Grasha}, {Salvato},
  {Daddi}, {Davies}, {Jin}, {Liu}, {Schinnerer}, {Vaccari}, \& {COSMOS
  Collaboration}}]{battisti2019}
{Battisti}, A.~J., {da Cunha}, E., {Grasha}, K., {et~al.} 2019, \apj, 882, 61,
  \dodoi{10.3847/1538-4357/ab345d}

\bibitem[{{Bell} \& {de Jong}(2001)}]{bell2001}
{Bell}, E.~F., \& {de Jong}, R.~S. 2001, \apj, 550, 212, \dodoi{10.1086/319728}

\bibitem[{{Bellstedt} {et~al.}(2020){Bellstedt}, {Robotham}, {Driver},
  {Thorne}, {Davies}, {Lagos}, {Stevens}, {Taylor}, {Baldry}, {Moffett},
  {Hopkins}, \& {Phillipps}}]{bellstedt2020}
{Bellstedt}, S., {Robotham}, A. S.~G., {Driver}, S.~P., {et~al.} 2020, \mnras,
  498, 5581, \dodoi{10.1093/mnras/staa2620}

\bibitem[{{Bellstedt} {et~al.}(2021){Bellstedt}, {Robotham}, {Driver},
  {Thorne}, {Davies}, {Holwerda}, {Hopkins}, {Lara-Lopez},
  {L{\'o}pez-S{\'a}nchez}, \& {Phillipps}}]{bellstedt2021}
---. 2021, \mnras, 503, 3309, \dodoi{10.1093/mnras/stab550}

\bibitem[{Berta {et~al.}(2013)Berta, Lutz, Santini, Wuyts, Rosario, Brisbin,
  Cooray, Franceschini, Gruppioni, Hatziminaoglou, \& et~al.}]{berta2013}
Berta, S., Lutz, D., Santini, P., {et~al.} 2013, Astronomy \& Astrophysics,
  551, A100, \dodoi{10.1051/0004-6361/201220859}

\bibitem[{{Berti} {et~al.}(2021){Berti}, {Coil}, {Hearin}, \&
  {Behroozi}}]{berti2021}
{Berti}, A.~M., {Coil}, A.~L., {Hearin}, A.~P., \& {Behroozi}, P.~S. 2021, \aj,
  161, 49, \dodoi{10.3847/1538-3881/abcc6a}

\bibitem[{{Boogaard} {et~al.}(2018){Boogaard}, {Brinchmann}, {Bouch{\'e}},
  {Paalvast}, {Bacon}, {Bouwens}, {Contini}, {Gunawardhana}, {Inami}, {Marino},
  {Maseda}, {Mitchell}, {Nanayakkara}, {Richard}, {Schaye}, {Schreiber},
  {Tacchella}, {Wisotzki}, \& {Zabl}}]{boogaard2018}
{Boogaard}, L.~A., {Brinchmann}, J., {Bouch{\'e}}, N., {et~al.} 2018, \aap,
  619, A27, \dodoi{10.1051/0004-6361/201833136}

\bibitem[{{Boquien} {et~al.}(2019){Boquien}, {Burgarella}, {Roehlly}, {Buat},
  {Ciesla}, {Corre}, {Inoue}, \& {Salas}}]{boquien2019}
{Boquien}, M., {Burgarella}, D., {Roehlly}, Y., {et~al.} 2019, \aap, 622, A103,
  \dodoi{10.1051/0004-6361/201834156}

\bibitem[{{Bowman} {et~al.}(2020){Bowman}, {Zeimann}, {Nagaraj}, {Ciardullo},
  {Gronwall}, {McCarron}, {Weiss}, {Molina}, {Belles}, \&
  {Schneider}}]{bowman2020}
{Bowman}, W.~P., {Zeimann}, G.~R., {Nagaraj}, G., {et~al.} 2020, \apj, 899, 7,
  \dodoi{10.3847/1538-4357/ab9f3c}

\bibitem[{{Brammer} {et~al.}(2012){Brammer}, {van Dokkum}, {Franx},
  {Fumagalli}, {Patel}, {Rix}, {Skelton}, {Kriek}, {Nelson}, {Schmidt},
  {Bezanson}, {da Cunha}, {Erb}, {Fan}, {F{\"o}rster Schreiber}, {Illingworth},
  {Labb{\'e}}, {Leja}, {Lundgren}, {Magee}, {Marchesini}, {McCarthy},
  {Momcheva}, {Muzzin}, {Quadri}, {Steidel}, {Tal}, {Wake}, {Whitaker}, \&
  {Williams}}]{brammer2012}
{Brammer}, G.~B., {van Dokkum}, P.~G., {Franx}, M., {et~al.} 2012, \apjs, 200,
  13, \dodoi{10.1088/0067-0049/200/2/13}

\bibitem[{{Brinchmann} {et~al.}(2004){Brinchmann}, {Charlot}, {White},
  {Tremonti}, {Kauffmann}, {Heckman}, \& {Brinkmann}}]{brinchmann2004}
{Brinchmann}, J., {Charlot}, S., {White}, S.~D.~M., {et~al.} 2004, \mnras, 351,
  1151, \dodoi{10.1111/j.1365-2966.2004.07881.x}

\bibitem[{{Bruzual} \& {Charlot}(2003)}]{BC03}
{Bruzual}, G., \& {Charlot}, S. 2003, \mnras, 344, 1000,
  \dodoi{10.1046/j.1365-8711.2003.06897.x}

\bibitem[{{Bruzual A.}(1983)}]{bruzual1983}
{Bruzual A.}, G. 1983, \apj, 273, 105, \dodoi{10.1086/161352}

\bibitem[{{Bruzual A.} \& {Charlot}(1993)}]{bruzual1993}
{Bruzual A.}, G., \& {Charlot}, S. 1993, \apj, 405, 538, \dodoi{10.1086/172385}

\bibitem[{{Buat} {et~al.}(2018){Buat}, {Boquien}, {Ma{\l}ek}, {Corre}, {Salas},
  {Roehlly}, {Shirley}, \& {Efstathiou}}]{buat2018}
{Buat}, V., {Boquien}, M., {Ma{\l}ek}, K., {et~al.} 2018, \aap, 619, A135,
  \dodoi{10.1051/0004-6361/201833841}

\bibitem[{{Buat} {et~al.}(2005){Buat}, {Iglesias-P{\'a}ramo}, {Seibert},
  {Burgarella}, {Charlot}, {Martin}, {Xu}, {Heckman}, {Boissier}, {Boselli},
  {Barlow}, {Bianchi}, {Byun}, {Donas}, {Forster}, {Friedman}, {Jelinski},
  {Lee}, {Madore}, {Malina}, {Milliard}, {Morissey}, {Neff}, {Rich},
  {Schiminovitch}, {Siegmund}, {Small}, {Szalay}, {Welsh}, \&
  {Wyder}}]{buat2005}
{Buat}, V., {Iglesias-P{\'a}ramo}, J., {Seibert}, M., {et~al.} 2005, \apjl,
  619, L51, \dodoi{10.1086/423241}

\bibitem[{Buchner(2016)}]{buchner2016}
Buchner, J. 2016, Statistics and Computing, 26, 383

\bibitem[{{Burgarella} {et~al.}(2005){Burgarella}, {Buat}, \&
  {Iglesias-P{\'a}ramo}}]{burgarella2005}
{Burgarella}, D., {Buat}, V., \& {Iglesias-P{\'a}ramo}, J. 2005, \mnras, 360,
  1413, \dodoi{10.1111/j.1365-2966.2005.09131.x}

\bibitem[{{Calistro Rivera} {et~al.}(2016){Calistro Rivera}, {Lusso},
  {Hennawi}, \& {Hogg}}]{calistrorivera2016}
{Calistro Rivera}, G., {Lusso}, E., {Hennawi}, J.~F., \& {Hogg}, D.~W. 2016,
  \apj, 833, \dodoi{10.3847/1538-4357/833/1/98}

\bibitem[{Calzetti(2001)}]{calzetti2001}
Calzetti, D. 2001, Publications of the Astronomical Society of the Pacific,
  113, 1449

\bibitem[{{Caplar} \& {Tacchella}(2019)}]{caplar2019}
{Caplar}, N., \& {Tacchella}, S. 2019, \mnras, 487, 3845,
  \dodoi{10.1093/mnras/stz1449}

\bibitem[{{Carnall} {et~al.}(2019{\natexlab{a}}){Carnall}, {Leja}, {Johnson},
  {McLure}, {Dunlop}, \& {Conroy}}]{carnall2019a}
{Carnall}, A.~C., {Leja}, J., {Johnson}, B.~D., {et~al.} 2019{\natexlab{a}},
  \apj, 873, 44, \dodoi{10.3847/1538-4357/ab04a2}

\bibitem[{{Carnall} {et~al.}(2018){Carnall}, {McLure}, {Dunlop}, \&
  {Dav{\'e}}}]{carnall2018}
{Carnall}, A.~C., {McLure}, R.~J., {Dunlop}, J.~S., \& {Dav{\'e}}, R. 2018,
  \mnras, 480, 4379, \dodoi{10.1093/mnras/sty2169}

\bibitem[{{Carnall} {et~al.}(2019{\natexlab{b}}){Carnall}, {McLure}, {Dunlop},
  {Cullen}, {McLeod}, {Wild}, {Johnson}, {Appleby}, {Dav{\'e}}, {Amorin},
  {Bolzonella}, {Castellano}, {Cimatti}, {Cucciati}, {Gargiulo}, {Garilli},
  {Marchi}, {Pentericci}, {Pozzetti}, {Schreiber}, {Talia}, \&
  {Zamorani}}]{carnall2019b}
{Carnall}, A.~C., {McLure}, R.~J., {Dunlop}, J.~S., {et~al.}
  2019{\natexlab{b}}, \mnras, 490, 417, \dodoi{10.1093/mnras/stz2544}

\bibitem[{{Chabrier}(2003)}]{chabrier2003}
{Chabrier}, G. 2003, \pasp, 115, 763, \dodoi{10.1086/376392}

\bibitem[{{Chang} {et~al.}(2017{\natexlab{a}}){Chang}, {Le Floc'h}, {Juneau},
  {da Cunha}, {Salvato}, {Civano}, {Marchesi}, {Gabor}, {Ilbert}, {Laigle},
  {McCracken}, {Hsieh}, \& {Capak}}]{chang2017a}
{Chang}, Y.-Y., {Le Floc'h}, E., {Juneau}, S., {et~al.} 2017{\natexlab{a}},
  \mnras, 466, L103, \dodoi{10.1093/mnrasl/slw247}

\bibitem[{{Chang} {et~al.}(2017{\natexlab{b}}){Chang}, {Le Floc'h}, {Juneau},
  {da Cunha}, {Salvato}, {Civano}, {Marchesi}, {Ilbert}, {Toba}, {Lim}, {Tang},
  {Wang}, {Ferraro}, {Urry}, {Griffiths}, \& {Kartaltepe}}]{chang2017b}
---. 2017{\natexlab{b}}, \apjs, 233, 19, \dodoi{10.3847/1538-4365/aa97da}

\bibitem[{{Charlot} \& {Fall}(2000)}]{charlot2000}
{Charlot}, S., \& {Fall}, S.~M. 2000, \apj, 539, 718, \dodoi{10.1086/309250}

\bibitem[{{Charlot} \& {Longhetti}(2001)}]{charlot2001}
{Charlot}, S., \& {Longhetti}, M. 2001, \mnras, 323, 887,
  \dodoi{10.1046/j.1365-8711.2001.04260.x}

\bibitem[{{Chevallard} \& {Charlot}(2016)}]{chevallard2016}
{Chevallard}, J., \& {Charlot}, S. 2016, \mnras, 462, 1415,
  \dodoi{10.1093/mnras/stw1756}

\bibitem[{{Chevallard} {et~al.}(2017){Chevallard}, {Curtis-Lake}, {Charlot},
  {Ferruit}, {Giardino}, {Franx}, {Maseda}, {Amorin}, {Arribas}, {Bunker},
  {Carniani}, {Husemann}, {Jakobsen}, {Maiolino}, {Pforr}, {Rawle}, {Rix},
  {Smit}, \& {Willott}}]{chevallard2017}
{Chevallard}, J., {Curtis-Lake}, E., {Charlot}, S., {et~al.} 2017, ArXiv
  e-prints, arXiv:1711.07481.
\newblock \doarXiv{1711.07481}

\bibitem[{{Ciesla} {et~al.}(2017){Ciesla}, {Elbaz}, \& {Fensch}}]{ciesla2017}
{Ciesla}, L., {Elbaz}, D., \& {Fensch}, J. 2017, \aap, 608, A41,
  \dodoi{10.1051/0004-6361/201731036}

\bibitem[{{Conroy}(2013)}]{conroy2013}
{Conroy}, C. 2013, \araa, 51, 393, \dodoi{10.1146/annurev-astro-082812-141017}

\bibitem[{{Conroy} {et~al.}(2009){Conroy}, {Gunn}, \& {White}}]{conroy2009}
{Conroy}, C., {Gunn}, J.~E., \& {White}, M. 2009, \apj, 699, 486,
  \dodoi{10.1088/0004-637X/699/1/486}

\bibitem[{{Curtis-Lake} {et~al.}(2021){Curtis-Lake}, {Chevallard}, {Charlot},
  \& {Sandles}}]{curtislake2021}
{Curtis-Lake}, E., {Chevallard}, J., {Charlot}, S., \& {Sandles}, L. 2021,
  \mnras, 503, 4855, \dodoi{10.1093/mnras/stab698}

\bibitem[{{da Cunha} {et~al.}(2008){da Cunha}, {Charlot}, \&
  {Elbaz}}]{dacunha2008}
{da Cunha}, E., {Charlot}, S., \& {Elbaz}, D. 2008, \mnras, 388, 1595,
  \dodoi{10.1111/j.1365-2966.2008.13535.x}

\bibitem[{{da Cunha} {et~al.}(2015){da Cunha}, {Walter}, {Smail}, {Swinbank},
  {Simpson}, {Decarli}, {Hodge}, {Weiss}, {van der Werf}, {Bertoldi},
  {Chapman}, {Cox}, {Danielson}, {Dannerbauer}, {Greve}, {Ivison}, {Karim}, \&
  {Thomson}}]{dacunha2015}
{da Cunha}, E., {Walter}, F., {Smail}, I.~R., {et~al.} 2015, \apj, 806,
  \dodoi{10.1088/0004-637X/806/1/110}

\bibitem[{{Davidzon} {et~al.}(2019){Davidzon}, {Laigle}, {Capak}, {Ilbert},
  {Masters}, {Hemmati}, {Apostolakos}, {Coupon}, {de la Torre}, {Devriendt},
  {Dubois}, {Kashino}, {Paltani}, \& {Pichon}}]{davidzon2019}
{Davidzon}, I., {Laigle}, C., {Capak}, P.~L., {et~al.} 2019, \mnras, 489, 4817,
  \dodoi{10.1093/mnras/stz2486}

\bibitem[{{Davies} {et~al.}(2017){Davies}, {Baes}, {Bianchi}, {Jones},
  {Madden}, {Xilouris}, {Bocchio}, {Casasola}, {Cassara}, {Clark}, {De Looze},
  {Evans}, {Fritz}, {Galametz}, {Galliano}, {Lianou}, {Mosenkov}, {Smith},
  {Verstocken}, {Viaene}, {Vika}, {Wagle}, \& {Ysard}}]{davies2017}
{Davies}, J.~I., {Baes}, M., {Bianchi}, S., {et~al.} 2017, \pasp, 129, 044102,
  \dodoi{10.1088/1538-3873/129/974/044102}

\bibitem[{{Davies} {et~al.}(2021){Davies}, {Thorne}, {Robotham}, {Bellstedt},
  {Driver}, {Adams}, {Bilicki}, {Bowler}, {Bravo}, {Cortese}, {Foster},
  {Grootes}, {H{\"a}u{\ss}ler}, {Hashemizadeh}, {Holwerda}, {Hurley}, {Jarvis},
  {Lidman}, {Maddox}, {Meyer}, {Paolillo}, {Phillipps}, {Radovich}, {Siudek},
  {Vaccari}, \& {Windhorst}}]{davies2021}
{Davies}, L.~J.~M., {Thorne}, J.~E., {Robotham}, A.~S.~G., {et~al.} 2021,
  \mnras, 506, 256, \dodoi{10.1093/mnras/stab1601}

\bibitem[{{Driver} {et~al.}(2011){Driver}, {Hill}, {Kelvin}, {Robotham},
  {Liske}, {Norberg}, {Baldry}, {Bamford}, {Hopkins}, {Loveday}, {Peacock},
  {Andrae}, {Bland-Hawthorn}, {Brough}, {Brown}, {Cameron}, {Ching}, {Colless},
  {Conselice}, {Croom}, {Cross}, {de Propris}, {Dye}, {Drinkwater}, {Ellis},
  {Graham}, {Grootes}, {Gunawardhana}, {Jones}, {van Kampen}, {Maraston},
  {Nichol}, {Parkinson}, {Phillipps}, {Pimbblet}, {Popescu}, {Prescott},
  {Roseboom}, {Sadler}, {Sansom}, {Sharp}, {Smith}, {Taylor}, {Thomas},
  {Tuffs}, {Wijesinghe}, {Dunne}, {Frenk}, {Jarvis}, {Madore}, {Meyer},
  {Seibert}, {Staveley-Smith}, {Sutherland}, \& {Warren}}]{driver2011}
{Driver}, S.~P., {Hill}, D.~T., {Kelvin}, L.~S., {et~al.} 2011, \mnras, 413,
  971, \dodoi{10.1111/j.1365-2966.2010.18188.x}

\bibitem[{{Drouart} \& {Falkendal}(2018)}]{drouart2018}
{Drouart}, G., \& {Falkendal}, T. 2018, \mnras, 477, 4981,
  \dodoi{10.1093/mnras/sty831}

\bibitem[{{Du} \& {McGaugh}(2020)}]{du2020}
{Du}, W., \& {McGaugh}, S.~S. 2020, \aj, 160, 122,
  \dodoi{10.3847/1538-3881/aba754}

\bibitem[{{Eldridge} \& {Stanway}(2016)}]{eldridge2016}
{Eldridge}, J.~J., \& {Stanway}, E.~R. 2016, \mnras, 462, 3302,
  \dodoi{10.1093/mnras/stw1772}

\bibitem[{{Eufrasio}(2017)}]{eufrasio2017}
{Eufrasio}, R.~T. 2017, {Lightning: SED Fitting Package}.
\newblock \doeprint{1711.009}

\bibitem[{{Fioc} \& {Rocca-Volmerange}(2019)}]{fioc2019}
{Fioc}, M., \& {Rocca-Volmerange}, B. 2019, \aap, 623, A143,
  \dodoi{10.1051/0004-6361/201833556}

\bibitem[{{Fontana} {et~al.}(2000){Fontana}, {D'Odorico}, {Poli}, {Giallongo},
  {Arnouts}, {Cristiani}, {Moorwood}, \& {Saracco}}]{fontana2000}
{Fontana}, A., {D'Odorico}, S., {Poli}, F., {et~al.} 2000, \aj, 120, 2206,
  \dodoi{10.1086/316803}

\bibitem[{{Foreman-Mackey} {et~al.}(2013){Foreman-Mackey}, {Hogg}, {Lang}, \&
  {Goodman}}]{foreman2013}
{Foreman-Mackey}, D., {Hogg}, D.~W., {Lang}, D., \& {Goodman}, J. 2013, \pasp,
  125, 306, \dodoi{10.1086/670067}

\bibitem[{{Gilda} {et~al.}(2021){Gilda}, {Lower}, \& {Narayanan}}]{gilda2021}
{Gilda}, S., {Lower}, S., \& {Narayanan}, D. 2021, \apj, 916, 43,
  \dodoi{10.3847/1538-4357/ac0058}

\bibitem[{{Goddard} {et~al.}(2017){Goddard}, {Thomas}, {Maraston}, {Westfall},
  {Etherington}, {Riffel}, {Mallmann}, {Zheng}, {Argudo-Fern{\'a}ndez}, {Lian},
  {Bershady}, {Bundy}, {Drory}, {Law}, {Yan}, {Wake}, {Weijmans}, {Bizyaev},
  {Brownstein}, {Lane}, {Maiolino}, {Masters}, {Merrifield}, {Nitschelm},
  {Pan}, {Roman-Lopes}, {Storchi-Bergmann}, \& {Schneider}}]{goddard2017}
{Goddard}, D., {Thomas}, D., {Maraston}, C., {et~al.} 2017, \mnras, 466, 4731,
  \dodoi{10.1093/mnras/stw3371}

\bibitem[{{Grogin} {et~al.}(2011){Grogin}, {Kocevski}, {Faber}, {Ferguson},
  {Koekemoer}, {Riess}, {Acquaviva}, {Alexander}, {Almaini}, {Ashby}, {Barden},
  {Bell}, {Bournaud}, {Brown}, {Caputi}, {Casertano}, {Cassata}, {Castellano},
  {Challis}, {Chary}, {Cheung}, {Cirasuolo}, {Conselice}, {Roshan Cooray},
  {Croton}, {Daddi}, {Dahlen}, {Dav{\'e}}, {de Mello}, {Dekel}, {Dickinson},
  {Dolch}, {Donley}, {Dunlop}, {Dutton}, {Elbaz}, {Fazio}, {Filippenko},
  {Finkelstein}, {Fontana}, {Gardner}, {Garnavich}, {Gawiser}, {Giavalisco},
  {Grazian}, {Guo}, {Hathi}, {H{\"a}ussler}, {Hopkins}, {Huang}, {Huang},
  {Jha}, {Kartaltepe}, {Kirshner}, {Koo}, {Lai}, {Lee}, {Li}, {Lotz}, {Lucas},
  {Madau}, {McCarthy}, {McGrath}, {McIntosh}, {McLure}, {Mobasher},
  {Moustakas}, {Mozena}, {Nandra}, {Newman}, {Niemi}, {Noeske}, {Papovich},
  {Pentericci}, {Pope}, {Primack}, {Rajan}, {Ravindranath}, {Reddy}, {Renzini},
  {Rix}, {Robaina}, {Rodney}, {Rosario}, {Rosati}, {Salimbeni}, {Scarlata},
  {Siana}, {Simard}, {Smidt}, {Somerville}, {Spinrad}, {Straughn}, {Strolger},
  {Telford}, {Teplitz}, {Trump}, {van der Wel}, {Villforth}, {Wechsler},
  {Weiner}, {Wiklind}, {Wild}, {Wilson}, {Wuyts}, {Yan}, \& {Yun}}]{grogin2011}
{Grogin}, N.~A., {Kocevski}, D.~D., {Faber}, S.~M., {et~al.} 2011, \apjs, 197,
  35, \dodoi{10.1088/0067-0049/197/2/35}

\bibitem[{{Guiderdoni} \& {Rocca-Volmerange}(1987)}]{guiderdoni1987}
{Guiderdoni}, B., \& {Rocca-Volmerange}, B. 1987, \aap, 186, 1

\bibitem[{{Guo} {et~al.}(2013){Guo}, {Ferguson}, {Giavalisco}, {Barro},
  {Willner}, {Ashby}, {Dahlen}, {Donley}, {Faber}, {Fontana}, {Galametz},
  {Grazian}, {Huang}, {Kocevski}, {Koekemoer}, {Koo}, {McGrath}, {Peth},
  {Salvato}, {Wuyts}, {Castellano}, {Cooray}, {Dickinson}, {Dunlop}, {Fazio},
  {Gardner}, {Gawiser}, {Grogin}, {Hathi}, {Hsu}, {Lee}, {Lucas}, {Mobasher},
  {Nandra}, {Newman}, \& {van der Wel}}]{guo2013}
{Guo}, Y., {Ferguson}, H.~C., {Giavalisco}, M., {et~al.} 2013, \apjs, 207, 24,
  \dodoi{10.1088/0067-0049/207/2/24}

\bibitem[{{Han} \& {Han}(2019)}]{han2019}
{Han}, Y., \& {Han}, Z. 2019, \apjs, 240, 3, \dodoi{10.3847/1538-4365/aaeffa}

\bibitem[{Harris {et~al.}(2020)Harris, Millman, van~der Walt, Gommers,
  Virtanen, Cournapeau, Wieser, Taylor, Berg, Smith, Kern, Picus, Hoyer, van
  Kerkwijk, Brett, Haldane, del R{\'{i}}o, Wiebe, Peterson,
  G{\'{e}}rard-Marchant, Sheppard, Reddy, Weckesser, Abbasi, Gohlke, \&
  Oliphant}]{harris2020}
Harris, C.~R., Millman, K.~J., van~der Walt, S.~J., {et~al.} 2020, Nature, 585,
  357, \dodoi{10.1038/s41586-020-2649-2}

\bibitem[{{Hunt} {et~al.}(2019){Hunt}, {De Looze}, {Boquien}, {Nikutta},
  {Rossi}, {Bianchi}, {Dale}, {Granato}, {Kennicutt}, {Silva}, {Ciesla},
  {Rela{\~n}o}, {Viaene}, {Brandl}, {Calzetti}, {Croxall}, {Draine},
  {Galametz}, {Gordon}, {Groves}, {Helou}, {Herrera-Camus}, {Hinz}, {Koda},
  {Salim}, {Sandstrom}, {Smith}, {Wilson}, \& {Zibetti}}]{hunt2019}
{Hunt}, L.~K., {De Looze}, I., {Boquien}, M., {et~al.} 2019, \aap, 621, A51,
  \dodoi{10.1051/0004-6361/201834212}

\bibitem[{Hunter(2007)}]{hunter2007}
Hunter, J.~D. 2007, Computing in Science \& Engineering, 9, 90,
  \dodoi{10.1109/MCSE.2007.55}

\bibitem[{{Ilbert} {et~al.}(2006){Ilbert}, {Arnouts}, {McCracken},
  {Bolzonella}, {Bertin}, {Le F{\`e}vre}, {Mellier}, {Zamorani}, {Pell{\`o}},
  {Iovino}, {Tresse}, {Le Brun}, {Bottini}, {Garilli}, {Maccagni}, {Picat},
  {Scaramella}, {Scodeggio}, {Vettolani}, {Zanichelli}, {Adami}, {Bardelli},
  {Cappi}, {Charlot}, {Ciliegi}, {Contini}, {Cucciati}, {Foucaud}, {Franzetti},
  {Gavignaud}, {Guzzo}, {Marano}, {Marinoni}, {Mazure}, {Meneux}, {Merighi},
  {Paltani}, {Pollo}, {Pozzetti}, {Radovich}, {Zucca}, {Bondi}, {Bongiorno},
  {Busarello}, {de La Torre}, {Gregorini}, {Lamareille}, {Mathez}, {Merluzzi},
  {Ripepi}, {Rizzo}, \& {Vergani}}]{ilbert2006}
{Ilbert}, O., {Arnouts}, S., {McCracken}, H.~J., {et~al.} 2006, \aap, 457, 841,
  \dodoi{10.1051/0004-6361:20065138}

\bibitem[{{Iyer} \& {Gawiser}(2017)}]{iyer2017}
{Iyer}, K., \& {Gawiser}, E. 2017, \apj, 838, \dodoi{10.3847/1538-4357/aa63f0}

\bibitem[{{Iyer} {et~al.}(2019){Iyer}, {Gawiser}, {Faber}, {Ferguson},
  {Kartaltepe}, {Koekemoer}, {Pacifici}, \& {Somerville}}]{iyer2019}
{Iyer}, K.~G., {Gawiser}, E., {Faber}, S.~M., {et~al.} 2019, \apj, 879, 116,
  \dodoi{10.3847/1538-4357/ab2052}

\bibitem[{{Johnson} {et~al.}(2021){Johnson}, {Leja}, {Conroy}, \&
  {Speagle}}]{johnson2021}
{Johnson}, B.~D., {Leja}, J., {Conroy}, C., \& {Speagle}, J.~S. 2021, \apjs,
  254, 22, \dodoi{10.3847/1538-4365/abef67}

\bibitem[{Kass \& Raftery(1995)}]{kass1995}
Kass, R.~E., \& Raftery, A.~E. 1995, Journal of the American Statistical
  Association, 90, 773.
\newblock \url{http://www.jstor.org/stable/2291091}

\bibitem[{{Katsianis} {et~al.}(2019){Katsianis}, {Zheng}, {Gonzalez}, {Blanc},
  {Lagos}, {Davies}, {Camps}, {Tr{\v{c}}ka}, {Baes}, {Schaye}, {Trayford},
  {Theuns}, \& {Stalevski}}]{katsianis2019}
{Katsianis}, A., {Zheng}, X., {Gonzalez}, V., {et~al.} 2019, \apj, 879, 11,
  \dodoi{10.3847/1538-4357/ab1f8d}

\bibitem[{{Kennicutt}(1989)}]{kennicutt1989}
{Kennicutt}, Robert~C., J. 1989, \apj, 344, 685, \dodoi{10.1086/167834}

\bibitem[{{Kennicutt}(1998)}]{kennicutt1998}
---. 1998, \araa, 36, 189, \dodoi{10.1146/annurev.astro.36.1.189}

\bibitem[{{Kennicutt} {et~al.}(2011){Kennicutt}, {Calzetti}, {Aniano},
  {Appleton}, {Armus}, {Beir{\~a}o}, {Bolatto}, {Brandl}, {Crocker}, {Croxall},
  {Dale}, {Donovan Meyer}, {Draine}, {Engelbracht}, {Galametz}, {Gordon},
  {Groves}, {Hao}, {Helou}, {Hinz}, {Hunt}, {Johnson}, {Koda}, {Krause},
  {Leroy}, {Li}, {Meidt}, {Montiel}, {Murphy}, {Rahman}, {Rix}, {Roussel},
  {Sandstrom}, {Sauvage}, {Schinnerer}, {Skibba}, {Smith}, {Srinivasan},
  {Vigroux}, {Walter}, {Wilson}, {Wolfire}, \& {Zibetti}}]{kennicutt2011}
{Kennicutt}, R.~C., {Calzetti}, D., {Aniano}, G., {et~al.} 2011, \pasp, 123,
  1347, \dodoi{10.1086/663818}

\bibitem[{{Kochanek} {et~al.}(2012){Kochanek}, {Eisenstein}, {Cool},
  {Caldwell}, {Assef}, {Jannuzi}, {Jones}, {Murray}, {Forman}, {Dey}, {Brown},
  {Eisenhardt}, {Gonzalez}, {Green}, \& {Stern}}]{kochanek2012}
{Kochanek}, C.~S., {Eisenstein}, D.~J., {Cool}, R.~J., {et~al.} 2012, \apjs,
  200, 8, \dodoi{10.1088/0067-0049/200/1/8}

\bibitem[{{Kodra} {et~al.}(2022){Kodra}, {Andrews}, {Newman}, {Finkelstein},
  {Fontana}, {Hathi}, {Salvato}, {Wiklind}, {Wuyts}, {Broussard}, {Chartab},
  {Conselice}, {Cooper}, {Dekel}, {Dickinson}, {Ferguson}, {Gawiser}, {Grogin},
  {Iyer}, {Kartaltepe}, {Kassin}, {Koekemoer}, {Koo}, {Lucas}, {Mantha},
  {McIntosh}, {Mobasher}, {Pacifici}, {P{\'e}rez-Gonz{\'a}lez}, \&
  {Santini}}]{kodra2022}
{Kodra}, D., {Andrews}, B.~H., {Newman}, J.~A., {et~al.} 2022, arXiv e-prints,
  arXiv:2210.01140.
\newblock \doarXiv{2210.01140}

\bibitem[{{Koekemoer} {et~al.}(2011){Koekemoer}, {Faber}, {Ferguson}, {Grogin},
  {Kocevski}, {Koo}, {Lai}, {Lotz}, {Lucas}, {McGrath}, {Ogaz}, {Rajan},
  {Riess}, {Rodney}, {Strolger}, {Casertano}, {Castellano}, {Dahlen},
  {Dickinson}, {Dolch}, {Fontana}, {Giavalisco}, {Grazian}, {Guo}, {Hathi},
  {Huang}, {van der Wel}, {Yan}, {Acquaviva}, {Alexander}, {Almaini}, {Ashby},
  {Barden}, {Bell}, {Bournaud}, {Brown}, {Caputi}, {Cassata}, {Challis},
  {Chary}, {Cheung}, {Cirasuolo}, {Conselice}, {Roshan Cooray}, {Croton},
  {Daddi}, {Dav{\'e}}, {de Mello}, {de Ravel}, {Dekel}, {Donley}, {Dunlop},
  {Dutton}, {Elbaz}, {Fazio}, {Filippenko}, {Finkelstein}, {Frazer}, {Gardner},
  {Garnavich}, {Gawiser}, {Gruetzbauch}, {Hartley}, {H{\"a}ussler},
  {Herrington}, {Hopkins}, {Huang}, {Jha}, {Johnson}, {Kartaltepe},
  {Khostovan}, {Kirshner}, {Lani}, {Lee}, {Li}, {Madau}, {McCarthy},
  {McIntosh}, {McLure}, {McPartland}, {Mobasher}, {Moreira}, {Mortlock},
  {Moustakas}, {Mozena}, {Nandra}, {Newman}, {Nielsen}, {Niemi}, {Noeske},
  {Papovich}, {Pentericci}, {Pope}, {Primack}, {Ravindranath}, {Reddy},
  {Renzini}, {Rix}, {Robaina}, {Rosario}, {Rosati}, {Salimbeni}, {Scarlata},
  {Siana}, {Simard}, {Smidt}, {Snyder}, {Somerville}, {Spinrad}, {Straughn},
  {Telford}, {Teplitz}, {Trump}, {Vargas}, {Villforth}, {Wagner}, {Wandro},
  {Wechsler}, {Weiner}, {Wiklind}, {Wild}, {Wilson}, {Wuyts}, \&
  {Yun}}]{koekemoer2011}
{Koekemoer}, A.~M., {Faber}, S.~M., {Ferguson}, H.~C., {et~al.} 2011, \apjs,
  197, 36, \dodoi{10.1088/0067-0049/197/2/36}

\bibitem[{{Kriek} \& {Conroy}(2013)}]{kriek2013}
{Kriek}, M., \& {Conroy}, C. 2013, \apjl, 775, L16,
  \dodoi{10.1088/2041-8205/775/1/L16}

\bibitem[{{Kriek} {et~al.}(2015){Kriek}, {Shapley}, {Reddy}, {Siana}, {Coil},
  {Mobasher}, {Freeman}, {de Groot}, {Price}, {Sanders}, {Shivaei}, {Brammer},
  {Momcheva}, {Skelton}, {van Dokkum}, {Whitaker}, {Aird}, {Azadi}, {Kassis},
  {Bullock}, {Conroy}, {Dav{\'e}}, {Kere{\v{s}}}, \& {Krumholz}}]{kriek2015}
{Kriek}, M., {Shapley}, A.~E., {Reddy}, N.~A., {et~al.} 2015, \apjs, 218, 15,
  \dodoi{10.1088/0067-0049/218/2/15}

\bibitem[{{Lacy} {et~al.}(2007){Lacy}, {Petric}, {Sajina}, {Canalizo},
  {Storrie-Lombardi}, {Armus}, {Fadda}, \& {Marleau}}]{lacy2007}
{Lacy}, M., {Petric}, A.~O., {Sajina}, A., {et~al.} 2007, \aj, 133, 186,
  \dodoi{10.1086/509617}

\bibitem[{{Lacy} {et~al.}(2004){Lacy}, {Storrie-Lombardi}, {Sajina},
  {Appleton}, {Armus}, {Chapman}, {Choi}, {Fadda}, {Fang}, {Frayer},
  {Heinrichsen}, {Helou}, {Im}, {Marleau}, {Masci}, {Shupe}, {Soifer},
  {Surace}, {Teplitz}, {Wilson}, \& {Yan}}]{lacy2004}
{Lacy}, M., {Storrie-Lombardi}, L.~J., {Sajina}, A., {et~al.} 2004, \apjs, 154,
  166, \dodoi{10.1086/422816}

\bibitem[{{Lee} {et~al.}(2018){Lee}, {Giavalisco}, {Whitaker}, {Williams},
  {Ferguson}, {Acquaviva}, {Koekemoer}, {Straughn}, {Guo}, {Kartaltepe},
  {Lotz}, {Pacifici}, {Croton}, {Somerville}, \& {Lu}}]{lee2018}
{Lee}, B., {Giavalisco}, M., {Whitaker}, K., {et~al.} 2018, \apj, 853, 131,
  \dodoi{10.3847/1538-4357/aaa40f}

\bibitem[{{Leitherer} \& {Heckman}(1995)}]{leitherer1995}
{Leitherer}, C., \& {Heckman}, T.~M. 1995, \apjs, 96, 9, \dodoi{10.1086/192112}

\bibitem[{{Leja} {et~al.}(2019){Leja}, {Carnall}, {Johnson}, {Conroy}, \&
  {Speagle}}]{leja2019}
{Leja}, J., {Carnall}, A.~C., {Johnson}, B.~D., {Conroy}, C., \& {Speagle},
  J.~S. 2019, \apj, 876, 3, \dodoi{10.3847/1538-4357/ab133c}

\bibitem[{{Leja} {et~al.}(2017){Leja}, {Johnson}, {Conroy}, {van Dokkum}, \&
  {Byler}}]{leja2017}
{Leja}, J., {Johnson}, B.~D., {Conroy}, C., {van Dokkum}, P.~G., \& {Byler}, N.
  2017, \apj, 837, 170, \dodoi{10.3847/1538-4357/aa5ffe}

\bibitem[{{Leja} {et~al.}(2021){Leja}, {Speagle}, {Ting}, {Johnson}, {Conroy},
  {Whitaker}, {Nelson}, {van Dokkum}, \& {Franx}}]{leja2021}
{Leja}, J., {Speagle}, J.~S., {Ting}, Y.-S., {et~al.} 2021, arXiv e-prints,
  arXiv:2110.04314.
\newblock \doarXiv{2110.04314}

\bibitem[{{Liu} {et~al.}(2018){Liu}, {Daddi}, {Dickinson}, {Owen}, {Pannella},
  {Sargent}, {B{\'e}thermin}, {Magdis}, {Gao}, {Shu}, {Wang}, {Jin}, \&
  {Inami}}]{liu2018}
{Liu}, D., {Daddi}, E., {Dickinson}, M., {et~al.} 2018, \apj, 853, 172,
  \dodoi{10.3847/1538-4357/aaa600}

\bibitem[{{Lower} {et~al.}(2020){Lower}, {Narayanan}, {Leja}, {Johnson},
  {Conroy}, \& {Dav{\'e}}}]{lower2020}
{Lower}, S., {Narayanan}, D., {Leja}, J., {et~al.} 2020, \apj, 904, 33,
  \dodoi{10.3847/1538-4357/abbfa7}

\bibitem[{{Madau} \& {Dickinson}(2014)}]{madau2014}
{Madau}, P., \& {Dickinson}, M. 2014, \araa, 52, 415,
  \dodoi{10.1146/annurev-astro-081811-125615}

\bibitem[{{Magnelli} {et~al.}(2013){Magnelli}, {Popesso}, {Berta}, {Pozzi},
  {Elbaz}, {Lutz}, {Dickinson}, {Altieri}, {Andreani}, {Aussel},
  {B{\'e}thermin}, {Bongiovanni}, {Cepa}, {Charmandaris}, {Chary}, {Cimatti},
  {Daddi}, {F{\"o}rster Schreiber}, {Genzel}, {Gruppioni}, {Harwit}, {Hwang},
  {Ivison}, {Magdis}, {Maiolino}, {Murphy}, {Nordon}, {Pannella}, {P{\'e}rez
  Garc{\'\i}a}, {Poglitsch}, {Rosario}, {Sanchez-Portal}, {Santini}, {Scott},
  {Sturm}, {Tacconi}, \& {Valtchanov}}]{magnelli2013}
{Magnelli}, B., {Popesso}, P., {Berta}, S., {et~al.} 2013, \aap, 553, A132,
  \dodoi{10.1051/0004-6361/201321371}

\bibitem[{{Maiolino} {et~al.}(2008){Maiolino}, {Nagao}, {Grazian}, {Cocchia},
  {Marconi}, {Mannucci}, {Cimatti}, {Pipino}, {Ballero}, {Calura}, {Chiappini},
  {Fontana}, {Granato}, {Matteucci}, {Pastorini}, {Pentericci}, {Risaliti},
  {Salvati}, \& {Silva}}]{maiolino2008}
{Maiolino}, R., {Nagao}, T., {Grazian}, A., {et~al.} 2008, \aap, 488, 463,
  \dodoi{10.1051/0004-6361:200809678}

\bibitem[{{Ma{\l}ek} {et~al.}(2018){Ma{\l}ek}, {Buat}, {Roehlly}, {Burgarella},
  {Hurley}, {Shirley}, {Duncan}, {Efstathiou}, {Papadopoulos}, {Vaccari},
  {Farrah}, {Marchetti}, \& {Oliver}}]{malek2018}
{Ma{\l}ek}, K., {Buat}, V., {Roehlly}, Y., {et~al.} 2018, \aap, 620, A50,
  \dodoi{10.1051/0004-6361/201833131}

\bibitem[{{Maraston}(1998)}]{maraston1998}
{Maraston}, C. 1998, \mnras, 300, 872, \dodoi{10.1046/j.1365-8711.1998.01947.x}

\bibitem[{{Matthee} \& {Schaye}(2018)}]{matthee2018}
{Matthee}, J., \& {Schaye}, J. 2018, \mnras, 479, L34,
  \dodoi{10.1093/mnrasl/sly093}

\bibitem[{{McCracken} {et~al.}(2012){McCracken}, {Milvang-Jensen}, {Dunlop},
  {Franx}, {Fynbo}, {Le F{\`e}vre}, {Holt}, {Caputi}, {Goranova}, {Buitrago},
  {Emerson}, {Freudling}, {Hudelot}, {L{\'o}pez-Sanjuan}, {Magnard}, {Mellier},
  {M{\o}ller}, {Nilsson}, {Sutherland}, {Tasca}, \& {Zabl}}]{mccracken2012}
{McCracken}, H.~J., {Milvang-Jensen}, B., {Dunlop}, J., {et~al.} 2012, \aap,
  544, A156, \dodoi{10.1051/0004-6361/201219507}

\bibitem[{{McLure} {et~al.}(2018){McLure}, {Pentericci}, {Cimatti}, {Dunlop},
  {Elbaz}, {Fontana}, {Nandra}, {Amorin}, {Bolzonella}, {Bongiorno}, {Carnall},
  {Castellano}, {Cirasuolo}, {Cucciati}, {Cullen}, {De Barros}, {Finkelstein},
  {Fontanot}, {Franzetti}, {Fumana}, {Gargiulo}, {Garilli}, {Guaita},
  {Hartley}, {Iovino}, {Jarvis}, {Juneau}, {Karman}, {Maccagni}, {Marchi},
  {M{\'a}rmol-Queralt{\'o}}, {Pompei}, {Pozzetti}, {Scodeggio}, {Sommariva},
  {Talia}, {Almaini}, {Balestra}, {Bardelli}, {Bell}, {Bourne}, {Bowler},
  {Brusa}, {Buitrago}, {Caputi}, {Cassata}, {Charlot}, {Citro}, {Cresci},
  {Cristiani}, {Curtis-Lake}, {Dickinson}, {Fazio}, {Ferguson}, {Fiore},
  {Franco}, {Fynbo}, {Galametz}, {Georgakakis}, {Giavalisco}, {Grazian},
  {Hathi}, {Jung}, {Kim}, {Koekemoer}, {Khusanova}, {Le F{\`e}vre}, {Lotz},
  {Mannucci}, {Maltby}, {Matsuoka}, {McLeod}, {Mendez-Hernandez},
  {Mendez-Abreu}, {Mignoli}, {Moresco}, {Mortlock}, {Nonino}, {Pannella},
  {Papovich}, {Popesso}, {Rosario}, {Salvato}, {Santini}, {Schaerer},
  {Schreiber}, {Stark}, {Tasca}, {Thomas}, {Treu}, {Vanzella}, {Wild},
  {Williams}, {Zamorani}, \& {Zucca}}]{mclure2018}
{McLure}, R.~J., {Pentericci}, L., {Cimatti}, A., {et~al.} 2018, \mnras, 479,
  25, \dodoi{10.1093/mnras/sty1213}

\bibitem[{{Merlin} {et~al.}(2019){Merlin}, {Fortuni}, {Torelli}, {Santini},
  {Castellano}, {Fontana}, {Grazian}, {Pentericci}, {Pilo}, \&
  {Schmidt}}]{merlin2019}
{Merlin}, E., {Fortuni}, F., {Torelli}, M., {et~al.} 2019, \mnras, 490, 3309,
  \dodoi{10.1093/mnras/stz2615}

\bibitem[{{Momcheva} {et~al.}(2016){Momcheva}, {Brammer}, {van Dokkum},
  {Skelton}, {Whitaker}, {Nelson}, {Fumagalli}, {Maseda}, {Leja}, {Franx},
  {Rix}, {Bezanson}, {Da Cunha}, {Dickey}, {F{\"o}rster Schreiber},
  {Illingworth}, {Kriek}, {Labb{\'e}}, {Ulf Lange}, {Lundgren}, {Magee},
  {Marchesini}, {Oesch}, {Pacifici}, {Patel}, {Price}, {Tal}, {Wake}, {van der
  Wel}, \& {Wuyts}}]{momcheva2016}
{Momcheva}, I.~G., {Brammer}, G.~B., {van Dokkum}, P.~G., {et~al.} 2016, \apjs,
  225, 27, \dodoi{10.3847/0067-0049/225/2/27}

\bibitem[{{Muzzin} {et~al.}(2013){Muzzin}, {Marchesini}, {Stefanon}, {Franx},
  {Milvang-Jensen}, {Dunlop}, {Fynbo}, {Brammer}, {Labb{\'e}}, \& {van
  Dokkum}}]{muzzin2013}
{Muzzin}, A., {Marchesini}, D., {Stefanon}, M., {et~al.} 2013, \apjs, 206, 8,
  \dodoi{10.1088/0067-0049/206/1/8}

\bibitem[{{Newman} {et~al.}(2013){Newman}, {Cooper}, {Davis}, {Faber}, {Coil},
  {Guhathakurta}, {Koo}, {Phillips}, {Conroy}, {Dutton}, {Finkbeiner}, {Gerke},
  {Rosario}, {Weiner}, {Willmer}, {Yan}, {Harker}, {Kassin}, {Konidaris},
  {Lai}, {Madgwick}, {Noeske}, {Wirth}, {Connolly}, {Kaiser}, {Kirby},
  {Lemaux}, {Lin}, {Lotz}, {Luppino}, {Marinoni}, {Matthews}, {Metevier}, \&
  {Schiavon}}]{newman2013}
{Newman}, J.~A., {Cooper}, M.~C., {Davis}, M., {et~al.} 2013, \apjs, 208, 5,
  \dodoi{10.1088/0067-0049/208/1/5}

\bibitem[{{Noeske} {et~al.}(2007){Noeske}, {Weiner}, {Faber}, {Papovich},
  {Koo}, {Somerville}, {Bundy}, {Conselice}, {Newman}, {Schiminovich}, {Le
  Floc'h}, {Coil}, {Rieke}, {Lotz}, {Primack}, {Barmby}, {Cooper}, {Davis},
  {Ellis}, {Fazio}, {Guhathakurta}, {Huang}, {Kassin}, {Martin}, {Phillips},
  {Rich}, {Small}, {Willmer}, \& {Wilson}}]{noeske2007}
{Noeske}, K.~G., {Weiner}, B.~J., {Faber}, S.~M., {et~al.} 2007, \apjl, 660,
  L43, \dodoi{10.1086/517926}

\bibitem[{{Ocvirk} {et~al.}(2006){Ocvirk}, {Pichon}, {Lan{\c{c}}on}, \&
  {Thi{\'e}baut}}]{ocvirk2006}
{Ocvirk}, P., {Pichon}, C., {Lan{\c{c}}on}, A., \& {Thi{\'e}baut}, E. 2006,
  \mnras, 365, 46, \dodoi{10.1111/j.1365-2966.2005.09182.x}

\bibitem[{{Pacifici} {et~al.}(2012){Pacifici}, {Charlot}, {Blaizot}, \&
  {Brinchmann}}]{pacifici2012}
{Pacifici}, C., {Charlot}, S., {Blaizot}, J., \& {Brinchmann}, J. 2012, \mnras,
  421, 2002, \dodoi{10.1111/j.1365-2966.2012.20431.x}

\bibitem[{{Pacifici} {et~al.}(2013){Pacifici}, {Kassin}, {Weiner}, {Charlot},
  \& {Gardner}}]{pacifici2013}
{Pacifici}, C., {Kassin}, S.~A., {Weiner}, B., {Charlot}, S., \& {Gardner},
  J.~P. 2013, \apjl, 762, L15, \dodoi{10.1088/2041-8205/762/1/L15}

\bibitem[{{Pacifici} {et~al.}(2015){Pacifici}, {da Cunha}, {Charlot}, {Rix},
  {Fumagalli}, {Wel}, {Franx}, {Maseda}, {van Dokkum}, {Brammer}, {Momcheva},
  {Skelton}, {Whitaker}, {Leja}, {Lundgren}, {Kassin}, \& {Yi}}]{pacifici2015}
{Pacifici}, C., {da Cunha}, E., {Charlot}, S., {et~al.} 2015, \mnras, 447, 786,
  \dodoi{10.1093/mnras/stu2447}

\bibitem[{pandas~development team(2020)}]{pandas2020}
pandas~development team, T. 2020, pandas-dev/pandas: Pandas, latest,  Zenodo,
  \dodoi{10.5281/zenodo.3509134}

\bibitem[{{Pandya} {et~al.}(2017){Pandya}, {Brennan}, {Somerville}, {Choi},
  {Barro}, {Wuyts}, {Taylor}, {Behroozi}, {Kirkpatrick}, {Faber}, {Primack},
  {Koo}, {McIntosh}, {Kocevski}, {Bell}, {Dekel}, {Fang}, {Ferguson}, {Grogin},
  {Koekemoer}, {Lu}, {Mantha}, {Mobasher}, {Newman}, {Pacifici}, {Papovich},
  {van der Wel}, \& {Yesuf}}]{pandya2017}
{Pandya}, V., {Brennan}, R., {Somerville}, R.~S., {et~al.} 2017, \mnras, 472,
  2054, \dodoi{10.1093/mnras/stx2027}

\bibitem[{{Papovich} {et~al.}(2001){Papovich}, {Dickinson}, \&
  {Ferguson}}]{papovich2001}
{Papovich}, C., {Dickinson}, M., \& {Ferguson}, H.~C. 2001, \apj, 559, 620,
  \dodoi{10.1086/322412}

\bibitem[{{Pappalardo} {et~al.}(2021){Pappalardo}, {Bendo}, {Boquien}, {Baes},
  {Viaene}, {Bianchi}, \& {Fritz}}]{pappalardo2021}
{Pappalardo}, C., {Bendo}, G.~J., {Boquien}, M., {et~al.} 2021, \aap, 655,
  A104, \dodoi{10.1051/0004-6361/202141678}

\bibitem[{{Pentericci} {et~al.}(2018){Pentericci}, {McLure}, {Garilli},
  {Cucciati}, {Franzetti}, {Iovino}, {Amorin}, {Bolzonella}, {Bongiorno},
  {Carnall}, {Castellano}, {Cimatti}, {Cirasuolo}, {Cullen}, {De Barros},
  {Dunlop}, {Elbaz}, {Finkelstein}, {Fontana}, {Fontanot}, {Fumana},
  {Gargiulo}, {Guaita}, {Hartley}, {Jarvis}, {Juneau}, {Karman}, {Maccagni},
  {Marchi}, {Marmol-Queralto}, {Nandra}, {Pompei}, {Pozzetti}, {Scodeggio},
  {Sommariva}, {Talia}, {Almaini}, {Balestra}, {Bardelli}, {Bell}, {Bourne},
  {Bowler}, {Brusa}, {Buitrago}, {Caputi}, {Cassata}, {Charlot}, {Citro},
  {Cresci}, {Cristiani}, {Curtis-Lake}, {Dickinson}, {Fazio}, {Ferguson},
  {Fiore}, {Franco}, {Fynbo}, {Galametz}, {Georgakakis}, {Giavalisco},
  {Grazian}, {Hathi}, {Jung}, {Kim}, {Koekemoer}, {Khusanova}, {Le F{\`e}vre},
  {Lotz}, {Mannucci}, {Maltby}, {Matsuoka}, {McLeod}, {Mendez-Hernandez},
  {Mendez-Abreu}, {Mignoli}, {Moresco}, {Mortlock}, {Nonino}, {Pannella},
  {Papovich}, {Popesso}, {Rosario}, {Salvato}, {Santini}, {Schaerer},
  {Schreiber}, {Stark}, {Tasca}, {Thomas}, {Treu}, {Vanzella}, {Wild},
  {Williams}, {Zamorani}, \& {Zucca}}]{pentericci2018}
{Pentericci}, L., {McLure}, R.~J., {Garilli}, B., {et~al.} 2018, \aap, 616,
  A174, \dodoi{10.1051/0004-6361/201833047}

\bibitem[{{Popesso} {et~al.}(2019{\natexlab{a}}){Popesso}, {Concas},
  {Morselli}, {Schreiber}, {Rodighiero}, {Cresci}, {Belli}, {Erfanianfar},
  {Mancini}, {Inami}, {Dickinson}, {Ilbert}, {Pannella}, \&
  {Elbaz}}]{popesso2019a}
{Popesso}, P., {Concas}, A., {Morselli}, L., {et~al.} 2019{\natexlab{a}},
  \mnras, 483, 3213, \dodoi{10.1093/mnras/sty3210}

\bibitem[{{Popesso} {et~al.}(2019{\natexlab{b}}){Popesso}, {Morselli},
  {Concas}, {Schreiber}, {Rodighiero}, {Cresci}, {Belli}, {Ilbert},
  {Erfanianfar}, {Mancini}, {Inami}, {Dickinson}, {Pannella}, \&
  {Elbaz}}]{popesso2019b}
{Popesso}, P., {Morselli}, L., {Concas}, A., {et~al.} 2019{\natexlab{b}},
  \mnras, 490, 5285, \dodoi{10.1093/mnras/stz2635}

\bibitem[{{Qiu} \& {Kang}(2021)}]{qui2021}
{Qiu}, Y., \& {Kang}, X. 2021, arXiv e-prints, arXiv:2112.14434.
\newblock \doarXiv{2112.14434}

\bibitem[{{Rosario}(2019)}]{rosario2019}
{Rosario}, D.~J. 2019, {FortesFit: Flexible spectral energy distribution
  modelling with a Bayesian backbone}.
\newblock \doeprint{1904.011}

\bibitem[{{Rujopakarn} {et~al.}(2013){Rujopakarn}, {Rieke}, {Weiner},
  {P{\'e}rez-Gonz{\'a}lez}, {Rex}, {Walth}, \& {Kartaltepe}}]{rujopakarn2013}
{Rujopakarn}, W., {Rieke}, G.~H., {Weiner}, B.~J., {et~al.} 2013, \apj, 767,
  73, \dodoi{10.1088/0004-637X/767/1/73}

\bibitem[{{Salim} {et~al.}(2018){Salim}, {Boquien}, \& {Lee}}]{salim2018}
{Salim}, S., {Boquien}, M., \& {Lee}, J.~C. 2018, \apj, 859, 11,
  \dodoi{10.3847/1538-4357/aabf3c}

\bibitem[{{Salim} \& {Narayanan}(2020)}]{salim2020}
{Salim}, S., \& {Narayanan}, D. 2020, \araa, 58, 529,
  \dodoi{10.1146/annurev-astro-032620-021933}

\bibitem[{{Salmon} {et~al.}(2015){Salmon}, {Papovich}, {Finkelstein}, {Tilvi},
  {Finlator}, {Behroozi}, {Dahlen}, {Dav{\'e}}, {Dekel}, {Dickinson},
  {Ferguson}, {Giavalisco}, {Long}, {Lu}, {Mobasher}, {Reddy}, {Somerville}, \&
  {Wechsler}}]{salmon2015}
{Salmon}, B., {Papovich}, C., {Finkelstein}, S.~L., {et~al.} 2015, \apj, 799,
  \dodoi{10.1088/0004-637X/799/2/183}

\bibitem[{{Salmon} {et~al.}(2016){Salmon}, {Papovich}, {Long}, {Willner},
  {Finkelstein}, {Ferguson}, {Dickinson}, {Duncan}, {Faber}, {Hathi},
  {Koekemoer}, {Kurczynski}, {Newman}, {Pacifici}, {P{\'e}rez- Gonz{\'a}lez},
  \& {Pforr}}]{salmon2016}
{Salmon}, B., {Papovich}, C., {Long}, J., {et~al.} 2016, \apj, 827,
  \dodoi{10.3847/0004-637X/827/1/20}

\bibitem[{{S{\'a}nchez} {et~al.}(2016){S{\'a}nchez}, {P{\'e}rez},
  {S{\'a}nchez-Bl{\'a}zquez}, {Gonz{\'a}lez}, {Ros{\'a}les-Ortega},
  {Cano-D{\'\i}az}, {L{\'o}pez-Cob{\'a}}, {Marino}, {Gil de Paz}, {Moll{\'a}},
  {L{\'o}pez-S{\'a}nchez}, {Ascasibar}, \& {Barrera-Ballesteros}}]{sanchez2016}
{S{\'a}nchez}, S.~F., {P{\'e}rez}, E., {S{\'a}nchez-Bl{\'a}zquez}, P., {et~al.}
  2016, \rmxaa, 52, 21.
\newblock \doarXiv{1509.08552}

\bibitem[{{Santini} {et~al.}(2015){Santini}, {Ferguson}, {Fontana}, {Mobasher},
  {Barro}, {Castellano}, {Finkelstein}, {Grazian}, {Hsu}, {Lee}, {Lee},
  {Pforr}, {Salvato}, {Wiklind}, {Wuyts}, {Almaini}, {Cooper}, {Galametz},
  {Weiner}, {Amorin}, {Boutsia}, {Conselice}, {Dahlen}, {Dickinson},
  {Giavalisco}, {Grogin}, {Guo}, {Hathi}, {Kocevski}, {Koekemoer},
  {Kurczynski}, {Merlin}, {Mortlock}, {Newman}, {Paris}, {Pentericci},
  {Simons}, \& {Willner}}]{santini2015}
{Santini}, P., {Ferguson}, H.~C., {Fontana}, A., {et~al.} 2015, \apj, 801, 97,
  \dodoi{10.1088/0004-637X/801/2/97}

\bibitem[{{Schreiber} {et~al.}(2015){Schreiber}, {Pannella}, {Elbaz},
  {B{\'e}thermin}, {Inami}, {Dickinson}, {Magnelli}, {Wang}, {Aussel}, {Daddi},
  {Juneau}, {Shu}, {Sargent}, {Buat}, {Faber}, {Ferguson}, {Giavalisco},
  {Koekemoer}, {Magdis}, {Morrison}, {Papovich}, {Santini}, \&
  {Scott}}]{schreiber2015}
{Schreiber}, C., {Pannella}, M., {Elbaz}, D., {et~al.} 2015, \aap, 575, A74,
  \dodoi{10.1051/0004-6361/201425017}

\bibitem[{{Scodeggio} {et~al.}(2018){Scodeggio}, {Guzzo}, {Garilli}, {Granett},
  {Bolzonella}, {de la Torre}, {Abbas}, {Adami}, {Arnouts}, {Bottini}, {Cappi},
  {Coupon}, {Cucciati}, {Davidzon}, {Franzetti}, {Fritz}, {Iovino}, {Krywult},
  {Le Brun}, {Le F{\`e}vre}, {Maccagni}, {Ma{\l}ek}, {Marchetti}, {Marulli},
  {Polletta}, {Pollo}, {Tasca}, {Tojeiro}, {Vergani}, {Zanichelli}, {Bel},
  {Branchini}, {De Lucia}, {Ilbert}, {McCracken}, {Moutard}, {Peacock},
  {Zamorani}, {Burden}, {Fumana}, {Jullo}, {Marinoni}, {Mellier}, {Moscardini},
  \& {Percival}}]{scodeggio2018}
{Scodeggio}, M., {Guzzo}, L., {Garilli}, B., {et~al.} 2018, \aap, 609, A84,
  \dodoi{10.1051/0004-6361/201630114}

\bibitem[{{Scoville} {et~al.}(2007){Scoville}, {Aussel}, {Brusa}, {Capak},
  {Carollo}, {Elvis}, {Giavalisco}, {Guzzo}, {Hasinger}, {Impey}, {Kneib},
  {LeFevre}, {Lilly}, {Mobasher}, {Renzini}, {Rich}, {Sanders}, {Schinnerer},
  {Schminovich}, {Shopbell}, {Taniguchi}, \& {Tyson}}]{scoville2007}
{Scoville}, N., {Aussel}, H., {Brusa}, M., {et~al.} 2007, \apjs, 172, 1,
  \dodoi{10.1086/516585}

\bibitem[{Seabold \& Perktold(2010)}]{seabold2010}
Seabold, S., \& Perktold, J. 2010, in 9th Python in Science Conference

\bibitem[{{Simet} {et~al.}(2021){Simet}, {Chartab}, {Lu}, \&
  {Mobasher}}]{simet2021}
{Simet}, M., {Chartab}, N., {Lu}, Y., \& {Mobasher}, B. 2021, \apj, 908, 47,
  \dodoi{10.3847/1538-4357/abd179}

\bibitem[{Skilling(2004)}]{skilling2004}
Skilling, J. 2004in , American Institute of Physics, 395--405

\bibitem[{Skilling(2006)}]{skilling2006}
Skilling, J. 2006, Bayesian analysis, 1, 833

\bibitem[{{Smit} {et~al.}(2014){Smit}, {Bouwens}, {Labb{\'e}}, {Zheng},
  {Bradley}, {Donahue}, {Lemze}, {Moustakas}, {Umetsu}, {Zitrin}, {Coe},
  {Postman}, {Gonzalez}, {Bartelmann}, {Ben{\'\i}tez}, {Broadhurst}, {Ford},
  {Grillo}, {Infante}, {Jimenez-Teja}, {Jouvel}, {Kelson}, {Lahav}, {Maoz},
  {Medezinski}, {Melchior}, {Meneghetti}, {Merten}, {Molino}, {Moustakas},
  {Nonino}, {Rosati}, \& {Seitz}}]{smit2014}
{Smit}, R., {Bouwens}, R.~J., {Labb{\'e}}, I., {et~al.} 2014, \apj, 784, 58,
  \dodoi{10.1088/0004-637X/784/1/58}

\bibitem[{{Speagle} {et~al.}(2014){Speagle}, {Steinhardt}, {Capak}, \&
  {Silverman}}]{speagle2014}
{Speagle}, J.~S., {Steinhardt}, C.~L., {Capak}, P.~L., \& {Silverman}, J.~D.
  2014, \apjs, 214, 15, \dodoi{10.1088/0067-0049/214/2/15}

\bibitem[{{Spinrad} \& {Taylor}(1969)}]{spinrad1969}
{Spinrad}, H., \& {Taylor}, B.~J. 1969, \apj, 157, 1279, \dodoi{10.1086/150154}

\bibitem[{{Steidel} {et~al.}(1996){Steidel}, {Giavalisco}, {Pettini},
  {Dickinson}, \& {Adelberger}}]{steidel1996}
{Steidel}, C.~C., {Giavalisco}, M., {Pettini}, M., {Dickinson}, M., \&
  {Adelberger}, K.~L. 1996, \apjl, 462, L17, \dodoi{10.1086/310029}

\bibitem[{{Straatman} {et~al.}(2016){Straatman}, {Spitler}, {Quadri},
  {Labb{\'e}}, {Glazebrook}, {Persson}, {Papovich}, {Tran}, {Brammer},
  {Cowley}, {Tomczak}, {Nanayakkara}, {Alcorn}, {Allen}, {Broussard}, {van
  Dokkum}, {Forrest}, {van Houdt}, {Kacprzak}, {Kawinwanichakij}, {Kelson},
  {Lee}, {McCarthy}, {Mehrtens}, {Monson}, {Murphy}, {Rees}, {Tilvi}, \&
  {Whitaker}}]{straatman2016}
{Straatman}, C. M.~S., {Spitler}, L.~R., {Quadri}, R.~F., {et~al.} 2016, \apj,
  830, 51, \dodoi{10.3847/0004-637X/830/1/51}

\bibitem[{{Tacchella} {et~al.}(2016){Tacchella}, {Dekel}, {Carollo},
  {Ceverino}, {DeGraf}, {Lapiner}, {Mandelker}, \& {Primack
  Joel}}]{tacchella2016}
{Tacchella}, S., {Dekel}, A., {Carollo}, C.~M., {et~al.} 2016, \mnras, 457,
  2790, \dodoi{10.1093/mnras/stw131}

\bibitem[{{Thorne} {et~al.}(2021){Thorne}, {Robotham}, {Davies}, {Bellstedt},
  {Driver}, {Bravo}, {Bremer}, {Holwerda}, {Hopkins}, {Lagos}, {Phillipps},
  {Siudek}, {Taylor}, \& {Wright}}]{thorne2021}
{Thorne}, J.~E., {Robotham}, A. S.~G., {Davies}, L. J.~M., {et~al.} 2021,
  \mnras, 505, 540, \dodoi{10.1093/mnras/stab1294}

\bibitem[{{van der Wel} {et~al.}(2016){van der Wel}, {Noeske}, {Bezanson},
  {Pacifici}, {Gallazzi}, {Franx}, {Mu{\~n}oz-Mateos}, {Bell}, {Brammer},
  {Charlot}, {Chauk{\'e}}, {Labb{\'e}}, {Maseda}, {Muzzin}, {Rix}, {Sobral},
  {van de Sande}, {van Dokkum}, {Wild}, \& {Wolf}}]{vanderwel2016}
{van der Wel}, A., {Noeske}, K., {Bezanson}, R., {et~al.} 2016, \apjs, 223, 29,
  \dodoi{10.3847/0067-0049/223/2/29}

\bibitem[{{Vazdekis}(1999)}]{vazdekis1999}
{Vazdekis}, A. 1999, \apj, 513, 224, \dodoi{10.1086/306843}

\bibitem[{{Wang} {et~al.}(2018){Wang}, {Kassin}, {Pacifici}, {Barro}, {de la
  Vega}, {Simons}, {Faber}, {Salmon}, {Ferguson}, {P{\'e}rez-Gonz{\'a}lez},
  {Snyder}, {Gordon}, {Chen}, \& {Kodra}}]{wang2018}
{Wang}, W., {Kassin}, S.~A., {Pacifici}, C., {et~al.} 2018, \apj, 869, 161,
  \dodoi{10.3847/1538-4357/aaef79}

\bibitem[{Waskom(2021)}]{waskom2021}
Waskom, M.~L. 2021, Journal of Open Source Software, 6, 3021,
  \dodoi{10.21105/joss.03021}

\bibitem[{{Weaver} {et~al.}(2022){Weaver}, {Kauffmann}, {Ilbert}, {McCracken},
  {Moneti}, {Toft}, {Brammer}, {Shuntov}, {Davidzon}, {Hsieh}, {Laigle},
  {Anastasiou}, {Jespersen}, {Vinther}, {Capak}, {Casey}, {McPartland},
  {Milvang-Jensen}, {Mobasher}, {Sanders}, {Zalesky}, {Arnouts}, {Aussel},
  {Dunlop}, {Faisst}, {Franx}, {Furtak}, {Fynbo}, {Gould}, {Greve}, {Gwyn},
  {Kartaltepe}, {Kashino}, {Koekemoer}, {Kokorev}, {Le F{\`e}vre}, {Lilly},
  {Masters}, {Magdis}, {Mehta}, {Peng}, {Riechers}, {Salvato}, {Sawicki},
  {Scarlata}, {Scoville}, {Shirley}, {Silverman}, {Sneppen}, {Smolc?i{\'c}},
  {Steinhardt}, {Stern}, {Tanaka}, {Taniguchi}, {Teplitz}, {Vaccari}, {Wang},
  \& {Zamorani}}]{weaver2022}
{Weaver}, J.~R., {Kauffmann}, O.~B., {Ilbert}, O., {et~al.} 2022, \apjs, 258,
  11, \dodoi{10.3847/1538-4365/ac3078}

\bibitem[{{Whitaker} {et~al.}(2012){Whitaker}, {van Dokkum}, {Brammer}, \&
  {Franx}}]{whitaker2012}
{Whitaker}, K.~E., {van Dokkum}, P.~G., {Brammer}, G., \& {Franx}, M. 2012,
  \apjl, 754, L29, \dodoi{10.1088/2041-8205/754/2/L29}

\bibitem[{{Whitaker} {et~al.}(2014){Whitaker}, {Franx}, {Leja}, {van Dokkum},
  {Henry}, {Skelton}, {Fumagalli}, {Momcheva}, {Brammer}, {Labb{\'e}},
  {Nelson}, \& {Rigby}}]{whitaker2014}
{Whitaker}, K.~E., {Franx}, M., {Leja}, J., {et~al.} 2014, \apj, 795, 104,
  \dodoi{10.1088/0004-637X/795/2/104}

\bibitem[{{Wilkinson} {et~al.}(2017){Wilkinson}, {Maraston}, {Goddard},
  {Thomas}, \& {Parikh}}]{wilkinson2017}
{Wilkinson}, D.~M., {Maraston}, C., {Goddard}, D., {Thomas}, D., \& {Parikh},
  T. 2017, \mnras, 472, 4297, \dodoi{10.1093/mnras/stx2215}

\bibitem[{{Williams} {et~al.}(2018){Williams}, {Curtis-Lake}, {Hainline},
  {Chevallard}, {Robertson}, {Charlot}, {Endsley}, {Stark}, {Willmer},
  {Alberts}, {Amorin}, {Arribas}, {Baum}, {Bunker}, {Carniani}, {Crandall},
  {Egami}, {Eisenstein}, {Ferruit}, {Husemann}, {Maseda}, {Maiolino}, {Rawle},
  {Rieke}, {Smit}, {Tacchella}, \& {Willott}}]{williams2018}
{Williams}, C.~C., {Curtis-Lake}, E., {Hainline}, K.~N., {et~al.} 2018, The
  Astrophysical Journal Supplement Series, 236, 33,
  \dodoi{10.3847/1538-4365/aabcbb}

\bibitem[{{Worthey}(1994)}]{worthey1994}
{Worthey}, G. 1994, \apjs, 95, 107, \dodoi{10.1086/192096}

\bibitem[{{Yang} {et~al.}(2020){Yang}, {Boquien}, {Buat}, {Burgarella},
  {Ciesla}, {Duras}, {Stalevski}, {Brandt}, \& {Papovich}}]{yang2020}
{Yang}, G., {Boquien}, M., {Buat}, V., {et~al.} 2020, \mnras, 491, 740,
  \dodoi{10.1093/mnras/stz3001}

\bibitem[{{Yang} {et~al.}(2022){Yang}, {Boquien}, {Brandt}, {Buat},
  {Burgarella}, {Ciesla}, {Lehmer}, {Ma{\l}ek}, {Mountrichas}, {Papovich},
  {Pons}, {Stalevski}, {Theul{\'e}}, \& {Zhu}}]{yang2022}
{Yang}, G., {Boquien}, M., {Brandt}, W.~N., {et~al.} 2022, arXiv e-prints,
  arXiv:2201.03718.
\newblock \doarXiv{2201.03718}

\bibitem[{{York} {et~al.}(2000){York}, {Adelman}, {Anderson}, {Anderson},
  {Annis}, {Bahcall}, {Bakken}, {Barkhouser}, {Bastian}, {Berman}, {Boroski},
  {Bracker}, {Briegel}, {Briggs}, {Brinkmann}, {Brunner}, {Burles}, {Carey},
  {Carr}, {Castander}, {Chen}, {Colestock}, {Connolly}, {Crocker}, {Csabai},
  {Czarapata}, {Davis}, {Doi}, {Dombeck}, {Eisenstein}, {Ellman}, {Elms},
  {Evans}, {Fan}, {Federwitz}, {Fiscelli}, {Friedman}, {Frieman}, {Fukugita},
  {Gillespie}, {Gunn}, {Gurbani}, {de Haas}, {Haldeman}, {Harris}, {Hayes},
  {Heckman}, {Hennessy}, {Hindsley}, {Holm}, {Holmgren}, {Huang}, {Hull},
  {Husby}, {Ichikawa}, {Ichikawa}, {Ivezi{\'c}}, {Kent}, {Kim}, {Kinney},
  {Klaene}, {Kleinman}, {Kleinman}, {Knapp}, {Korienek}, {Kron}, {Kunszt},
  {Lamb}, {Lee}, {Leger}, {Limmongkol}, {Lindenmeyer}, {Long}, {Loomis},
  {Loveday}, {Lucinio}, {Lupton}, {MacKinnon}, {Mannery}, {Mantsch}, {Margon},
  {McGehee}, {McKay}, {Meiksin}, {Merelli}, {Monet}, {Munn}, {Narayanan},
  {Nash}, {Neilsen}, {Neswold}, {Newberg}, {Nichol}, {Nicinski}, {Nonino},
  {Okada}, {Okamura}, {Ostriker}, {Owen}, {Pauls}, {Peoples}, {Peterson},
  {Petravick}, {Pier}, {Pope}, {Pordes}, {Prosapio}, {Rechenmacher}, {Quinn},
  {Richards}, {Richmond}, {Rivetta}, {Rockosi}, {Ruthmansdorfer}, {Sandford},
  {Schlegel}, {Schneider}, {Sekiguchi}, {Sergey}, {Shimasaku}, {Siegmund},
  {Smee}, {Smith}, {Snedden}, {Stone}, {Stoughton}, {Strauss}, {Stubbs},
  {SubbaRao}, {Szalay}, {Szapudi}, {Szokoly}, {Thakar}, {Tremonti}, {Tucker},
  {Uomoto}, {Vanden Berk}, {Vogeley}, {Waddell}, {Wang}, {Watanabe},
  {Weinberg}, {Yanny}, {Yasuda}, \& {SDSS Collaboration}}]{york2000}
{York}, D.~G., {Adelman}, J., {Anderson}, John~E., J., {et~al.} 2000, \aj, 120,
  1579, \dodoi{10.1086/301513}

\end{thebibliography}
\end{document}